\documentclass[11pt]{article}
\usepackage{etoolbox}
\usepackage[english]{babel}
\usepackage[utf8]{inputenc}
\usepackage{authblk} 
\usepackage{comment}
\usepackage{subcaption}
\usepackage{fancyhdr}
\usepackage{extramarks}
\usepackage{amsfonts}
\usepackage{amsmath}
\usepackage{dsfont}
\usepackage{amsthm}
\usepackage{amssymb}
\usepackage{braket}
\usepackage{mathtools}
\usepackage{physics} 
\usepackage{bbm}

\usepackage{breqn} 
\usepackage[ruled]{algorithm2e}

\usepackage{graphicx}
\usepackage{adjustbox}
\usepackage{hyperref}
\hypersetup{colorlinks=true,citecolor=teal,urlcolor=teal,linkcolor=teal}

\usepackage{tikz}
\usetikzlibrary{positioning} 
\usepackage{blkarray}

\usepackage{enumitem}
\usepackage{bm}
\usepackage{nicefrac}
\usepackage{threeparttable}
\usepackage{natbib}
\usepackage{rotating}
\usepackage{geometry}
\usepackage{textcomp}
\usepackage{upgreek}
\usepackage[nottoc]{tocbibind}
\usepackage{multirow}
\usepackage{tabularx}

\usepackage{tocloft}
\usepackage{placeins}
\usepackage{lipsum,booktabs} 
\usepackage{tabulary}
\usepackage{ragged2e} 
\usepackage{stackengine}
\usepackage{floatrow}
\usepackage{longtable}
\usepackage{wasysym}
\usepackage{tabto}
\usepackage{abstract}
\usepackage{titlesec}
\usepackage[toc,page]{appendix}

\makeatletter
\newsavebox\myboxA
\newsavebox\myboxB
\newlength\mylenA

\newcommand*\xoverline[2][0.75]{%
    \sbox{\myboxA}{$\m@th#2$}%
    \setbox\myboxB\null
    \ht\myboxB=\ht\myboxA%
    \dp\myboxB=\dp\myboxA%
    \wd\myboxB=#1\wd\myboxA
    \sbox\myboxB{$\m@th\overline{\copy\myboxB}$}
    \setlength\mylenA{\the\wd\myboxA}
    \addtolength\mylenA{-\the\wd\myboxB}%
    \ifdim\wd\myboxB<\wd\myboxA%
       \rlap{\hskip 0.5\mylenA\usebox\myboxB}{\usebox\myboxA}%
    \else
        \hskip -0.5\mylenA\rlap{\usebox\myboxA}{\hskip 0.5\mylenA\usebox\myboxB}%
    \fi}
\makeatother

\usepackage[mathlines]{lineno}

\title{Reconstructing firm-level input-output networks from partial information}
\date{\vspace{-6ex}}

\author[1,2,*]{Andrea Bacilieri}
\author[3,4]{Pablo Austudillo-Estevez}
\affil[1]{\footnotesize Institute for New Economic Thinking, University of Oxford, Oxford, OX1 3UQ, UK}
\affil[2]{\footnotesize Smith School of Enterprise and Environment, University of Oxford, Oxford, OX1 3QY, UK}
\affil[3]{\footnotesize School of Geography and the Environment, University of Oxford, Oxford, OX1 3QY, UK}
\affil[4]{\footnotesize Universidad San Francisco de Quito \& School of Economics, Quito, 170902, Ecuador}
\affil[*]{Corresponding author. \textit{E-mail address:} andrea.bacilieri@sant.ox.ac.uk (A. Bacilieri).}

\let\oldequation\equation
\let\oldendequation\endequation

\renewenvironment{equation}
  {\linenomathNonumbers\oldequation}
  {\oldendequation\endlinenomath}

\begin{document}
\floatsetup[table]{capposition=bottom}
\floatsetup[figure]{capposition=bottom}

\maketitle 

\begin{abstract}
\noindent There is a large consensus on the fundamental role of firm-level supply chain networks in macroeconomics. However, data on supply chains at the fine-grained, firm level are scarce and frequently incomplete. For listed firms, some commercial datasets exist but only contain information about the existence of a trade relationship between two companies, not the value of the monetary transaction. We use a recently developed maximum entropy method to reconstruct the values of the transactions based on information about their existence and aggregate information disclosed by firms in financial statements. We test the method on the administrative dataset of Ecuador and reconstruct a commercial dataset (FactSet). We test the method's performance on the weights, the technical and allocation coefficients (microscale quantities), two measures of firms' systemic importance and GDP volatility. The method reconstructs the distribution of microscale quantities reasonably well but shows diverging results for the measures of firms' systemic importance. Due to the network structure of supply chains and the sampling process of firms and links, quantities relying on the number of customers firms have (out-degrees) are harder to reconstruct. We also reconstruct the input-output table of globally listed firms and merge it with a global input-output table at the sector level (the WIOD). Differences in accounting standards between national accounts and firms' financial statements significantly reduce the quality of the reconstruction.

\bigbreak
\bigbreak
\end{abstract}

\newpage
\section{Introduction}
Recent events, such as the war in Ukraine and the evermore frequent natural disasters, have highlighted the fragility of global supply chains. Shocks to individual firms or a cluster of firms, sometimes located in a particular geographical region, can quickly spread through the network with severe repercussions on the global economy. Most of the research has so far been conducted at the sector level \citep{acemoglu2012network,carvalho2014MicroToMacro,pichler2021simultaneous}. However, analysing such shocks at the level of sectors can lead to misleading results \citep{diem2021quantifying}. The value of firm-level data is thus increasingly recognised, but research efforts are constrained by data availability. 

Due to confidentiality and the data collection process, supply chain data are scarce, hard to access, and frequently incomplete. Countries collect the best data sources through VAT filings, but only a handful of countries collect them \citep[for a comprehensive review and discussion of the different datasets][]{bacil2022emprical}. Some datasets of broader global coverage derived from US disclosure requirements are available. Given their wider breadth of coverage and relatively easier access, such datasets are used in many studies \citep[e.g.,][]{wu2016firm,wu2016shock,pankratz2019climate,taschereau2020cascades,boehm2020vertical,barrot2016input,atalay2011networkStructure,kelly2013firm}. In contrast to national datasets, which report the monetary values of the transactions (i.e., the weighted network), global datasets do not provide this valuable information.

Reconstructing firm-level production networks is thus an important topic, showing growing research interest. The reconstruction problem concerns mainly two features of the production network: supplier-customer relations and transaction values. Several studies develop methods to infer both links and transaction values \citep{reisch2021inferring,ialongo2021reconstructing,hooijmaaijers2019methodology,Hillman2021newFirm-levelModel} or links only \citep{brintrup2018predicting,mungo2022reconstructing,kosasih2022machine}, and two focus on inferring weights given the binary topology \citep{inoue2019firm,welburn2020systemic}. The assessment of the quality of the weights reconstruction is usually missing and sometimes carried out on aggregate quantities or compared to empirical facts about another country's network.

In this paper, we focus on reconstructing the transaction values using partial information on the supply chain relations and aggregate information about firms' revenues and expenditures. We provide a rigorous assessment of a recently developed maximum entropy reconstruction method \citep{parisi2020faster} using the administrative dataset of Ecuador. We evaluate the method on microscale, higher-order and macroscale quantities that are widely used in economic input-output (I-O) models. 

We assess the method's performance at recovering the link weights, and the technical and allocation coefficients (i.e., normalised weights). We also use two indicators of firms' systemic importance, the output multipliers and the influence vector, that are prominent in macroeconomic I-O models of shock propagation. While the reconstruction method reproduces the weights (normalised or not) rather poorly, it reconstructs their distributions reasonably well. In contrast, the reconstruction shows diverging results for higher-orders quantities: the output multipliers are in remarkable agreement with the empirical values, while the influence vector is overestimated. We then use a general equilibrium I-O model \citep{acemoglu2012network} to assess how shocks to firms' total factor productivity (TFP) propagate through the network, ultimately affecting aggregate GDP fluctuations. We show that aggregate volatility is overestimated by the reconstruction method we employ. 

Our results suggest that quantities relying more prominently on the number of customers firms have are more adversely affected by missing firms and links due to the structure of supply chain networks \citep{bacil2022emprical} and the sampling process underpinning the observed firms and links. We also find that including a proxy node, to represent the rest of the economy that is not captured by the network, is of help in predicting microscale and higher-order quantities but not for predicting aggregate volatility.

An additional contribution we make in this paper is to construct the I-O table of globally listed firms using the dataset collected by FactSet. We merge FactSet with the World Input-Output Database (WIOD). Key challenges related to differences in accounting standards between national accounts and firms' financial statements prevent us from (1) merging the two datasets at the desired country-sector or even sector level, and (2) carrying out an accurate quantification (at the firm level) of the key variables making up an I-O table. We then reconstruct and compute weights, coefficients and higher-order quantities for FactSet as well. The inability to accurately quantify the variables of the I-O table at the firm level dramatically reduces the quality of the final dataset and thus of the reconstruction. 

The remainder of the paper is organised as follows. In Section~\ref{sec:firm-level_prodNet}, we explain the notation and define the production network at the firm level. In Section~\ref{sec:data}, we discuss the two datasets we use. In Section~\ref{sec:Net_reconstruction}, we review the literature on network reconstruction, briefly describe the maximum entropy method we employ and the metrics we use to assess the performance of the reconstruction method. We then show and discuss the results for microscale, higher-order and macroscale properties for Ecuador and FactSet. Section~\ref{sec:differ_number_unknowns} shows results for different numbers of unknown links and Section~\ref{sec:conclusions} concludes.
\section{Firm-level input-output tables}\label{sec:firm-level_prodNet}
This section describes firm-level production networks and gives an example of the supply chain network of publicly listed firms we aim to reconstruct. We then explain I-O tables and outline key differences between I-O tables at the sector and firm level. 

The production or supply chain network is composed of $N$ firms (nodes) and links between firms indicate yearly trading relationships. Links may be weighted, where each weight $w_{ij}$ represents firm $j$'s intermediate input expenditure on goods produced by $i$. We label the weighted adjacency $\bm{W}$ and $\bm{A}$ the binary adjacency matrix. Figure~\ref{fig:adjacMat_example_binary_weighted} shows the binary (left) and weighted (right) adjacency matrices depicting the empirical data collected by FactSet. Both matrices have on the $i$-th row the customers of firm $i$, while column $j$ lists the suppliers of the $j$-th firm. The unknown weights are labelled as question marks. Given $\bm{A}$ (and other aggregate information about firms that we outline below), we aim to reconstruct $\bm{W}$.

\begin{figure}[!htbp]
\begin{subfigure}{0.36\textwidth}
\[\bm{A} = 
\begin{bmatrix}
    0 & 1 & 0 & 1 \\
    0 & 0 & 1 & 0 \\
    1 & 0 & 0 & 1\\
    1 & 1 & 0 & 0 \\
\end{bmatrix}
\]
\end{subfigure}%
\begin{subfigure}{0.36\textwidth}
\[\bm{W} =
\begin{bmatrix}
    0 & ? & 0 & ? \\
    0 & 0 & ? & 0 \\
    ? & 0 & 0 & ?\\
    ? & ? & 0 & 0 \\
\end{bmatrix}
\]
\end{subfigure}
\caption{Example of the data we aim to reconstruct. \textbf{Left:} Binary directed adjacency matrix. \textbf{Right:} Weighted directed adjacency matrix.}
\label{fig:adjacMat_example_binary_weighted}
\end{figure}

For each firm, we can define its (total) intermediate expenditure and sales as, respectively, the column and row sums of the weighted adjacency matrix. The column and row sums are also called the weighted in- and out-degrees or in- and out-strengths; they are given by

\begin{equation}\label{eq:column_sums}
\bm{s}^{\text{in}} = \bm{W}^{\top}\mathds{1},\; \text{and}
\end{equation} 

\begin{equation}\label{eq:row_sums}
\bm{s}^{\text{out}} = \bm{W}\mathds{1},
\end{equation}
where $\mathds{1}$ is a vector of ones of appropriate size.

The supply chain network just described captures only a part of the economic activity, namely firm-to-firm trades. Transactions with other economic actors (e.g., households) are captured in an input-output table, usually at the sector level. One can also define an input-output table at the firm level, however with notable differences. For a more in-depth discussion, we refer to Appendix~\ref{sec:proxy_nodes_rebalanc} and \cite{bacil2022emprical}.

In I-O studies of production networks, the weighted adjacency matrix is usually normalised using firms' total costs instead of their in-strengths. A firm's total costs are the costs of intermediate inputs plus value-added, which is itself composed of labour costs, depreciation, amortisation and profit (see Appendix~\ref{app:value_added}). The weights so normalised are called the \textit{technical coefficients} and represent the percentage of inputs firm $j$ buys from firm $i$. The technical coefficients are given by
\begin{equation}\label{eq:technical_coeff}
    T_{ij} = \frac{W_{ij}}{\sum_{i} W_{ij} + y_j}\;,
\end{equation}
where $y_j$ is value-added of firm $j$.

Similar to the technical coefficients, one can define the \textit{allocation coefficients}, $\bm{B}$. $B_{ij}$ tells us the percentage of output firm $i$ sells to firm $j$. Letting $f_i$ be the amount of final demand satisfied by firm $i$, the allocation coefficient is defined as
\begin{equation}\label{eq:ghosh_coeff}
    B_{ij} = \frac{W_{ij}}{\sum_{j}W_{ij}+f_i}\;.
\end{equation}
\section{Data}\label{sec:data}
FactSet is the global production network that we aim to reconstruct and for which we do not know the values of the monetary transactions. Therefore, 
we test the reconstruction method on the administrative dataset of Ecuador, for which we know the monetary values of the transactions.

\subsection{FactSet}
We use three primary data sources provided by FactSet: Fundamentals, Supply Chain Relationships and Supply Chain Shipping Transactions.\footnote{
The datasets were downloaded in April 2020.
} 
FactSet covers mainly listed firms around the world. The supply chain relationships of these companies are collected through two primary sources: company filings required by US Federal Accounting Standards (Supply Chain Relationships) and import and export declarations at ports from the US Bill of Lading (Supply Chain Shipping Transactions).\footnote{
The Statement of Financial Accounting Standards No. 131 requires publicly traded firms on US stock exchanges to report customers that account for 10\% or more of their annual revenues, formally called \textit{major customers}.
}
FactSet also collects information on supply chain relationships from investor presentations, company websites and press releases. Due to the nature of the data collection process, coverage is biased toward companies listed on US stock exchanges, large firms and large transactions. For a more detailed description of FactSet see Appendix~\ref{app:FactSet_description_dataset} and \cite{bacil2022emprical}.

We aggregate customer-supplier relations within a fiscal year to ensure time consistency between the formation of supplier-customer relations and financial statements.\footnote{
The fiscal year goes from June to May, meaning that if a company's fiscal year end-month falls between January and May, the fiscal year is the current calendar year minus one; otherwise, it is the current calendar year.
}
We further aggregate all three datasets at the parent company level. For each company, we also have information on the sector (NACE Rev.2 codes at the 4-digit level) and the country where the company's headquarters are located.

We use several variables from companies' income statements (FactSet Fundamentals): revenues, the cost of goods sold, labour expenses, earnings before interest and taxes (EBIT), depreciation and amortisation. We convert all the variables to USD using the currency conversion tables provided by FactSet. We define value-added as the sum of labour expenses, EBIT, amortisation and depreciation (Appendix~\ref{app:value_added}). Some firms do not disclose their labour costs and include them in the costs of goods sold; we estimate these firms' labour expenses (see Appendix~\ref{app:constr_labor_exp}).

For simplicity, we limit ourselves to the 2014 network, which we call henceforth ``FactSet''. We keep firms with positive sales, intermediate expenses and value-added, and with non-negative labour costs (see Appendix~\ref{sec:cleaning_financials}). We exclude firms in financial and insurance, extraterritorial organisations and bodies and activities of households as employers. The number of firms in the 2014 cleaned dataset is 5,442; these are involved in 15,916 trading relations. The average degree is 2.9.

\paragraph{Evaluation of coverage and proxy node.} 
In 2014, FactSet captures around 16.4\% of world gross output as reported in the WIOD (see Appendix~\ref{sec:evaluation_grossOutput}). To capture the rest of the economic activity that we do not capture in FactSet, we introduce a ``proxy'' node in the network to which all firms are connected. We construct the proxy node's variables (gross output, intermediate sales and expenditure, etc.) using the WIOD aggregated at the world level. 

A good approach would be to integrate FactSet with the WIOD at the country and sector level. However, due to differences between national accounting standards and firms' financial statements, we had to aggregate the WIOD at the world level. We refer to Appendix~\ref{sec:proxy_nodes_rebalanc} for a detailed discussion of how we integrated the two datasets.

\subsection{Ecuador}\label{sec:data_ec}
Ecuador collects customer-supplier relations through VAT filings, which are mandatory for firms and natural persons. We do not have access to firms' financial statements, so we do not know their revenues, labour costs or profit, but we know the sectors' firms are in (ISIC Rev. 4 codes). The Ecuador dataset was provided by Ecuador’s government to one of the authors. We refer to \cite{astudillo2021} and \cite{bacil2022emprical} for more information about the dataset. 

\paragraph{Constructing the test network.}
While Ecuador's dataset has comprehensive coverage, FactSet does not. Therefore, we construct a test network that mimics the missing firms and links in FactSet. To mimic the missing firms in FactSet, we keep the same number of firms in Ecuador that we have in FactSet. We choose to keep the largest firms (in terms of out-strength) since we observed predominantly large firms in FactSet. We also require firms to have positive in-strengths, meaning that firms need to buy some inputs from the other firms in the subgraph.\footnote{
4 firms are further dropped because they do not have any links with the other firms in the sampled subgraph.
} The resulting network, which we call ``test network'' (third column in Table~\ref{tab:Ecuador_summary_stats}), has a much higher average degree compared to FactSet, meaning that the average Ecuadorian firm is connected to many more firms than the average firm in FactSet. 

To mimic the missing links in FactSet, we eliminate links at random in the ``test network'' until we match FactSet's average degree. Since FactSet's supply chain relations mostly cover customers that account for 10\% or more of a firm's annual revenues, we delete links with a smaller weight with a higher probability: we set the link deletion probability to be inversely proportional to the link weight $p_{ij} \propto 1/W_{ij}$. We do this procedure 50 times and reconstruct each of the 50 randomised networks. The summary statistics for these networks, which we call ``trimmed test network'', are shown in the last column of Table~\ref{tab:Ecuador_summary_stats}. In deleting links to match FactSet's average degree, 96\% of the links among firms in the test network are deleted. Consequently, our results can be interpreted as an approximate lower bound on the quality of the reconstruction. 

We aggregate the firms and transactions left out of the test network in one proxy node representing the rest of the economy. As done for FactSet, we establish an incoming and outgoing link between each firm and the proxy node.

\begin{table}[!htbp]
\centering
\def\arraystretch{1.2}
\begin{tabular}{l *{3}{r}}
\toprule
Summary statistics & Full network & Test network  & Trimmed test network\\ 
\midrule
N. nodes & 84,978 & 5,440 & 5,440\\
N. edges &  3,439,975 & 432,910 & 15,776\\
Average degree & 40.5 & 79.6 & 2.9\\
\bottomrule
\end{tabular}

\caption{Summary statistics for the 2014 Ecuador network.  The first column ``Full network'' is the network composed of all the firms except those in sectors: finance, insurance, activities of households as employers, activities of extraterritorial organisations and bodies, and those that have no sectoral code. The third column, ``Test network'', refers to our test network, which is composed of the top 5,440 firms with the largest total intermediate sales and positive intermediate expenses. The last column ``Trimmed test network'' refers to the network built from the ``Test network'' by further deleting links at random to match FactSet's average degree. The summary statistics of the test and trimmed test network do not include the proxy node.}
\label{tab:Ecuador_summary_stats}
\end{table}

\paragraph{Inferring missing data.}
For Ecuador, we do not have information on final demand, revenues and the variables that compose value-added (i.e., labour costs, depreciation, amortisation and profits). To carry out the analyses described in Section~\ref{sec:def_measure_comparison}, we need final demand and value-added of each firm. Therefore, we simulate final demand and value-added using the 2014 I-O table of Ecuador at the sector level.\footnote{
The sector-level I-O table is available at \url{https://contenido.bce.fin.ec/documentos/PublicacionesNotas/Catalogo/CuentasNacionales/Anuales/Dolares/MenuMatrizInsumoProducto.htm}.
} Consider value-added (a similar procedure is done for final demand), for each sector $s$, we calculate the ratio of value-added to intermediate expenditure $\nu_s = y_s/s^{\text{in}}_s$. Assuming that a firm's ratio is the same as that of the sector the firm is in, a firm's value-added is given by $y_i = \nu_s\cdot s_i^{\text{in}}$.
\section{Network reconstruction}\label{sec:Net_reconstruction}

Methods for reconstructing networks with missing information have mostly been developed for financial or trade networks \citep[e.g.,][]{moussa2011contagion,mastrandrea2014enhanced,cimini2015systemic,gandy2016bayesian,anand2015filling} and I-O tables at the sector level \citep[e.g.,][]{golan1994recovering,robinson2001updating,lenzen2009matrix}. Only a few studies develop methods for reconstructing firm-level production networks \citep{inoue2019firm,welburn2020systemic,reisch2021inferring,hooijmaaijers2019methodology,ialongo2021reconstructing,Hillman2021newFirm-levelModel}. We start with a brief overview of the different reconstruction methods developed in the literature, which we divide into deterministic and ensemble methods, and subsequently give a more detailed account of the methods developed to reconstruct firm-level networks. We refer to \cite{squartini2018reconstruction} and \cite{cimini2021reconstructing} for reviews on reconstruction methods developed mostly for financial and trade networks, and to \cite{miller2009input}, \cite{mcdougall1999entropy} and \cite{lahr2004biproportional} for reviews on sector-level reconstruction methods and matrix balancing problems. 

The network reconstruction problem, being it for financial, trade or production networks, boils down to inferring a matrix of bilateral flows among entities (e.g., banks, countries or sectors) given constraints on the total in- and out-flows of each entity and other information when available (e.g., degrees or a prior bilateral flows matrix). Most of the reconstruction methods are based on the maximum entropy principle, of which there are two strands: deterministic and ensemble methods. Deterministic methods yield a single reconstruction of the weighted network while meeting the constraints exactly. Instead, ensemble methods sample many networks from a distribution that is constructed to respect the constraints on average. Therefore, while ensemble methods generate a probability distribution over the likely networks, deterministic methods assign a probability of one to the reconstructed network and a zero probability to all the other networks -- which likely include the true network \citep{parisi2020faster}. The shortcomings that most deterministic and ensemble methods share are that they tend to create a fully connected network with weights distributed as uniformly as possible given the imposed constraints on the in- and out-flows. To create sparser networks, algorithms with tunable \citep{moussa2011contagion,upper2011simulation,mastromatteo2012reconstruction} or exact network density \citep{mastrandrea2014enhanced,cimini2015systemic} have been developed.

The most well-known deterministic method is known as \textit{MaxEnt}. It maximises an entropy-like functional subject to constraints on the in- and out-strength of each node. The solution to this maximisation yields the well-known gravity model (without distance) in the international trade literature (first proposed by \citealp{tinbergen1962shaping} and \citealp{poyhonen1963tentative}; see also \citealp{squartini2014jan} for a discussion). MaxEnt displays all the shortcomings mentioned above: it generates a fully connected network and the weights are distributed as equally as possible given the constraints. To enhance the MaxEnt reconstruction, if some prior information about the binary topology or the weights is available, it can be integrated using a cross-entropy method \citep{golan1994recovering,digiangi2018assessing,upper2011simulation,wells2004financial}. The cross-entropy method reconstructs a network that has minimum distance to the prior while accounting for the imposed constraints. The cross-entropy method is equivalent to the iterative proportional fitting (IPF) algorithm, which iteratively distributes the weights (coming from the MaxEnt solution or any other prior) among the non-zero edges until the row and column sums are satisfied. The IPF algorithm is also known as the RAS technique in the I-O literature \citep{miller2009input}. If the network is fully connected, the IPF algorithm is equivalent to MaxEnt.

There are different ensemble methods depending on the information used for the reconstruction (e.g., in- and out-degrees or strengths sequences). We discuss the method developed by \cite{cimini2015systemic} since it is one of the best performing \citep{anand2018missing,lebacher2019search}. To enhance the reconstruction of the weighted network, \cite{cimini2015systemic} impose constraints on both the in- and out-strength sequences and on the degrees. Their strategy is motivated by recent results showing that the strengths do not encode information about the binary topology (although they are correlated with degrees) and that the degrees are ``fundamental'' local structural properties of weighted networks \citep{mastrandrea2014enhanced}. Combining constraints on the degrees and strengths thus greatly enhances the reconstruction of weighted networks since the degrees provide information about the binary topology that strengths do not, enabling to identify better the matrix of link probabilities as well as higher-order properties \citep{mastrandrea2014enhanced,gandy2016bayesian}.

The method proposed by \cite{mastrandrea2014enhanced} requires knowledge of the degrees and strengths of all the nodes, which are not always available. \cite{cimini2015systemic} note that in financial networks, the in- and out-strengths are usually known while the degrees might be known for a subset of nodes only. To account for these two pieces of information, \cite{cimini2015systemic} restore to the \textit{fitness} ansatz. The fitnesses are nodes' non-topological futures that relate to the ability of nodes to establish connections: nodes with higher fitness attract more connections and are thus likely to become hubs \citep{squartini2018reconstruction,mazzarisi2017methods}. Given the empirical correlation frequently observed among strengths and degrees, strengths are often used as a proxy for nodes' fitnesses. To estimate the binary topology, \cite{cimini2015systemic} thus develop a fitness-induced configuration model that estimates link probabilities using the nodes' fitnesses and the degrees of only a few nodes. The weights are estimated in a second step using a degree-corrected gravity model that accounts for the sparsity of the adjacency matrix inferred in the first step.

How does one choose which reconstruction method to use? \cite{anand2018missing}, \cite{lebacher2019search} and \cite{ramadiah2020reconstructing} find that the choice of the reconstruction method ultimately depends on the feature of the network one aims to reconstruct, which usually boils down to connectivity structure versus weights. If one cares about inferring links, methods that focus on reconstructing a sparse connectivity structure are better suited. \cite{anand2018missing}, \cite{lebacher2019search} and \cite{mazzarisi2017methods} conclude that the best-performing ones are the degree-corrected gravity model \citep{cimini2015systemic}, the minimum density \citep{anand2015filling} and the Bayesian hierarchical fitness \citep{gandy2016bayesian}. If one cares about inferring the link weights, methods based on MaxEnt or the IPF algorithm perform best. Since financial networks have many link weights of relatively equal size \citep[something that is less likely to be the case in firm-level networks;][]{bacil2022emprical}, they score well on weight-based similarity measures \citep{anand2015filling}. In their horse races, \cite{anand2018missing} and \cite{lebacher2019search} identify the methods developed by \cite{cimini2015systemic} and \cite{baral2012estimation} to be the best performing. Altogether, the reconstruction method proposed by \cite{cimini2015systemic} seems to be the best in reconstructing both binary and weighted topological features.

\paragraph{Firm-level network reconstruction.}
\cite{welburn2020systemic} reconstruct the network of listed firms in the US available through a dataset similar to FactSet but covering only their major customers.\footnote{
They retrieved the customer-supplier relations from firms' filings available through the EDGAR database.
} Therefore, they know the binary topology only partially. To reconstruct the weighted network, they use a two-step procedure. In the first step, they infer missing links using a logistic regression. In the second step, they develop a linear programming method to reconstruct the unknown weights given the links. Since they do not observe the whole economy, they introduce a proxy node that captures the rest of the economy and to which each firm is linked. The cumulative in- and out-flows of the proxy node are then minimised subject to constraints on firms' revenues and the cost of goods sold. While there are almost 6,000 firms in their network, they only reconstruct the network composed of 1,000 firms due to the high computational complexity of the second step of their procedure. \cite{inoue2019firm} reconstruct the weighted network of Japanese firms given the binary topology.\footnote{
The network is collected by a private company. Although the coverage is extensive, comprising almost 890,000 firms, it is not exhaustive.
} Firstly, they assume that the link weight is proportional to the supplier and customer's sales. Subsequently, they re-adjust the estimated weights using the sector-level I-O tables to ensure that if the firm-level network is aggregated at the sector level, it is consistent with national accounts. In a similar fashion and using the same Japanese dataset, \cite{carvalho2021japan} assume that the technical coefficient between two firms is proportional to the technical coefficient between the sectors those two firms are in.

\cite{Hillman2021newFirm-levelModel} reconstruct the global network of private and public firms in the ORBIS database. As done in other reconstruction methods, they build the firm-level network so that it is consistent with sector-level I-O data (OECD). In each step, a firm $i$ is chosen at random and its sales are split into $n$ units that are then sold to $n$ different customers. $i$'s customers are chosen according to the industry they are in based on sector-level I-O tables, meaning that if firm $i$ is in industry $s$, its customers need to be in one of the industries to which $s$ sells (in the sector-level I-O table). They further calibrate their model on the observation that larger firms tend to have more customers \citep{bernard2019origins}. This feature can be controlled by changing how, at the beginning of each step, firm $i$'s sales are split into $n$ units. For computational reasons, they only use 5,000 firms among the more than 200 million firms in the Orbis database. \cite{hooijmaaijers2019methodology} reconstruct the supply chain network of the Netherlands using several microdata sources available to the Office of National Statistics. They describe their method as being akin to maximum entropy methods with exact link density \citep[as classified in][]{squartini2018reconstruction}, but they pose additional constraints thanks to the richness of their microdata and to findings in the literature about empirical facts of firm-level supply chain networks. A novel feature of their reconstruction is the disaggregation of firms' output into different goods. 

None of the studies just described can assess how well their method recovers the empirical weighted network because none of them has access to it. Two studies assess their reconstruction method, at least to some extent. First, \cite{reisch2021inferring} reconstruct a firm-level production network using mobile communication data. To guarantee anonymity, the company providing the data and the country are not disclosed. Roughly speaking, links are inferred by assuming that if two firms communicate with each other, they are involved in a supply-chain relationship. To determine the link direction, they use the national I-O table at the sector level and information about the sectors the customer and supplier are in. A gravity model is then used to estimate the link weights, where a firm's size is given by its total assets. To assess the reconstruction of the binary topology, they compare to the Hungarian network, whereas to assess the performance of the method regarding the weights reconstruction, they use the Economic Systemic Risk Index and compare with results obtained for Hungary by \cite{diem2021quantifying}. They find similarities between results obtained for Hungary and their reconstructed network. Second, \cite{ialongo2021reconstructing} develop the stripe-corrected gravity model, which builds on the degree-corrected gravity model \citep{cimini2015estimating} by adding constraints on the in-strength of each industry. They test their method on two transaction data made available by two Dutch banks. The assessment is carried out on the degree and strength distributions, degree-strength correlations and average nearest neighbour strength.

\subsection{Method}

To reconstruct the weighted network given the binary topology, we use the conditional maximum entropy ensemble reconstruction method developed by \cite{parisi2020faster}. We do not use any of the previously developed methods for reconstructing firm-level networks because either they are too computationally expensive \citep{welburn2020systemic,Hillman2021newFirm-levelModel}, demand too many data inputs that we do not have \citep{hooijmaaijers2019methodology} or would imply constraining the firm-level network with sector-level I-O tables \citep{inoue2019firm,ialongo2021reconstructing}. We disregard the latter methodologies because we think using sector-level data in the way proposed by \cite{inoue2019firm} or in the spirit of \cite{ialongo2021reconstructing} could bias the reconstruction in unwanted ways given the underlying differences in accounting standards between national I-O tables and firms' financial statements \citep[see Appendix~\ref{sec:proxy_nodes_rebalanc}, but also][]{bacil2022emprical}.

\cite{parisi2020faster} develop a maximum entropy method that reconstructs an ensemble of likely weighted networks given some prior information about the binary ensemble and aggregate information about each node. The procedure is flexible in that it allows to use of an observed binary topology or to infer it in a previous step and account for the additional uncertainty. They proposed two methods that use different constraints. One method constrains the in- and out-strength sequences, while the other one constrains the expected link weights (and the in- and out-strengths indirectly). We choose the second method for two reasons. First, it is computationally more efficient since it involves solving $m$ (the number of links) decoupled equations, while the method constraining the in- and out-strengths require solving $2N$ coupled equations, where $N$ is the number of firms. Second, the authors show that it predicts the weights better compared to the model constraining the in- and out-strengths.

The method consists of two steps. The first step constrains the in- and out-strengths (total intermediate expenditure and sales) of each firm and determines the values of the expected link weights used as constraints in the second step. As discussed in Section~\ref{sec:Net_reconstruction}, MaxEnt reconstructs the weights best among the other methods \citep{anand2018missing,lebacher2019search}; thus, MaxEnt is used in the first step. The second step allows us to account for any prior information about the binary topology and generates an ensemble of weighted networks compatible with this information and with the constraints on the expected link weights.

As discussed, MaxEnt assumes a fully connected network; however, the conditional maximum entropy method assumes no self-loops. Additionally, in our case, we know the binary topology. To redistribute the weights corresponding to $A_{ij} = 0, \forall\; (i, j) \notin\mathcal{E}$, where $\mathcal{E}$ is the edge set, we employ the IPF algorithm. The IPF algorithm redistributes the weights in an iterative procedure until the constraints on the in- and out-strengths are met (see Appendix~\ref{app:derivation_parisi}). \citeauthor{parisi2020faster}'s \citeyearpar{parisi2020faster} method turns the IPF algorithm into a probabilistic method and allows calculating confidence intervals around each reconstructed weight. We give a brief outline of the method below and describe it in detail in Appendix~\ref{app:derivation_parisi}.

\paragraph{First step.} The method derives values for the expected link weights to enforce as constraints in the second step. The values of the weights are derived by solving the MaxEnt problem, which maximises an entropy-like functional subject to constraints on intermediate sales and costs of each firm. The value of each weight is given by
\begin{linenomath}
\begin{equation*}
    W_{ij}^{\text{ME}} = \frac{s_i^{\text{out}^*} s_j^{\text{in}^*}}{W^{\text{tot}^*}}\;,
\end{equation*}
\end{linenomath}
where $W^{\text{tot}^*} = \sum_{i}s_i^{\text{out}^*} = \sum_{j}s_j^{\text{in}^*}$ is the total weight of the empirical network.
    
\paragraph{Second step.}
The method maximises the conditional entropy defined over the probability density function of the weighted networks compatible with the prior on the binary ensemble and subject to constraints on the expected weights. Solving the conditional maximum entropy problem yields that the probability of observing a weight $W_{ij} >0$ given that there is a link between $i$ and $j$ is of exponential form with parameter $\lambda_{ij}$ (which also corresponds to the Lagrange multiplier):
\begin{equation}\label{eq:CReMb_prob_functional_form}
Q_{ij}(W_{ij} \mid A_{ij} = 1) = \lambda_{ij} e^{- \lambda_{ij} W_{ij}},\; W_{ij} > 0.
\end{equation}
To find the values of the $\lambda_{ij}$'s, one maximises the log-likelihood function, which leads to the first order conditions
\begin{equation}
    \expval{W_{ij}} = \frac{p_{ij}}{\lambda_{ij}}\;,\;  \forall\; i\neq j.
\end{equation}
Since for each link the expected weight $\expval{W_{ij}} = W_{ij}^{\text{ME}}$, the Lagrange multipliers are given by
\begin{equation}
    \lambda_{ij}^* = p_{ij}\frac{W^{\text{tot}^*}}{s_i^{\text{out}^*} s_j^{\text{in}^*}}\;,\;  \forall\; i\neq j.
\end{equation}
We set $p_{ij} = 1, \forall (i, j) \in \mathcal{E}$.

\paragraph{Confidence interval on the expected edge weight.}
For each expected weight, the confidence interval is $[w^-, w^+]$. The lower bound is given by
\begin{equation}\label{eq:lower_bound_weights}
    w^- = - \frac{\ln[e^{-1} + q^-]}{\lambda_{ij}^*}\;,
\end{equation}
where $q^-$ is a desired confidence level and the upper bound is given by
\begin{equation}\label{eq:upper_bound_weights}
    w^+ = - \frac{\ln[e^{-1} - q^+]}{\lambda_{ij}^*}\;.
\end{equation}
We set $q^+ = q^- = 0.25$. We refer to Appendix E in \cite{parisi2020faster} for the derivation.

\subsection{Assessing the reconstruction}\label{sec:def_measure_comparison}
We assess how well the reconstruction method can recover the empirical network at three scales. First, we look at microscale quantities: weights, and technical and allocation coefficients. Second, we evaluate the reconstruction of higher-order properties (i.e., multipliers): the output multipliers and the influence vector. Third, we turn to macroscale properties and use a general equilibrium I-O model to study how the propagation of shocks through the network affects GDP volatility. We conclude this section by defining the statistical indicators used to compare the empirical and reconstructed quantities.

\subsubsection{Weights}\label{sec:microscale_prop}
As discussed in Section~\ref{sec:firm-level_prodNet}, technical and allocation coefficients (Equation~\ref{eq:technical_coeff} and~\ref{eq:ghosh_coeff}, respectively) are normalised weights frequently used in economic models. Therefore, we assess the reconstruction of both coefficients and the weights: their exact values and their distributions.

\subsubsection{Higher-order properties}\label{sec:higher_order}
We look at two higher-order properties widespread in the economic literature: the output multipliers and the influence vector. These are two centrality measures that quantify firms' contributions to economy-wide fluctuations.

\paragraph{The output multipliers.}
The output multipliers are derived from the Leontief model \citep{miller2009input} and are defined as
\begin{equation}\label{eq:output_mult_app}
    \mathcal{O} \equiv (\bm{I} - \bm{T}^\top)^{-1} \mathds{1}\;,
\end{equation}
where $\bm{I}$ is the identity matrix and $\mathds{1}$ a vector of ones, both of appropriate size. The output multiplier captures the upstream propagation channel of an exogenous shock to a firm's final demand and its economy-wide impacts  \citep[where final demand increases by one monetary unit;][]{miller2009input}. It can also be seen as the average length of a firm's production chain \citep{mcnerney2018production,fally2012production,miller2017downstreamn}. The higher the output multiplier is, the longer the production chain is on average and the greater the impact of a change in a firm's final demand is on the whole economy.

\paragraph{The influence vector.}
The influence vector is derived from the Cobb-Douglas model proposed by \cite{acemoglu2012network}. The influence vector is defined as

\begin{equation}\label{eq:influence_vector}
\bm{v} \equiv \frac{\alpha}{N}[\bm{I} - (1 - \alpha)\bm{\Omega}]^{-1}\mathds{1}\;,
\end{equation}
where $\bm{\Omega}$ is the matrix of input shares with $\omega_{ij} \in [0, 1]$ being the share of input $i$ used in $j$'s production process ($\sum_i \omega_{ij} = 1$), $\alpha \in (0, 1]$ is the share of labour and $N$ is the number of firms.

Contrary to the output multipliers, the influence vector captures the downstream propagation channel of TFP shocks and it gauges the contribution of firms to fluctuations in aggregate GDP. Positive TFP shocks can be thought of as firms' innovating their production processes and becoming more efficient in using their inputs. The influence vector is equivalent to a Reverse Weighted PageRank with a damping factor equal to $(1 - \alpha)$.

\subsubsection{Macroscale properties}
\label{sec:macroscale_prop}

As highlighted by recent crises and natural disasters, assessing how different shocks affect the economy is of paramount importance to be better prepared in preventing or alleviating crises. Therefore, after assessing the weights and the multipliers, we study how supply-side shocks at the firm level propagate through the network to downstream firms, ultimately leading to fluctuations in aggregate GDP.

\paragraph{Supply-side shocks and aggregate volatility.}
We model supply shocks as shocks to firms' TFP. We use the model developed by \cite{acemoglu2012network} and refer to the paper for a derivation. In the competitive equilibrium, aggregate volatility is given by
\begin{equation}\label{eq:gr_volatility_gdp}
    \sigma_{\Delta y} = \sqrt{\sum_i \text{Var}(\Delta\epsilon_{i}) v_{i}^2}\;, 
\end{equation}
where $v_i$ is the influence of firms $i$ (Equation~\ref{eq:influence_vector}), $\Delta\epsilon_{i}$ is the TFP shock of firm $i$ and $\epsilon_i$ is an i.i.d. random variable with mean zero and bounded variance.
Equation~\ref{eq:gr_volatility_gdp} shows that productivity shocks at the firm level affect aggregate value-added through the production network. Initially, the shock propagates to the customers of the affected firm and, subsequently, propagates downstream to the customers' customers and so on, potentially spreading through the whole supply chain network.

Note that the model gives a static representation of the economy in equilibrium and firms' input shares are exogenous; therefore, the network structure and hence the influence vector are constant over time. We use Equation~\ref{eq:gr_volatility_gdp} to assess the impact of firms' TFP shocks on GDP volatility using either the empirical or the reconstructed weighted network given the TFP shocks.

\paragraph{Simulating TFP shocks.}
To estimate TFP shocks, we cannot use an econometric technique \citep[e.g.,][]{magerman2016heterogeneous} because it requires knowledge of several variables that we do not observe for Ecuador. Besides requiring all the variables describing a firm's production function, the estimation of TFP requires a time series of these variables. Since our goal is not to empirically validate the economic model but to assess the discrepancy in the predicted GDP volatility when the reconstructed influence vector is used instead of the empirical one, we simulate firm-level TFP shocks. For FactSet, we do not perform this test since it would entail estimating TFP, which is outside of the scope of this paper.

\begin{figure}[!htbp]
    \centering
    \includegraphics{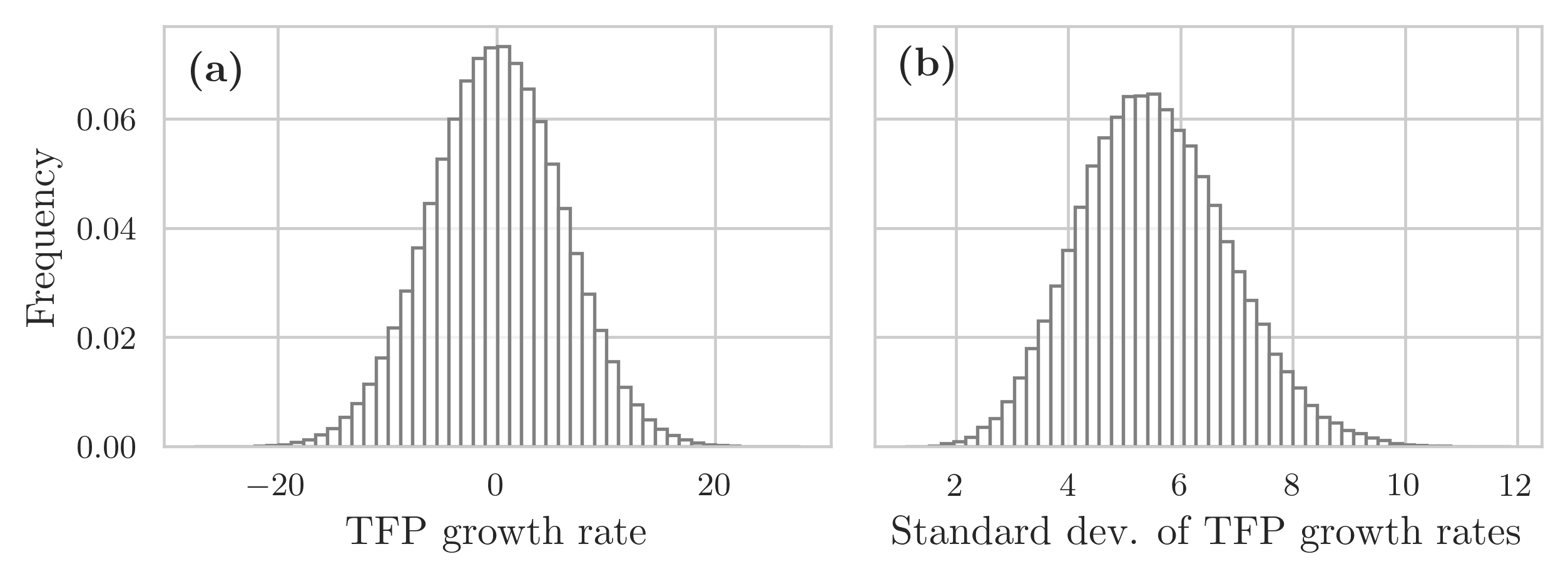}
    \caption{Distribution of TFP \textbf{(a)} growth rates and \textbf{(b)} their standard deviation. The growth rates are pooled over the 10 simulated years. We binned the data into 50 equally-spaced bins.}
    \label{fig:TFP_growthRate_volatility}
\end{figure}

We simulate TFP shocks from a normal distribution with mean zero and standard deviation of 6. We chose a zero mean in line with the empirical mean reported for the Belgian production network by \cite{magerman2016heterogeneous}. Since they do not report the variance, we set the variance to 6 so that GDP volatility, calculated using the model with the true network, matches the observed country-level volatility.\footnote{
To calculate the GDP volatility in Ecuador, we use data from the IMF; available at \url{https://data.imf.org/?sk=388DFA60-1D26-4ADE-B505-A05A558D9A42&sId=1479331931186}. We use nominal GDP since our data are not adjusted for inflation. Since we simulate 10 years, we calculate GDP volatility for the period 2005-2015, which is 6.35\%. 

We tried different parametrisations of the normal distribution, which do not match the volatility in the empirical data, and results do not change, at least qualitatively.
} We set the TFP shock of the proxy sector equal to the median of the TFP shocks of firms that were excluded from our test network. We simulate 10 time periods. Figure~\ref{fig:TFP_growthRate_volatility} shows the distribution of the simulated TFP growth rates in Panel (a) and the distribution of their standard deviation in Panel (b). Once we simulate the TFP volatilities, we use Equation~\ref{eq:gr_volatility_gdp} to predict the fluctuation in aggregate GDP using either the empirical or the reconstructed influence vector. We then compare the empirical volatility with that predicted by the reconstruction.

\subsubsection{Statistical indicators}\label{sec:comparison_metrics}
To compare how well the reconstruction method can recover the quantities defined in Section~\ref{sec:microscale_prop}, ~\ref{sec:higher_order} and~\ref{sec:macroscale_prop}, we employ metrics that are standard in the literature: the $L_1$-error, the root-mean-squared error (RMSE), the mean and median absolute error (MAE and MedAE, respectively) and the cosine similarity.

The $L_1$-error assesses the degree to which constraints on intermediate sales and costs are violated. It is defined as
\begin{linenomath}
\begin{equation*}
    L_1 = \sum_i |s_i^{\text{in}} - s_i^{\text{in}^*}| +  \sum_i|s_i^{\text{out}} - s_i^{\text{out}^*}|\;,
\end{equation*}
\end{linenomath}
$s_i^{\text{in}}$ is firm $i$'s in-strength in the reconstructed network and $s_i^{\text{out}}$ its out-strength in the reconstructed network; quantities with a $^*$ refer to observed, empirical values.

In what follows, we define the error measures using the technical coefficients, but they similarly apply to any other quantity of interest. We do not use the RMSE, the MAE and the MedAE to assess the raw weights because their distribution has heavy tails. For the technical and allocation coefficients, and the multipliers, we further normalise the metrics to allow their comparison across variables that have different scales. We rescale by $\phi$, the difference between the maximum and minimum value (excluding the zeros) of the empirical quantity of interest. Our rescaled measures compare the variation in the residuals to the range of the empirical data. For instance, a normalised RMSE of 0.1 means that the variation in the residuals is 10\% of the range of variation of the empirical data. The lower the normalised error metric is, the better the reconstruction is. The normalised root-mean-square error is given by
\begin{linenomath}
\begin{equation*}
\text{RMSE} = \frac{1}{\phi}\sqrt{\frac{1}{m} \sum_{ij} (T_{ij} - T_{ij}^*)^{2}}\;,
\end{equation*}
\end{linenomath}
where $m$ is the number of links. The normalised mean absolute error is given by
\begin{linenomath}
\begin{equation*}
\text{MAE} = \frac{1}{\phi m} \sum_{i,j}\lvert T_{ij} - T_{ij}^*\rvert\;.
\end{equation*}
\end{linenomath}
The normalised median absolute error is defined as
\begin{linenomath}
\begin{equation*}
\text{MedAE} = \frac{\text{Median}\Big(\lvert\bm{T} - \bm{T}^*\rvert\Big)}{\phi}\;.
\end{equation*}
\end{linenomath}
The cosine similarity is defined as
\begin{linenomath}
\begin{equation*}
    \vartheta = \frac{\sum_{ij} T^*_{ij} T_{ij}}{\sqrt{\sum_{ij} T^{*^2}_{ij}} \sqrt{\sum_{ij} T_{ij}^2}}\;.
\end{equation*}
\end{linenomath}

\subsection{Results: link weights}\label{sec:results}
In this section, we show the results of the reconstruction method for the weights, and the technical and allocation coefficients. We start by discussing the results for our test network, Ecuador, and then show the results for FactSet, for which we do not know the ground truth.

We compare weights (normalised or not) for firms only and not those of the proxy node since the proxy node can be thought of as a sink node and does not meaningfully represent either a firm or a sector. For parsimony, we show plots for one of the 50 randomised reconstructions (always the same one throughout the paper) since they all yield virtually identical results; the same holds for the summary statistics. Regarding the statistical indicators described in Section~\ref{sec:def_measure_comparison}, we compute them for each of the 50 randomised test networks and report the average value of each metric across the 50 randomised networks.

\subsubsection{Ecuador}\label{sec:results_ec}

\begin{figure}[H]
\centering
\includegraphics{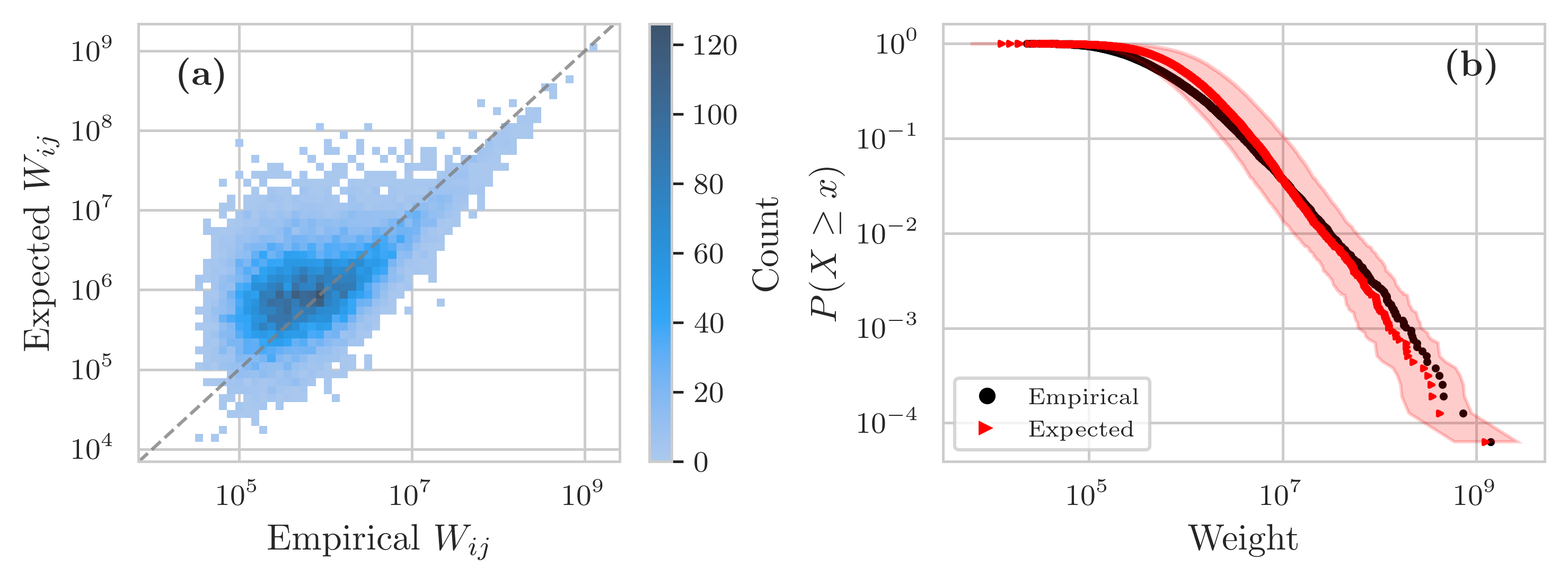}
\caption{\textbf{(a)} 2D histogram of the empirical (\textit{x}-axis) and expected (\textit{y}-axis) weights for Ecuador. We divide both axes into 50 log-spaced bins and then count the number of data points falling in each square. Perfect prediction is achieved when points lie on the identity line (dashed grey line). \textbf{(b)} CCDF of the empirical (black dots) and expected (red triangles) weights for Ecuador. The shaded area represents the 50\% confidence bounds. We compute the CCDF as $\Bar{F}_n(x) = \frac{1}{n} \sum_{i=1}^n \bm{1}(X_i \geq x)$, where $\bm{1}$ is the indicator function. Values are in USD.}
\label{fig:weights_fits_true_vs_hat_EC}
\end{figure}

The constraints on the intermediate sales and costs are always satisfied ($L_1$-error $= 10^{-4}$). The reconstruction of individual weights is rather poor, as shown in Figure~\ref{fig:weights_fits_true_vs_hat_EC}a; perfect prediction is achieved when points lie on the 45-degree line (grey dashed line). The reconstruction method tends to underpredict weights of high values and overpredict weights with intermediate or low values, although there is significant dispersion. (We show the histogram of the relative prediction errors and the empirical weights against their prediction errors in Figure~\ref{fig:pred_error_weights_ec}.) On average, 47\% of the weights fall in the 50\% confidence interval (CI). 

Although the reconstruction cannot recover individual weights particularly well, the weight distribution is recovered quite well. The weight distribution has heavy tails in both the empirical and the reconstructed networks (Figure~\ref{fig:weights_fits_true_vs_hat_EC}b; see Figure~\ref{fig:weight_ccdf_ec}a for a comparison of the weight distribution across the 50 randomised networks, the empirical test network and the full network). As expected from maximum entropy methods, the expected weight distribution is less heterogeneous than the empirical distribution. \cite{bacil2022emprical} find that the weight distribution is likely to follow a power-law with an exponent that varies between 1.0 and 1.4 depending on the country, year and estimation method. Therefore, we check whether we can recover a similar power-law exponent. We can recover it pretty well (see Figure~\ref{fig:weight_ccdf_ec}b). The power-law exponent is 1.1 for the empirical weight distribution and 1.3 for the reconstructed one.\footnote{
To fit a power-law distribution to our data, we use the method of \cite{clauset2009power} since (1) it yields a single exponent estimate and (2) it is the most widely used estimator in the literature. See \cite{bacil2022emprical} for an in-depth discussion. \cite{bacil2022emprical} use also the estimators developed by \cite{voitalov2019scale}, which are based on extreme value theory. We abstain from such an analysis in this paper.
} Although the exponent of the reconstructed distribution is slightly higher, it still implies a divergent second moment. 

In Table~\ref{tab:individual_weights_pred_error_metrics_similMeas}, we do not report the RMSE, MAE or MedAE for the weights since there is too much variation to obtain a meaningful metric; additionally, the weights likely have a diverging second moment. Therefore, we report only the cosine similarity, which is 0.93.

\begin{table}[!htbp]
\small
\centering
\def\arraystretch{1.2}
\begin{tabular}{l *{4}{c}}
\toprule
Type & RMSE & MAE & MedAE & Cosine similarity\\ 
\midrule
Weight    &    --   & -- & -- & 0.928\\
	& & & & (0.006) \\
Technical & 0.081 & 0.041 & 0.013 & 0.723 \\
			 & (0.001) & (0.000) & (0.000) & (0.004) \\
Allocation & 0.105 & 0.054 & 0.019 & 0.758 \\
			 & (0.001) & (0.000) & (0.000) & (0.003) \\
\bottomrule
\end{tabular}
\caption{Statistical indicators for the weights, and the technical and allocation coefficients for Ecuador. As defined in the main text, RMSE denotes the root mean squared error, MAE the mean absolute error and MedAE the median absolute error. For each metric, we show its mean value across the 50 randomised reconstructions. Below the mean value, the standard deviation in parenthesis. We excluded the proxy node from the calculations.}
\label{tab:individual_weights_pred_error_metrics_similMeas}
\end{table}

\begin{figure}[!htbp]
\centering
\includegraphics{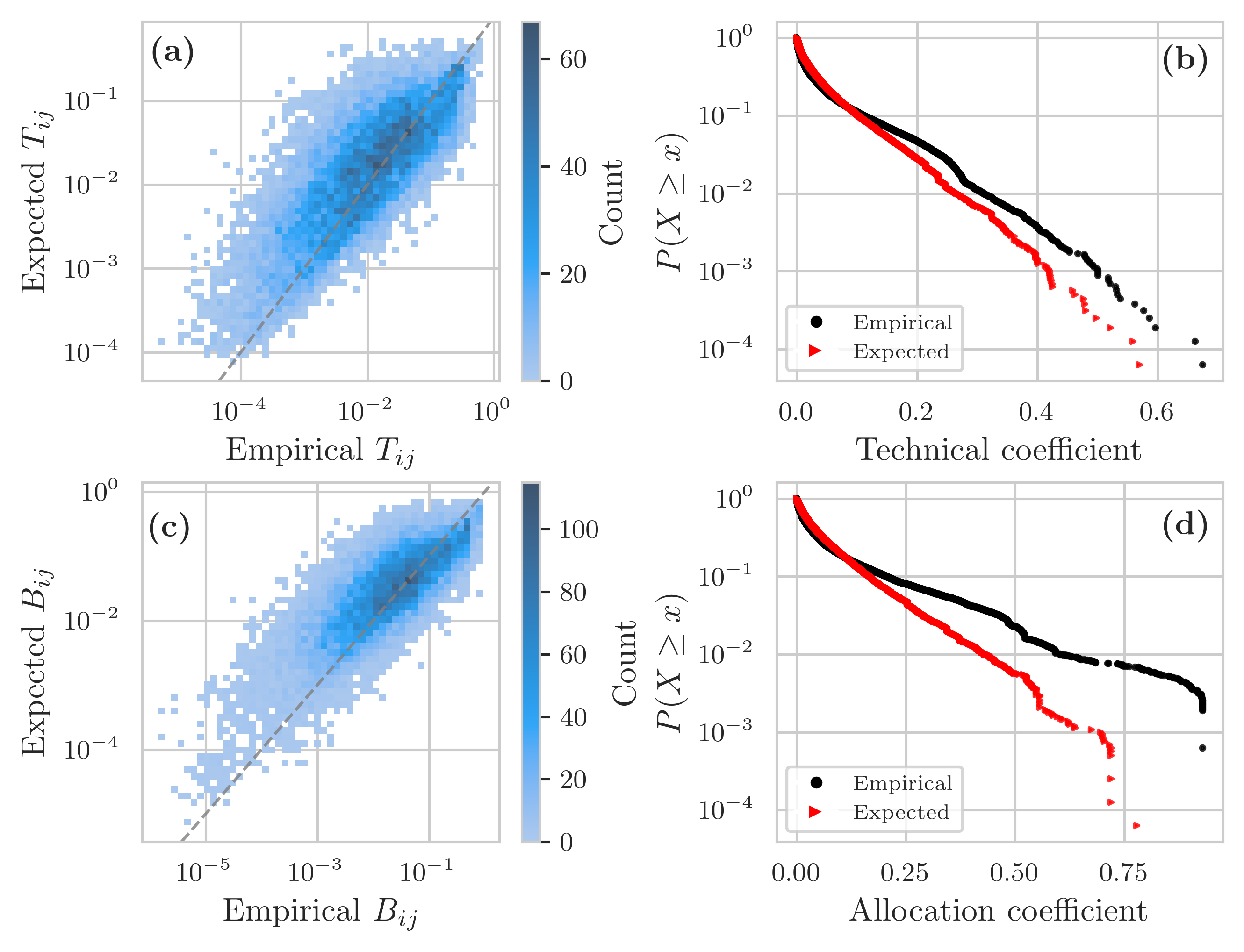}
\caption{\textbf{(a)}, \textbf{(c)} 2D histogram of the empirical (\textit{x}-axis) and the expected (\textit{y}-axis) technical and allocation coefficients, respectively, for Ecuador. We use 50 log-spaced bins and then count the number of data points falling in each square. \textbf{(b)}, \textbf{(d)} CCDF of the empirical (black dots) and expected (red triangles) technical and allocation coefficients, respectively, in semi-log scale for Ecuador.}
\label{fig:tech_coeff_true_vs_hat_EC}
\end{figure}

Figure~\ref{fig:tech_coeff_true_vs_hat_EC}a and~\ref{fig:tech_coeff_true_vs_hat_EC}c show, respectively, the empirical technical and allocation coefficients on the \textit{x}-axis and their expected values on the \textit{y}-axis. As for the weights, the reconstruction method does not perform particularly well in recovering either of the coefficients. Although less pronounced for the coefficients than for the weights, the method tends to overpredict coefficients of small values and underpredict coefficients with high values, again with considerable dispersion. This is further highlighted in Figure~\ref{fig:tech_coeff_true_vs_hat_EC}b and~\ref{fig:tech_coeff_true_vs_hat_EC}d, showing the empirical and reconstructed CCDF of the technical and allocation coefficients, respectively. Figure~\ref{fig:tech_coeff_true_vs_hat_EC}b and~\ref{fig:tech_coeff_true_vs_hat_EC}d further show that we can reconstruct the technical coefficients better, which also have smaller error metrics (Table~\ref{tab:individual_weights_pred_error_metrics_similMeas}) and for which we can recover the moments more accurately (Table~\ref{tab:individual_weights_summary_sats_ec}). However, the cosine similarity is slightly higher for the allocation coefficients: 0.76 compared to 0.72.

\begin{table}[!htbp]
\small
\centering
\def\arraystretch{1.1}
\begin{tabular}{l *{5}{c}}
\toprule
& \multicolumn{2}{c}{Technical coefficients} && \multicolumn{2}{c}{Allocation coefficients} \\
\cmidrule{2-3} \cmidrule{5-6}
& Empirical & Expected && Empirical & Expected\\
\midrule
Mean    & 0.037 & 0.038 && 0.071 & 0.065 \\
Median  &  0.010 & 0.016 && 0.019 & 0.031 \\
Standard dev. & 0.067 & 0.057 && 0.132 &  0.088 \\
\bottomrule
\end{tabular}
\caption{Summary statistics of the technical and allocation coefficients for Ecuador. For each coefficient, the first column reports summary statistics for the empirical coefficients and the second column for the reconstructed ones. We show results for one of the reconstructions; all three quantities have virtually the same summary statistics across the 50 randomised empirical and reconstructed networks. We excluded the proxy node from the calculations.}
\label{tab:individual_weights_summary_sats_ec}
\end{table}

\FloatBarrier
\subsubsection{FactSet}
\begin{figure}[!htbp]
\centering
\includegraphics{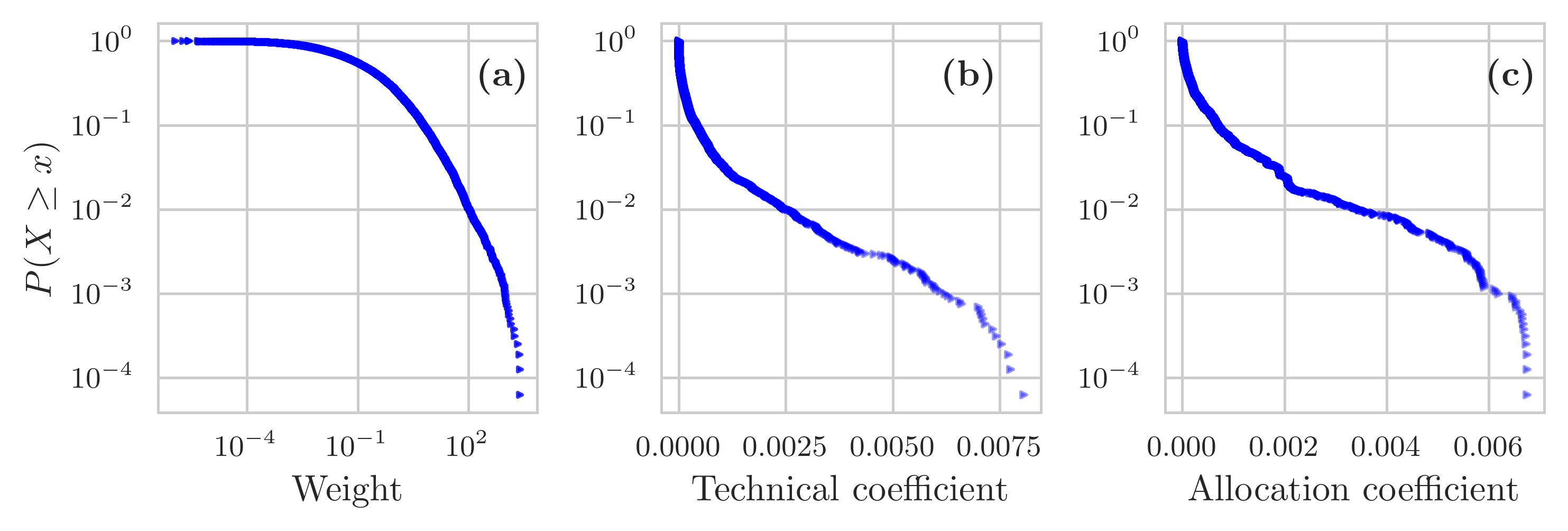}
\caption{\textbf{(a)} CCDF of the expected weights, values in thousand USD. \textbf{(b)}, \textbf{(c)} CCDF of, respectively, the expected technical and allocation coefficients in semi-log scale for FactSet.}
\label{fig:tech_alloc_coeff_factSet}
\end{figure}

The constraints on the total intermediate expenditure and sales are satisfied ($L_1$-error $= 6\times 10^{-4}$). Figure~\ref{fig:tech_alloc_coeff_factSet}a shows the reconstructed weight distribution, which visually appears to have heavy tails. As done for Ecuador, we fit a power-law distribution. The estimated power-law exponent is lower than that recovered for Ecuador (1.0 for FactSet and 1.3 for Ecuador, see Figure~\ref{fig:weight_ccdf_factset}) and on the lower end of what is observed for other supply chain networks \citep{bacil2022emprical}. The reconstructed technical coefficients (Figure~\ref{fig:tech_alloc_coeff_factSet}b) and allocation coefficients (Figure~\ref{fig:tech_alloc_coeff_factSet}c) have a narrow range of variation and tend to be much smaller than those reconstructed for Ecuador. For FactSet, the technical coefficients reach a maximum value of approximately 0.008, while for the expected Ecuadorian network, they can be as high as 0.570. Similarly, the allocation coefficients are not higher than $\sim$0.007 in FactSet, while in Ecuador, they reach a maximum value of 0.778. 

Given the data collection method of customer-supplier relations in FactSet, there is a bias towards observing links with customers that account for 10\% or more of a firm's annual revenues. Therefore, we would expect the CCDF of the allocation coefficients to have most of the mass around, or at the very least include this 10\% threshold. Instead, the maximum value is around 0.7\%, well below this threshold. The average expected allocation coefficient is $3\times10^{-4}$ while the median is $8\times10^{-5}$ (Table~\ref{tab:tech_coeff_summary_sats_factset}), both of which are three to four orders of magnitude smaller than what we would have expected given the data collection method. 

\begin{table}[!htbp]
\small
\centering
\def\arraystretch{1.1}
\begin{tabular}{l *{2}{c}}
\toprule
 & \multicolumn{2}{c}{Expected} \\
 \cmidrule{2-3}
& Technical coefficient &  Allocation coefficient\\
\midrule
Mean & 2$\times 10^{-4}$ & 3$\times 10^{-4}$ \\
Median & 3$\times 10^{-5}$ & 8$\times 10^{-5}$ \\
Standard dev. & 5$\times 10^{-4}$ & 6$\times 10^{-4 }$\\
\bottomrule
\end{tabular}
\caption{Summary statistics of the technical and allocation coefficients for FactSet.}
\label{tab:tech_coeff_summary_sats_factset}
\end{table}

\begin{figure}[!htbp]
\centering
\includegraphics{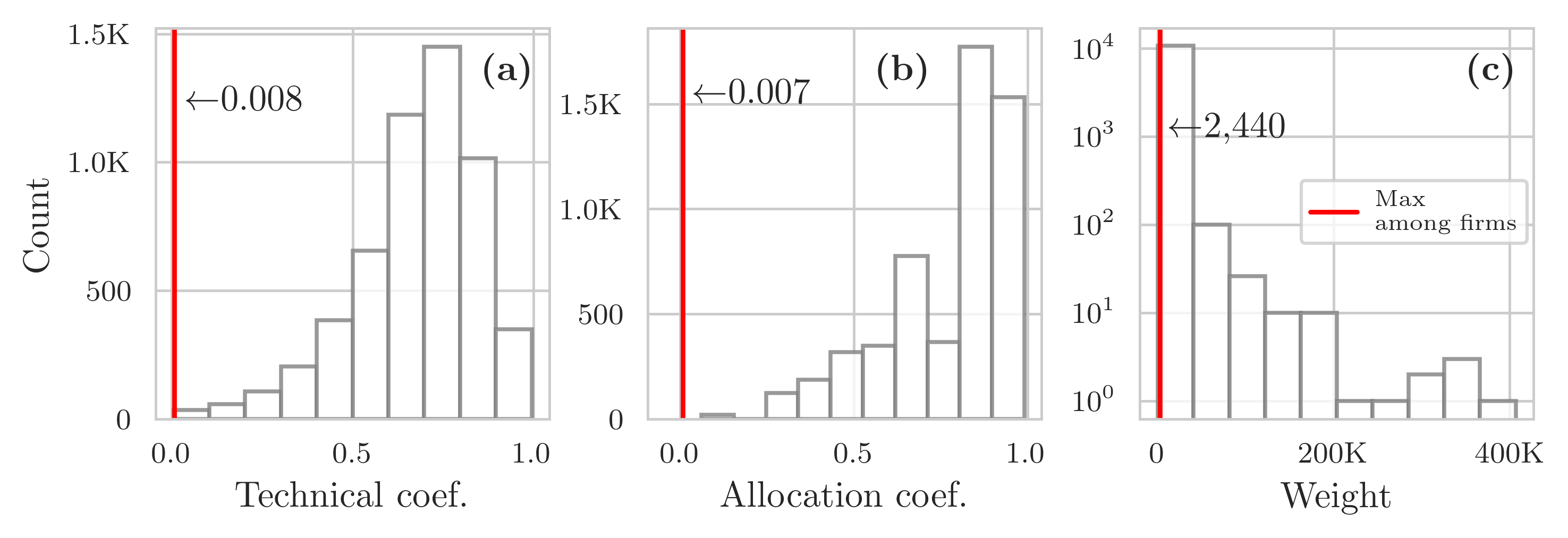}
\caption{Trades between firms and the proxy node. \textbf{(a)} Percentage of inputs that firms buy from the proxy node (i.e, their technical coefficients with the proxy node). \textbf{(b)} Percentage of sales that firms make to the proxy node (i.e., their allocation coefficients with the proxy node). \textbf{(c)} Weights among firms and the proxy node (in both directions); values in thousand USD. We do not show quantities regarding trades of the proxy node with itself. The red line marks the maximum value of the quantity among firms only.}
\label{fig:coeff_weight_among_firms_proxy_factset}
\end{figure}

All three quantities have a narrow range of variation and are smaller than expected because total intermediate sales and expenditure of the proxy node (which represent the constraints that need to be satisfied in the weights allocation) are much bigger than those of the other firms. The proxy node accounts for 80\% of intermediate expenditure and 75\% of intermediate sales. To compare, in Ecuador, the proxy node accounts for 29\% and 14\%, respectively. Therefore, in FactSet, firms make most of their trades with the proxy node (Figure~\ref{fig:coeff_weight_among_firms_proxy_factset}). The maximum transaction value for transactions between firms and the proxy node is \$407 million. But, the overall maximum is between the proxy node and itself, where it is \$48 billion, 3 orders of magnitude bigger than the maximum value among firms as well as among firms and the proxy node. Consequently, the coefficients representing trades among firms are much smaller than those representing the trades between firms and the proxy node, as shown in Figure~\ref{fig:coeff_weight_among_firms_proxy_factset}.

\FloatBarrier
\subsubsection{Discussion}\label{sec:discuss_weights}
Since the other studies that reconstruct firm-level production networks do not assess how their network reconstruction method performs on reconstructing weights, technical and allocation coefficients \citep{inoue2019firm,welburn2020systemic,hooijmaaijers2019methodology,reisch2021inferring,ialongo2021reconstructing,Hillman2021newFirm-levelModel}, we cannot compare our results with these studies. Moreover, although several studies assess network reconstruction methods on the international trade network (ITN) or financial networks, few assess the reconstruction of the weights. Most of the studies look at higher-order network properties or dynamic indicators of systemic risk, which we discuss in Section~\ref{sec:results_macroscale_prop}. 

Table~\ref{tab:comparison_literature_weights} shows a comparison of our results with those of the literature. \cite{parisi2020faster} test the reconstruction method on the ITN and the Electronic Market for Interbank Deposits (e-MID). They find a similar percentage of empirical weights that fall in the 50\% CI. For the e-MID network, depending on the year, between 35\% and 55\% of the empirical weights fall in the 50\% CI, while around 30\% of the weights fall in the 50\% CI for the ITN. They find that the empirical and expected link weights have a Pearson correlation of 0.50 for the e-MID and 0.75 for the ITN. The only other study we could find providing a comparison metric for link weights is \cite{ramadiah2020reconstructing}. The authors test several reconstruction methods on Japan's bipartite bank-firm credit network. For the most disaggregated network, their reconstruction yields a cosine similarity of around 0.68 for the MaxEnt method (for bi-partite networks) and of 0.63 for the configuration fitness model with weights allocated using the IPF algorithm; these are lower than what we find for the weights but similar to our results for the technical and allocation coefficients.

\begin{table}[!htbp]
\resizebox{\textwidth}{!}{%
\centering
\def\arraystretch{1.1}
\begin{tabular}{l l l r l c l}
\toprule
Data set & Method & Quantity & \multicolumn{1}{c}{Pct in 50\% CI} & Measure & Score & Source\\
\midrule
Ecuador SC & CReM & Weight &  47\% & Cosine & 0.93 & This paper \\
Ecuador SC & CReM & \begin{tabular}{@{}l@{}}Technical\\coefficient\end{tabular}&  & Cosine & 0.72 & This paper \\
Ecuador SC & CReM & \begin{tabular}{@{}l@{}}Allocation\\coefficient\end{tabular} & & Cosine & 0.76 & This paper \\
ITN & CReM & Weight &  30\% & Pearson & 0.75 & \cite{parisi2020faster}\\
e-MID & CReM & Weight & 35-55\% & Pearson & 0.50 & \cite{parisi2020faster}\\
\begin{tabular}{@{}l@{}}Japan's\\bank-firm credit\end{tabular} & MaxEnt & Weight & & Cosine & 0.68 & \cite{ramadiah2020reconstructing} \\
\begin{tabular}{@{}l@{}}Japan's\\bank-firm credit\end{tabular} & BFiCM + IPF & Weight & & Cosine & 0.63 & \cite{ramadiah2020reconstructing} \\
\bottomrule
\end{tabular}
}
\caption{Microscale quantities: comparison of our results with the literature.  In column ``Data set'', ``Ecuador SC'' stands for Ecuador supply chain network, ``ITN'' for international trade network and e-MID for electronic market for interbank deposits. In column ``Method'', CReM stands for the conditional reconstruction method that we employ in this paper \citep{parisi2020faster}.  Besides the MaxEnt method, \cite{ramadiah2020reconstructing} use a two-step procedure. In the first step, the binary topology is reconstructed using a bipartite fitness-induced configuration model (BFiCM) \citep{squartini2017enhanced}, which extends the fitness-induced configuration model of \cite{cimini2015systemic} to the bipartite case. In the second step,  weights are allocated using the iterative proportional fitting algorithm. For \cite{ramadiah2020reconstructing}, we show results for the most disaggregated network and only for the two methods with the highest cosine similarity.}
\label{tab:comparison_literature_weights}
\end{table}

While the power-law exponent of FactSet's weight distribution is in the ranges of what had been found for other supply chain networks \citep{bacil2022emprical}, the reconstructed weights and coefficients among firms are much smaller than expected. There are two main factors that deteriorate the quality of the reconstruction and act mainly through the constraints on intermediate sales and expenditure: the data cleaning procedure and the data imputations. On the one hand, the data cleaning procedure implies that we had to exclude many firms from the network (see Appendix~\ref{sec:cleaning_financials}); this leads to (1) a higher share of the proxy node in the economy and (2) excluding many existing supply-chain relations. On the other hand, the data imputations concerning final demand and labour costs affect the values of firms' intermediate sales and expenditures. Because firms report the cost of goods sold, which often includes labour costs, we do not always know intermediate costs exactly. Similarly, firms disclose their revenues (intermediate sales plus sales to final demand), so for all firms, we do not know their intermediate sales exactly. As shown in Appendix~\ref{app:constr_labor_exp}, labour costs tend to be overestimated. Although we cannot test whether final demand is over or underestimated, we think we are very likely overestimating it for the majority of firms. The main reason for overestimating final demand is the use of national I-O tables at the sector level, which have a very different treatment of the wholesale and retail sectors compared to firm-level data. National accounts treat wholesale and retail as ``pass-through'' sectors, accounting only for the trade margin they make and re-distribute the rest of their output among the other industries.\footnote{
If wholesale is trading service goods, then all of its output is distributed to other products.
}
The final demand of these other industries thus becomes (fictitiously) higher. In firm-level data quite the opposite happens, with many firms selling to retail and wholesale firms that then sell to final demand. We refer to Appendix~\ref{sec:proxy_nodes_rebalanc} and \cite{bacil2022emprical} for a longer discussion on differences between national accounts and firm-level data.

The data cleaning and data imputation problems just discussed imply that the constraints on intermediate sales and expenditure of the proxy sector are much bigger than those of other firms. Therefore, to satisfy the constraints posed in the maximum entropy procedure, the weights allocated to the proxy node need to be much bigger than those among firms, meaning that firms buy most of their inputs and sell most of their output to the proxy node. This deteriorates the reconstruction of the weights, and of the technical and allocation coefficients among firms. 

Lastly, the findings for both Ecuador and FactSet suggest that additional mechanisms, not captured by the constraints we pose, may be at work that are essential to the formation of network weights. Further unravelling what these mechanisms are could improve the reconstruction.
\subsection{Results: higher-order and macroscale properties}
\label{sec:results_macroscale_prop}

First, this section discusses the results for the output multipliers and the influence vector; we start with Ecuador and then FactSet. Second, we show the results for aggregate volatility. We do not assess aggregate volatility for FactSet as that would entail estimating TFP using an econometric technique such as that outlined in \cite{magerman2016heterogeneous}, which is outside of the scope of this paper. We conclude this section with a discussion.

\subsubsection{Multipliers}
\begin{figure}[!htbp]
\includegraphics{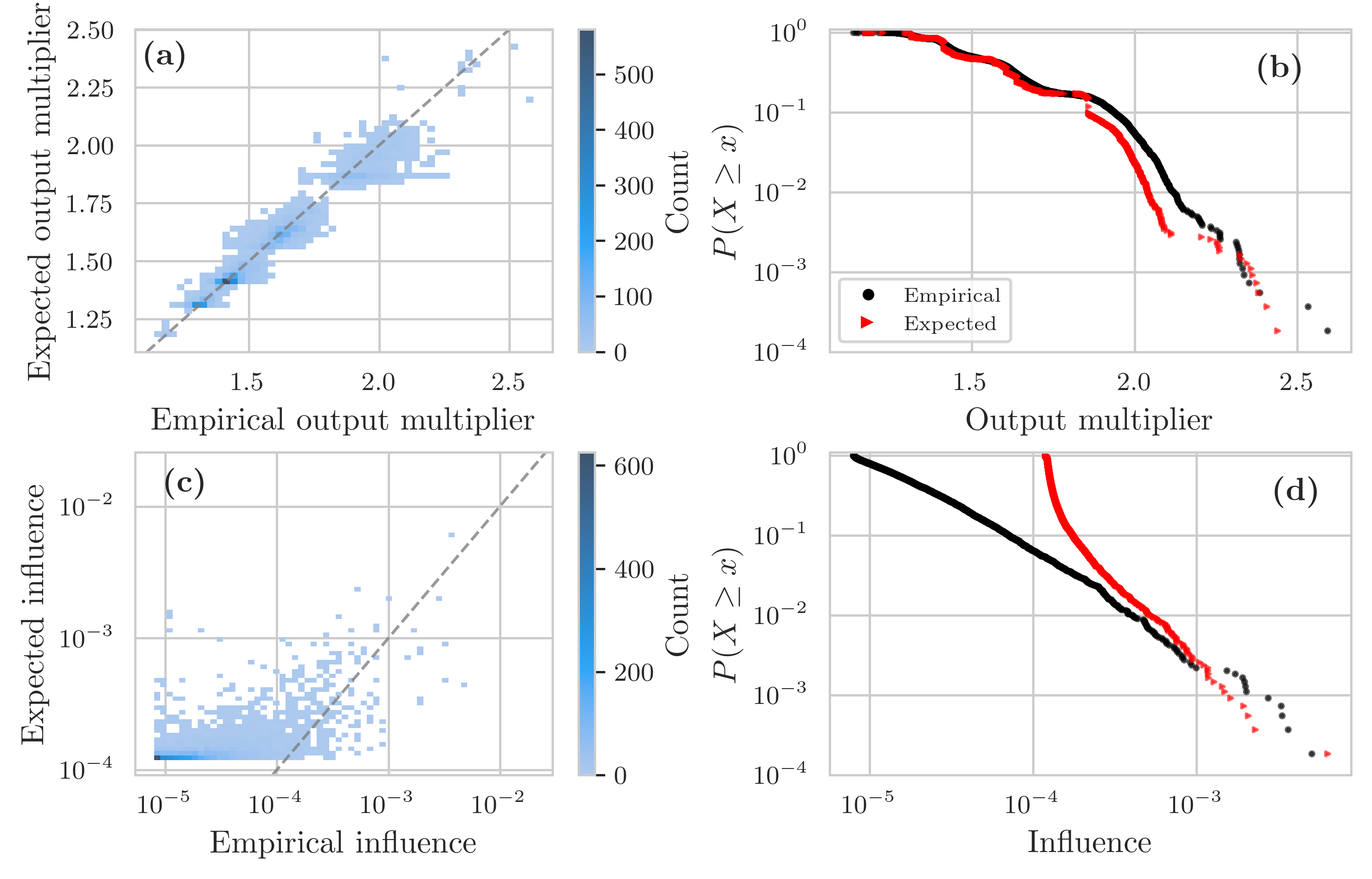}
\caption{\textbf{(a)}, \textbf{(c)} 2D histogram of the empirical (\textit{x}-axis) and reconstructed (\textit{y}-axis) output multipliers and influence vector, respectively, for Ecuador. Perfect prediction is achieved when points lie on the 45-degree line (dashed grey line). We use 50 log-spaced bins for both axes and count the number of points falling in each square. \textbf{(b)}, \textbf{(d)} CCDF of the output multipliers and influence vector, respectively, for Ecuador. Black dots refer to the empirical CCDF and red triangles to the CCDF of the expected network. For calculating the empirical multipliers, we use the full network and consider the multipliers of firms in our test network only.}
\label{fig:output_multip_ecuador}
\end{figure}

\paragraph{Ecuador.}
The reconstruction method performs well at reproducing the empirical output multipliers but not that well at reconstructing the influence vector. Figure~\ref{fig:output_multip_ecuador}a and~\ref{fig:output_multip_ecuador}b show the empirical (\textit{x}-axis) and the expected (\textit{y}-axis) output multipliers and influence vector, respectively. For the output multipliers, points cluster fairly tightly around the identity line (dashed grey line), while for the influence vector most points are located at the bottom left corner, suggesting that the influence is consistently overestimated. However, there are a few exceptions, which tend to be firms with higher influence (Figure~\ref{fig:pred_error_outputM_ec}). These findings are further confirmed in Figure~\ref{fig:output_multip_ecuador}c and~\ref{fig:output_multip_ecuador}d, showing the CCDF of the (empirical and reconstructed) output multipliers and influence vector, respectively. It also highlights that the minimum of the empirical influence vector is around one order of magnitude smaller than the reconstructed one. The cosine similarity of the output multipliers is higher than that of the influence vector (0.99 and 0.56, respectively); however, the other error metrics are less clear cut (Table~\ref{tab:higher_order_comparison_metrics}). The reconstruction method can recover the first two moments and the median of the output multipliers (Table~\ref{tab:higher_order_summary_sats_ec}). For the influence vector, we can recover only the standard deviation.

In Figure~\ref{fig:output_multip_ecuador}a clusters tend to form. Clusters arise because, for the output multipliers, we simulated firms' value-added using sector-level data. The interaction of firms' intermediate expenditures, sectoral $\nu_s$'s and the binary topology produces those clusters; see Appendix~\ref{sec:multipliers_sec} for more details.

\begin{table}[!htbp]
\small
\centering
\def\arraystretch{1.1}
\begin{tabular}{l*{5}{c}}
\toprule
& \multicolumn{2}{c}{Output multiplier}  && \multicolumn{2}{c}{Influence vector}\\
\cmidrule{2-3}\cmidrule{5-6}
& Empirical & Expected && Empirical & Expected \\
\midrule
Mean & 1.423 & 1.553 &  & 4$\times 10^{-5}$ & 1$\times 10^{-4}$ \\
Median & 1.420 & 1.475 &  & 2$\times 10^{-5}$ & 1$\times 10^{-4}$ \\
Standard dev. & 0.247 & 0.202 &  & 1$\times 10^{-4}$ & 1$\times 10^{-4}$ \\
\bottomrule
\end{tabular}
\caption{Summary statistics of the output multipliers and influence vector for Ecuador. For each multiplier, the first column reports summary statistics for the multipliers calculated on the empirical network while the second column on the reconstructed network. We show results for one of the reconstructions; all three quantities have virtually the same summary statistics across the 50 randomised empirical and reconstructed networks. We excluded the proxy node from the calculations of the expected multipliers. For calculating the empirical multipliers, we use the full network and consider the multipliers of firms in our test network only.}
\label{tab:higher_order_summary_sats_ec}
\end{table}


\begin{table}[!htbp]
\small
\centering
\def\arraystretch{1.1}
\begin{tabular}{l *{4}{c}}
\toprule
Type & RMSE & MAE & MedAE & Cosine similarity\\ 
\midrule
Output multipliers & 0.036 & 0.024 & 0.015 & 0.999 \\
 & (3$\times 10^{-4}$) & (1$\times 10^{-4}$) & (2$\times 10^{-4}$) & (8$\times 10^{-6}$) \\
Influence vector & 0.033 & 0.024 & 0.022 & 0.560 \\
 & (2$\times 10^{-4}$) & (2$\times 10^{-5}$) & (1$\times 10^{-5}$) & (3$\times 10^{-3}$) \\

\bottomrule
\end{tabular}
\caption{Comparison metrics for Ecuador's multipliers. As defined in the main text, RMSE denotes the root mean squared error, MAE the mean absolute error and MedAE the median absolute error. For each multiplier, we show its mean value across the 50 randomised reconstructions. Below the mean value, the standard deviation in parenthesis. We excluded the proxy node from the calculations of the expected multipliers. For calculating the empirical multipliers, we use the full network and consider the multipliers of firms in our test network only.}
\label{tab:higher_order_comparison_metrics}
\end{table}

\begin{figure}[!htbp]
\centering
\includegraphics{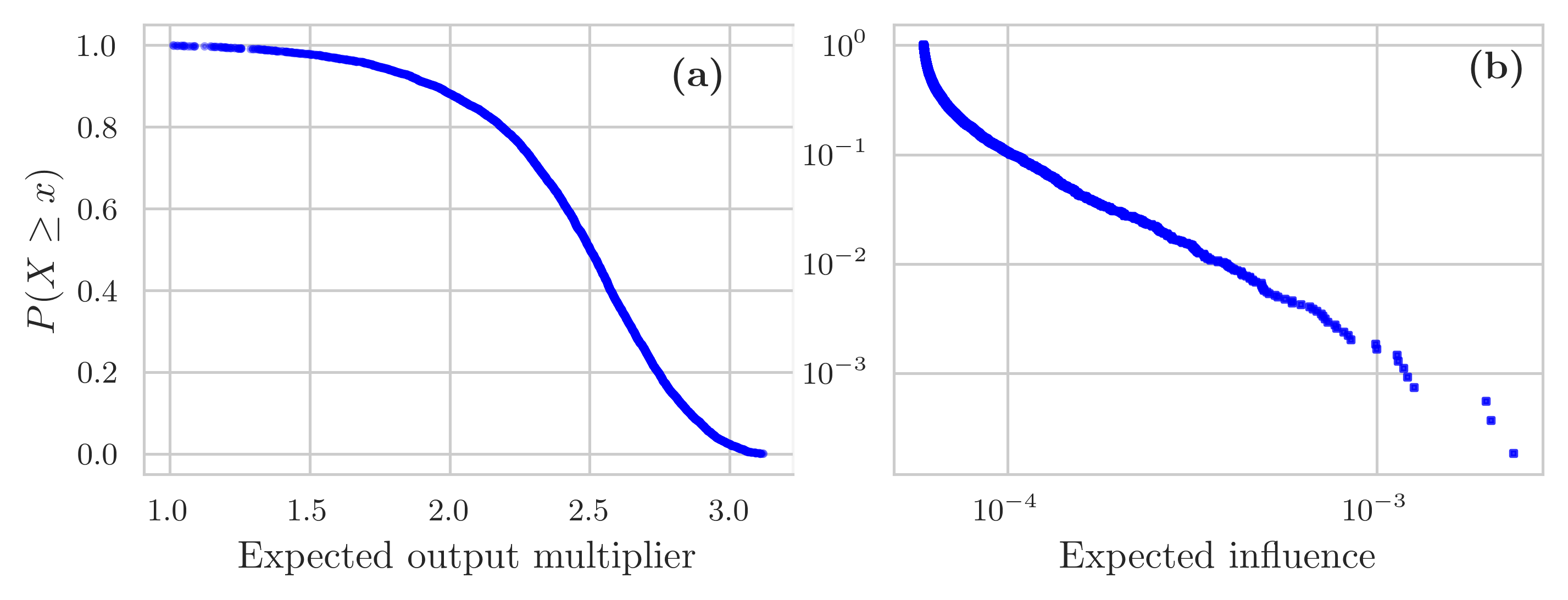}
\caption{CCDF of the expected \textbf{(a)} output multipliers and \textbf{(b)} influence vector for FactSet. The multipliers of the proxy node are excluded.}
\label{fig:multipliers_factset}
\end{figure}

\FloatBarrier
\paragraph{FactSet.}
Figure~\ref{fig:multipliers_factset}a and~\ref{fig:multipliers_factset}b show the CCDF of, respectively, the expected output multipliers and the expected influence vector for FactSet. While the CCDF of the influence vector displays heavy tails, that of the output multipliers does not. As done for the weight distribution, we compare with empirical findings of other networks where the influence vector is found to have heavy tails and likely follows a power-law with a divergent second moment \citep{bacil2022emprical}. The estimated power-law exponent is equal to 1.9, higher than what found for Belgium \citep[1.12,][]{magerman2016heterogeneous}, Hungary and Ecuador \citep[around 1.3-1.5 and 1.2-1.4, respectively,][]{bacil2022emprical}. We show the CCDF and its power-law fit in Appendix~\ref{app:infl_vec}.\footnote{We also fit a power-law distribution to Ecuador's influence vector and recover an exponent of 2.0. However, the fit was rather poor. We show results for Ecuador in Appendix~\ref{app:infl_vec}.} 
The output multipliers have a much higher median, first and second moment in FactSet (Table~\ref{tab:output_mult_summary_sats_factset}) compared to Ecuador (Table~\ref{tab:higher_order_summary_sats_ec}), while the influence vector has lower moments.

\begin{table}[!htbp]
\small
\centering
\def\arraystretch{1.1}
\begin{tabular}{l *{2}{c}}
\toprule
 & \multicolumn{2}{c}{Expected}\\
 \cmidrule{2-3}
 & Output multiplier & Influence vector \\
\midrule
Mean & 2.443 & 8$\times 10^{-5}$ \\
Median & 2.501 & 6$\times 10^{-5}$ \\
Standard dev. & 0.364 & 8$\times 10^{-5}$ \\
\bottomrule
\end{tabular}
\caption{Summary statistics of the output multipliers and the influence vector for FactSet. The multipliers of the proxy node are excluded.}
\label{tab:output_mult_summary_sats_factset}
\end{table}

\FloatBarrier
\subsubsection{Supply-side shocks and aggregate volatility}
We measure how firm-level TFP shocks affect macroeconomic output by propagating through the supply chain network using Equation~\ref{eq:gr_volatility_gdp}. We simulate TFP shocks as explained in Section~\ref{sec:macroscale_prop}. We then use the empirical influence vector to calculate the empirical volatility and the reconstructed influence vector to calculate the predicted volatility. The volatility predicted by our reconstructed network is much higher than the empirical one. Using the empirical network, GDP volatility is 6.4\%, while the mean across the 50 reconstructions is 102.3\% with a standard deviation of 0.002 (Table~\ref{tab:aggr_volatility}). 

To understand why we predict such a high GDP volatility, we do a variance decomposition analysis and look at the role of the proxy node in explaining aggregate volatility:
\begin{linenomath}
\begin{equation*}
    \sigma^2_{\Delta y} = \sum_{i\neq\kappa} \text{Var}(\Delta\epsilon_i)v_i^2 + \text{Var}(\Delta\epsilon_\kappa)v_\kappa^2\;,
\end{equation*}
\end{linenomath}
where the first term on the RHS is the contribution to total GDP volatility of the firms in the test network and the second term is the contribution of the proxy node, indexed by $\kappa$. We further look at shares
\begin{linenomath}
\begin{equation*}
    \Lambda_{S} = \frac{\sum_{i\in S} \text{Var}(\Delta\epsilon_i)v_i^2}{\text{Var}(\Delta\epsilon_\kappa)v_\kappa^2 + \sum_{j\neq\kappa} \text{Var}(\Delta\epsilon_j)v_j^2}\;.
\end{equation*}
\end{linenomath}
where $S$ is the set of observations for which we are computing the share of the variance. We use the variance to ensure that shares sum up to 1. The proxy node contributes to 99.3\% of the variance, while all the other firms to 0.7\%. If we calculate GDP volatility without including the proxy node, the predicted GDP volatility drops to 8.8\%. 

\begin{table}[!htbp]
\small
\centering
\def\arraystretch{1.1}
\begin{tabular}{l *{6}{c}}
\toprule
 & & \multicolumn{2}{c}{Reconstructed} && \multicolumn{2}{c}{Benchmark}\\
 \cmidrule{3-4} \cmidrule{6-7}
 & Empirical & Proxy node & No proxy node &&  Proxy node & No proxy node\\
\midrule
GDP volatility & 6.4\% & 102.3\% & 8.8\% && 94\% & 7.0\% \\
               &    & (0.0020) & (0.0010) && (0.0030) & (0.0002)\\ 
\bottomrule
\end{tabular}
\caption{Predicted aggregate GDP volatility using the empirical, the reconstructed and the benchmark influence vector (Equation~\ref{eq:gr_volatility_gdp}). For the reconstructed and benchmark, ``Proxy node'' reports the predicted aggregate volatility calculated including the proxy node, while ``No proxy node'' excludes the proxy node from the calculations. Values for the reconstructed and the benchmark show the average taken over each of the 50 randomised networks. Below the mean value, we show the standard deviation in parenthesis.}
\label{tab:aggr_volatility}
\end{table}

\paragraph{Benchmark.}
To benchmark our results, we first calculate the influence vector assuming that each firm buys the same proportion of inputs from its suppliers (i.e., we modify $\bm{\Omega}$ in Equation~\ref{eq:influence_vector}) and then compute aggregate volatility using Equation~\ref{eq:gr_volatility_gdp}. Assigning homogeneous input shares for each firm still satisfies the constraints on the intermediate costs, but the constraints on intermediate sales are not guaranteed to be satisfied. Our benchmark yields a volatility of 94\% when the proxy node is included and 7.0\%  when excluded (Table~\ref{tab:aggr_volatility}). The volatility predicted by the benchmark is 0.6 percentage points higher than the empirical volatility, while the volatility of the reconstruction is 2.4 percentage points higher than the empirical volatility.

\FloatBarrier
\subsubsection{Discussion}
As noticed in Section~\ref{sec:discuss_weights}, we cannot compare with previous results of other studies reconstructing firm-level production networks. For financial networks and the ITN, different reconstruction methods perform reasonably well in reconstructing higher-order network properties such as the weighted clustering coefficient \citep[e.g.,][]{mastrandrea2014enhanced,cimini2015systemic,cimini2015estimating,parisi2020faster}. Table~\ref{tab:comparison_literature_macroscale} reports some of the findings in the literature. For financial networks, \cite{ramadiah2020reconstructing} find that all the reconstruction methods they employ underestimate the level of systemic risk, except for a small region of the parameter space. \cite{anand2015filling} report similar findings for MaxEnt, but find that the minimum density method overestimates systemic risk. \cite{digiangi2018assessing} find that the cross-entropy capital asset pricing model can reproduce very well systemic risk while the other ensemble methods overestimate or underestimate it depending on the shock scenario. Differently, individual banks' systemic risk and indirect vulnerability, which could be seen as akin to the multipliers we test, are consistently underestimated across all the methods they assess. The degree-corrected gravity model can reproduce DebtRank (also akin to the multipliers), with a Person correlation equal to 1 \citep{cimini2015systemic}. 

\begin{table}[!htbp]
\resizebox{\textwidth}{!}{%
\centering
\begin{tabular}{p{.17\textwidth}p{.29\textwidth}p{.34\textwidth}p{.20\textwidth}}
\toprule
Dataset & Method & Finding & Source\\
\midrule
Ecuador SC & CReM & Overestimates influence vector, cosine sim. = 0.56 & This paper\\
Ecuador SC & CReM & Under/Overestimates output multipliers, cosine sim. = 1. & This paper\\
e-Mid & DcGM & DebtRank: correlation = 1 & \cite{cimini2015systemic}\\
ITN & DcGM & DebtRank: corrrelation = 1 & \cite{cimini2015systemic}\\
US bank-asset & CE CAPM, Max. entropy CAPM, BWCM, BECM & Underestimate banks' measures of systemic risk & \cite{digiangi2018assessing}\\ 
\midrule
Ecuador SC & CReM & Overestimates aggregate volatility & This paper\\
German banks & MaxEnt & Underestimates systemic risk & \cite{anand2015filling}\\
German banks & Minimum density & Overestimates systemic risk & \cite{anand2015filling}\\
US bank-asset & Cross-entropy CAP model & Reproduces well aggregate vulnerability & \cite{digiangi2018assessing}\\ 
US bank-asset & Max. entropy CAP model, BWCM, BECM & Over/underestimate aggregate vulnerability depending on the shock scenario & \cite{digiangi2018assessing}\\ 
Japan bank-firm credit & MaxEnt, Min. density, CF + IPF, BFiCM + IPF & Underestimate the average probability of default & \cite{ramadiah2020reconstructing} \\
\bottomrule
\end{tabular}
}
\caption{Higher-order and macroscale quantities: comparison of our results with the literature. In column ``Dataset'', ``Ecuador SC'' stands for Ecuador supply chain network, ``ITN'' for international trade network and ``e-MID'' for electronic market for interbank deposits. In column ``Method'', CReM stands for the conditional reconstruction method that we use in this paper \citep{parisi2020faster}. ``DcGM'' stands for degree-corrected gravity model \citep{cimini2015systemic}, ``CE CAPM'' for cross-entropy capital asset pricing model (it is a deterministic method), ``Max. entropy CAPM'' for maximum entropy capital asset pricing model, ``BWCM'' for bipartite weighted configuration model and ``BECM'' for bipartite enhanced configuration model (all of the last three methods are ensemble methods). ``Min. density'' is the minimum density method developed by \cite{anand2015filling}. ``CF'' is the bipartite maximum entropy configuration model and ``BFiCM'' is the bipartite fitness-induced configuration model \citep{squartini2017enhanced}; in both cases, weights are allocated in a second step using the iterative proportional fitting algorithm.}
\label{tab:comparison_literature_macroscale}
\end{table}

While the vast majority of the studies find that systemic risk is underestimated, for Ecuador, we find that the conditional maximum entropy reconstruction method we employ overestimates aggregate volatility, similar to what \cite{anand2018missing} find for the minimum density method. First, while including the proxy node in the calculations of aggregate volatility is supposed to capture the volatility of the firms that were excluded from the test network, there is a fundamental difference between the role the proxy node has in the network and that of the firms that we excluded. The proxy node is connected to all the other firms in the test network. In contrast, the firms excluded from the test network are less well connected in the empirical network. Additionally, the proxy node has the biggest influence, which equals 0.19. Therefore, when the proxy node is shocked, it immediately passes 19\% of the shock to all the other firms in the network. To compare, the firm with the biggest influence in the empirical network has an influence of 0.005, with the biggest influence among the firms excluded from the test network being much lower and equal to 0.0002. Noting also that the firms excluded from the test network contribute to a mere 7\% of aggregate volatility while the firms included in the test network to 93\%, it makes sense to exclude the proxy node from the calculation of aggregate volatility as it is not a good aggregate representation of the rest of the economy, at least in this context. In fact, while the inclusion of the proxy node degrades the prediction of aggregate volatility, it enhances the reconstruction of the multipliers; see Appendix~\ref{app:centrility_proxy_node}.

Second, aggregate volatility is still overestimated even when the proxy node is excluded because the influence vector tends to be overestimated. Notice that the benchmark can predict aggregate volatility more accurately since it tends to overestimate influences lightly less than the reconstruction (see Appendix~\ref{app:unif_reconstr_influence}). The reconstructed influence vector is overestimated for a combination of two factors, one of which is related to the binary topology and the other to the weighted topology. We can see how the binary topology affects the influence vector by looking at the out-degrees, which are highly affected by the selection of firms and the link deletion. In creating the test network, firms lose more customers than suppliers (see Appendix~\ref{app:in-out-degs}). Already at the first step (selecting the firms to keep in our test network), the maximum number of customers (out-degree) decreases almost 7 folds, while the maximum number of suppliers only 3 folds. Since the influence vector calculates the weighted sum of the number of walks from firm $i$ to firm $j$ (so following the outgoing edges starting at $i$) for walks of various lengths and the out-degrees have been considerably truncated, it cannot capture walks of longer lengths. Instead, the output multipliers, relying on the incoming edges and thus the in-degrees, are that not affected. The weighted topology comes into play because the reconstruction method tends to overestimate weighted quantities for intermediate and small values, which are numerous. Although higher (and other) weights can be underestimated, these are not abundant enough and the magnitude of the underestimation is not big enough compared to the quantity and the size of the overestimation of low and intermediate weights. 

For FactSet, a significant result is that the estimated power-law exponent of the distribution of the influence vector implies a divergent second moment. \cite{acemoglu2012network} show, at a theoretical level, that the influence vector affects aggregate volatility, so it is important that we can recover an exponent more or less in line with empirical observations.\footnote{
\citeauthor{acemoglu2012network}'s \citeyearpar{acemoglu2012network} theoretical result is exemplified in Equation~\ref{eq:gr_volatility_gdp}, which shows that aggregated volatility scales with the Euclidean norm of the influence vector. Luca's diversification argument implies that aggregate volatility decays as $N^{1/2}$. However, if the CCDF of the influence vector is Pareto distributed with parameter $\gamma \in (1, 2)$, the diversification argument no longer holds and aggregate volatility decays much more slowly, as $N^{1/(\gamma - 1)}$ \citep{carvalho2019production}.
}
\subsection{Results: different numbers of unknown links}\label{sec:differ_number_unknowns}
We now discuss how the reconstruction method performs when the number of unknown links in the test network changes; we keep the number of firms constant. We delete 0\%, 10\%, 20\% and so on up to 90\% of the links; we also show results for the reconstruction matching the mean degree of FactSet, which has 96\% of unknown links. For each percentage of unknown links, we simulate 50 randomised networks and investigate the number of weights that fall into the 50\% CI, and the median absolute error and the cosine similarity for microscale and higher-order quantities. We also assess aggregate volatility.

\begin{figure}[!htbp]
    \centering
\includegraphics{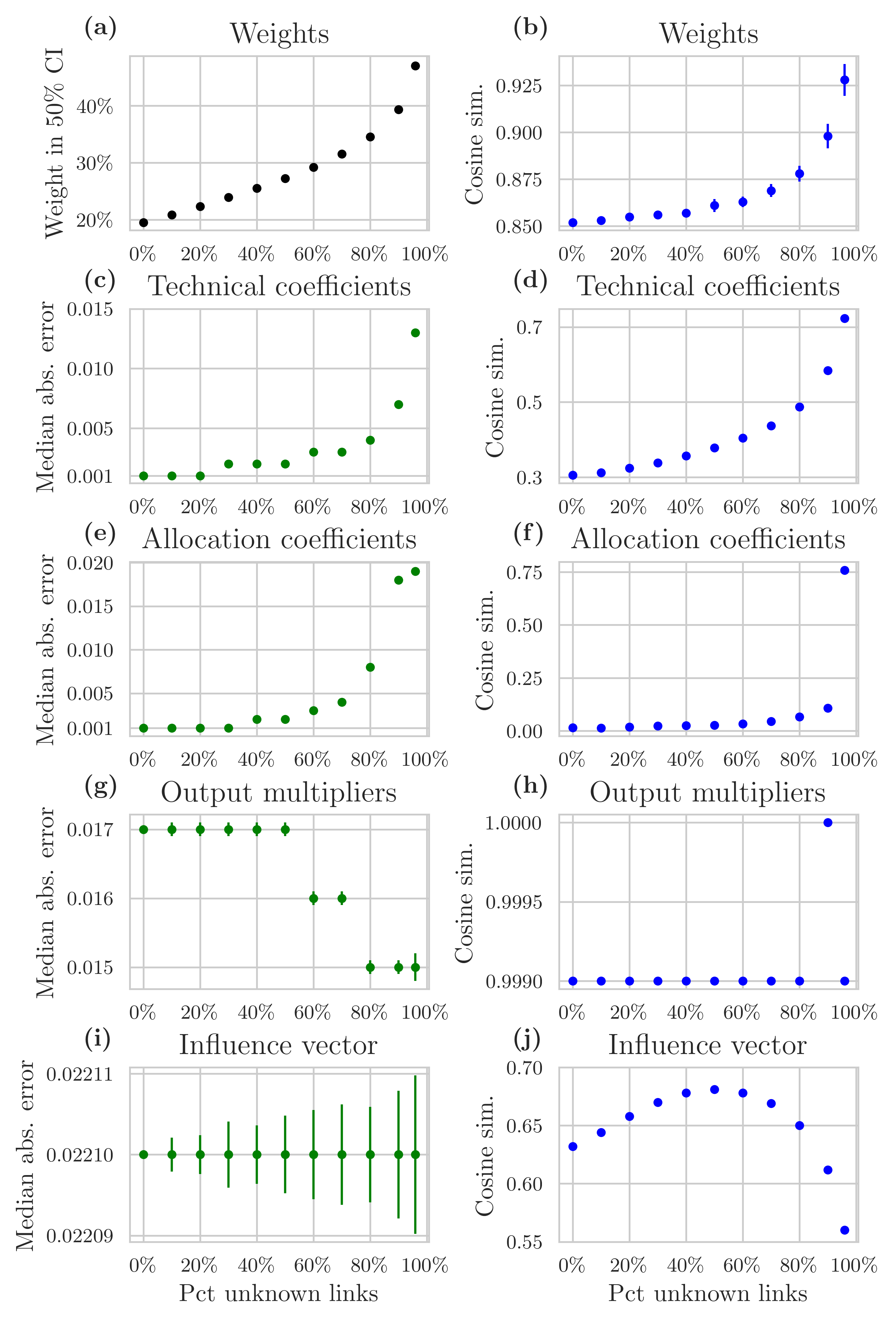}
    \caption{Error metrics and similarity measures as the number of unknown links increases. \textbf{(a)} Percentage of empirical weights in the 50\% confidence interval. \textbf{(b)}, \textbf{(d)}, \textbf{(f)}, \textbf{(h)}, \textbf{(j)} cosine similarity for, respectively, weights, technical and allocation coefficients, output multipliers and influence vector. \textbf{(c)}, \textbf{(e)}, \textbf{(g)}, \textbf{(i)}, Mean absolute error for, respectively, technical and allocation coefficients, output multipliers and influence vector. Bars show the standard deviation across the 50 randomised networks.}
    \label{fig:Nlinks_unknown_weights_coeffs_mult}
\end{figure}

\paragraph{Microscale and higher-order quantities.}
Figure~\ref{fig:Nlinks_unknown_weights_coeffs_mult} shows the different metrics used for assessing the quality of the reconstruction of the weights, and the technical and allocation coefficients, and the multipliers. For the weights, we show the percentage of weights that fall in the 50\% CI (Figure~\ref{fig:Nlinks_unknown_weights_coeffs_mult}a) and the cosine similarity (Figure~\ref{fig:Nlinks_unknown_weights_coeffs_mult}b). For all the other quantities, we show the median absolute error (left column) and the cosine similarity (right column). The overall trend is that as the number of unknown links increases, so do the metrics.

As the number of unknown links increases, the cosine similarity increases, with the two multipliers being the only exceptions. Such a counter-intuitive finding arises from the link deletion mechanism. We are deleting links with a smaller weight with a higher probability, hence as the number of unknown links increases, the empirical weights associated with the known links are less heterogeneous. Since maximum entropy methods allocate weights as uniformly as possible given the constraints, the higher the number of unknowns, the better it can reconstruct the weights associated with the known links. Consequently, also more weights fall into the 50\% confidence interval.

For the allocation coefficients, the cosine similarity is considerably lower than that of all the other quantities and it jumps from 0.05 to 0.76 when the number of unknown links increases from 90\% to 96\%. This jump is associated with a considerable decrease in the number of customers firms have when the number of unknown links increases from 90\% to 96\% (see Appendix~\ref{app:in-out-degs}). Therefore, the allocation coefficients, which gauge the percentage of total output firm $i$ sells to firm $j$, are highly affected by a drastic decrease in the number of customers (see Appendix~\ref{app:comparison_Nlinks_unknown}). However, we would have expected this to have a negative effect. Instead, it enhances the predictions. On the contrary, the technical coefficients, which measure the percentage of inputs that $j$ buys from $i$, do not experience such a jump because the same drastic change does not happen for the number of suppliers. Moreover, firms tend to have more customers than suppliers \citep[see Appendix~\ref{app:in-out-degs} and][]{bacil2022emprical}, making it easier to guess the weights correctly from the supply side and thus reconstruct the technical coefficients better than the allocation coefficients throughout. 

The output multipliers always have the same cosine similarity. Since the output multipliers are derived from the technical coefficient matrix and depend on the incoming edges, hence the number of suppliers firms have, they are not that affected by the number of unknown links. Instead, the influence vector relies on the number of customers firms have (out-degree), which we saw being more affected by the deletion of firms and links, so the influence vector has a much lower cosine similarity than the output multipliers. It is not so clear why the cosine similarity has an inverted U-shaped curve, with the maximum value of 0.68 reached when 50\% of the links are unknown. It is however the case that the cosine similarity increases from 0.63 when all links are known to 0.68 (50\% unknown links) and then decreases again until it drops to 0.56 when 96\% of the links are unknown. It might be that these are small fluctuations of no particular value and that the reconstruction of the influence vector starts deteriorating when more than 50\% of the links are unknown because the out-degrees start being too affected by the deletion of the links. 

The median absolute error slightly increases for the technical and allocation coefficients because higher errors (in absolute value) tend to occur for higher weights. The median absolute error is constant for the multipliers.

\begin{figure}[H]
    \centering\includegraphics{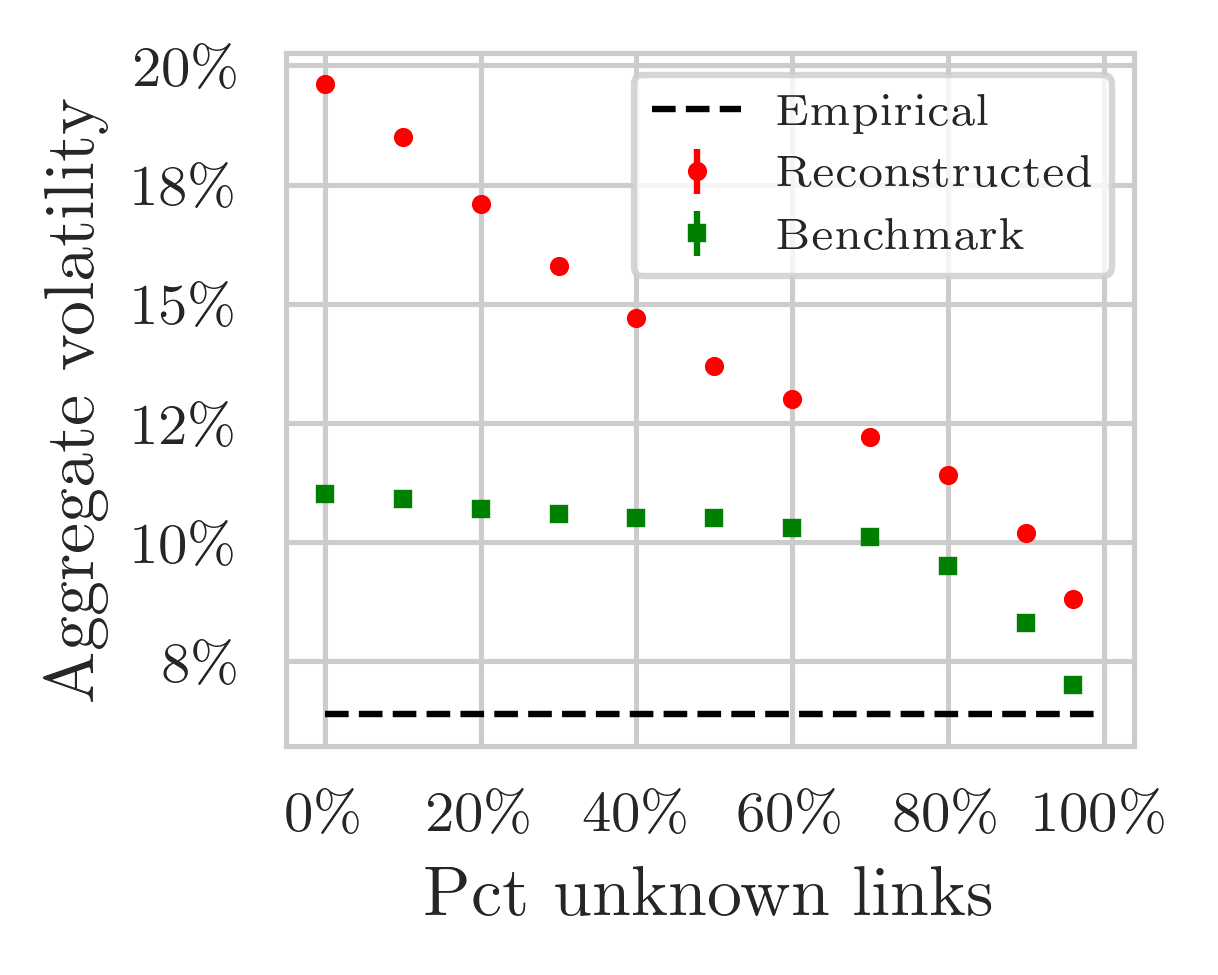}
    \caption{Aggregate volatility as the number of unknown links increases using the reconstructed network (red dots) and the benchmark (green squares). The black dashed line marks the empirical volatility. Bars show the standard deviation across the 50 randomised networks. We show the volatility calculated excluding the proxy node.}
    \label{fig:Nlinks_unknown_volatility}
\end{figure}

\paragraph{Aggregate volatility.}
Figure~\ref{fig:Nlinks_unknown_volatility} shows the predicted aggregated volatility (calculated excluding the proxy node) as the number of unknown links increases for the reconstruction (red dots), the benchmark (green squares) and the empirical volatility ( black dashed line). In line with the results for the case of 96\% of unknown links, the reconstruction always predicts a much higher volatility, as does the benchmark. Although the benchmark's predictions are closer to the empirical volatility. One would think that the higher the number of known links, the better we get at predicting aggregate volatility. However, this is not the case. In fact, the fewer links we know, the better we can predict GDP volatility. This is because the more links we know, the more we overestimate the influences (especially bigger influences, see Appendix~\ref{app:comparison_Nlinks_unknown}).

\FloatBarrier
\section{Conclusions}\label{sec:conclusions}
There is widespread interest in modelling the global economy from the bottom up. There is also widespread agreement that supplier-customer relations are an essential feature of such modelling efforts. However, data are scarce, have missing information and are not easily accessible. In this paper, we have made a couple of first steps in bringing this agenda forward.

First, we provided the first rigorous assessment of a network reconstruction method \citep{parisi2020faster} on the administrative dataset of Ecuador. We focused on reconstructing the weighted network given the binary topology when many links and firms are missing. An interesting finding is that the quality of the reconstruction of different quantities seems to depend on network features that are particularly sensitive to the sampling strategy of firms and links, something that future research should explore further. Second, we assessed whether a global dataset of listed firms, where many links and firms are missing, can be enhanced by merging it with sector-level data. We then used this ``augmented'' dataset for inferring the link weights using a conditional maximum entropy method \citep{parisi2020faster}. Our results show that further work needs to be done, especially in reconciling firms' financial accounts with national accounts, which is essential for better reconstructing the weighted production network, in particular when many firms are missing.

In our study, we assumed to know the binary topology (although partially) to cover a use case that could help reconstruct commercial datasets. A natural next step would be to predict links and then reconstruct weights, possibly with different degrees of knowledge of the production network. It is of particular value to understand the performance of reconstruction methods when one does not know the binary topology, as it could unlock the study of economies for which no such data is available. The reconstruction method we employed can easily accommodate the prediction of the binary topology, either partially or in its entirety.

Our assessment of macroscale quantities was rather limited and restricted to a standard general equilibrium I-O model \citep{acemoglu2012network}. Therefore, there is much research to be done on different models (and scenarios), especially agent-based models, where we believe that the accuracy of the reconstruction of microscale quantities matters more for the model's outcomes than for general equilibrium models. 

While many reconstruction methods are available, the high number of nodes and links in firm-level networks renders almost all of them infeasible. Future research is thus necessary to develop different reconstruction methods suited for large-scale firm-level networks. It is also important to conduct similar analyses on other firm-level datasets for which the ground truth network is available, as ours was only one of the first initial steps.

\subsubsection*{Aknowledgments}
We thank Fran\c{c}ois Lafond, Jose Moran, R. Maria del Rio-Chanona, Anton Pichler, Luca Mugo and Fabian Dablander for useful comments. We thank Doyne J. Farmer, Jangho Yang, Giancarlo Antonucci and the Complexity group at the Institute for New Economic Thinking at the Oxford Martin School for inspiring discussions. We have benefited from many comments from audiences at Cambridge IfM and CSH Vienna. We also thank the Servicio de Rentas Internas (SRI) and its Centro de Estudios Fiscales that provided the raw data of Ecuador for research purposes. This work was supported by the Oxford Martin Programme on the Post-Carbon Transition, the Institute for New Economic Thinking at the Oxford Martin School and Baillie Gifford. This research is based upon work supported in part by the Office of the Director of National Intelligence (ODNI), Intelligence Advanced Research Projects Activity (IARPA), via contract no. 2019-1902010003. The views and conclusions contained herein are those of the authors and should not be interpreted as necessarily representing the official policies, either expressed or implied, of ODNI, IARPA, or the U.S. Government. The U.S. Government is authorised to reproduce and distribute reprints for governmental purposes notwithstanding any copyright annotation therein. FactSet had the opportunity to review the paper.

\subsubsection*{Author contributions} A. Bacilieri suggested the research topic, designed the work, implemented the reconstruction method and associated tests, analysed the data and wrote the paper. Pablo Astudillo-Estevez provided the cleaned Ecuador dataset and relevant information about it. They also declare that there were no conflicts of interest.

\subsubsection*{Research data}
We do not have permission to share the two firm-level datasets. For FactSet because it is a commercial dataset, while for Ecuador because it is a confidential, administrative dataset with restricted access.

\renewcommand\bibname{References}
\bibliographystyle{apalike}
\bibliography{bibliography}

\newpage
\begin{appendices}
\setcounter{table}{0}
\numberwithin{table}{section}
\setcounter{figure}{0}
\numberwithin{figure}{section}
\setcounter{equation}{0}
\numberwithin{equation}{section}

\section{Data}
\subsection{FactSet: description of the dataset}
\label{app:FactSet_description_dataset}
We use three primary data sources provided by FactSet: Fundamentals, Supply Chain Relationships and Supply Chain Shipping Transactions. FactSet covers mostly listed firms around the world. We downloaded the datasets in April 2020. We refer to \cite{bacil2022emprical} for a more detailed description of the dataset. We remark only that due to the nature of the data collection process, coverage is biased towards companies listed on US stock exchanges, big firms and big transactions. 

The monetary values of customer-supplier transactions are rarely recorded. When recorded, the money flows are reported as a revenue percentage earned by the seller from a specific customer. However, it is unknown to what disclosed revenue figure that percentage refers. For example, it could be the quarterly or the annual income statement, but also some interim revenue disclosure not tied to any specific financial statement. Similarly, in the shipment dataset, the cumulative value of the goods shipped, when disclosed, is in most cases above the seller's revenues and/or the customers' costs \citep[for more information on why this is the case, see][]{bacil2022emprical}. Therefore, we dismiss this information when reconstructing the money flows. There are other, less comprehensive datasets, such as Compustat, which report the revenue percentage earned by the seller. These could be potentially merged to enhance the reconstruction. We leave this for future work.

We consider industrial firms only, hence we exclude firms in the financial and insurance sector, and firms classified as extraterritorial organisations. We aggregate customer-supplier relations within a year. We use the fiscal year instead of the calendar year to ensure time consistency between the formation of supplier-customer relations and financial statements. The fiscal year goes from June to May, meaning that if a company's fiscal year end-month falls between January and May, the fiscal year is the current calendar year minus one, otherwise it is the current calendar year.

 We further aggregate all three FactSet datasets at the parent company level, meaning that we use consolidated income statements and attribute subsidiaries' supplier-customer relations to the parent company. We delete self-loops (i.e., supply chain relations among the parent and its subsidiaries) as these stem from intra-group sales that cancel out in consolidated income statements. We rely on the latest available information on a company's ownership structure since it is impossible to know the evolution of companies' ownership structures (mergers, acquisitions, buy-backs, etc.). For each company, we also have information on the sector (NACE Rev.2 codes at the 4-digit level) and the country where the company's headquarters are located.

\subsubsection{Methods for writing the income statement}\label{app:methods_income_statem}

There are two main methods companies can follow to write their income statements: the \textit{function of expense} method or the \textit{nature of expense} method -- some companies adopt a hybrid method. As the names suggest, the nature of expense method lists expenses based on their nature, while the function of expense method lists expenses based on their function. Thus, the nature of expense method breaks down expenses based on the inputs used to perform the business activity, e.g., materials, delivery charges, changes in inventory, rent, labour expenses and employee benefits. On the other hand, the 
function of expense method allocates expenses based on the activity for which the expense arises. Therefore, ``if the expense did not contribute to the creation of the [good] or service that is the underlying source of sales revenue, they are not part of the cost of goods sold.'' \citep[~p. 208]{stolowy2006financial}.

\subsubsection{Labour expenses}\label{app:constr_labor_exp}

Depending on the method followed in writing the income statement (explained in Appendix~\ref{app:methods_income_statem}), companies may disclose or not their labour expenses as a separate line item. If a company follows the \textit{nature of expense} method, labour costs and the cost of intermediate inputs are broken down into two separate items on the income statement. However, if a company follows the \textit{function of expense} methods, labour and intermediate inputs expenses are lumped together in the cost of goods sold. Moreover, even if they follow a function of expenses method, some companies may disclose their labour costs in a footnote. 

FactSet collects the labour expenses disclosed in the footnote but without flagging the method followed by the company in writing its income statement. Therefore, there is no systematic way to know whether a company follows the nature of expense or the function of expense method. The only reasonable assumption is that if labour expenses are not observed in the dataset, it is because the company followed the nature of expenses method and thus the cost of goods sold correctly represents intermediate inputs costs. Consequently, it is impossible to extract labour costs from the cost of goods sold only when necessary.

\paragraph{Estimating missing labour expenses.}
For constructing the I-O table, we must discern labour and intermediate inputs expenses as they correspond to different parts of the I-O table. Labour costs are part of value-added, while intermediate input expenses correspond to in-strengths, i.e., the column sums of the weighted adjacency matrix (also referred to as inter-industry transaction matrix in sector-level data). Having an exact quantification of intermediate expenses is necessary to reconstruct the firm-to-firm money flows (i.e., network weights). 

To estimate labour expenses for those firms that do not report them, we use the cost of goods sold share of labour expenses disclosed by other firms in the same sector. For each sector, we calculate the sector-level cost of goods sold share of labour expenses using four different estimation strategies, all of which use a rolling window approach. In this paper, we use the method that minimises the root mean squared error, the average absolute error and the median absolute error. On the one hand, method 2a and 3 achieve the minimum RMSE and MAE while method 2b achieves the lowest MedAE regarding the cost of goods sold share of labour expenses (Table~\ref{tab:error_metrics_share_labour_exp}). On the other hand, if we look at the estimated labour costs and intermediate costs, it is method 2b that has the minimum error (Table~\ref{tab:error_metrics_labour_interm_exp}). Therefore, we use method 2b in this paper.

\begin{figure}[!htbp]
\centering
\includegraphics[scale=.34,keepaspectratio]{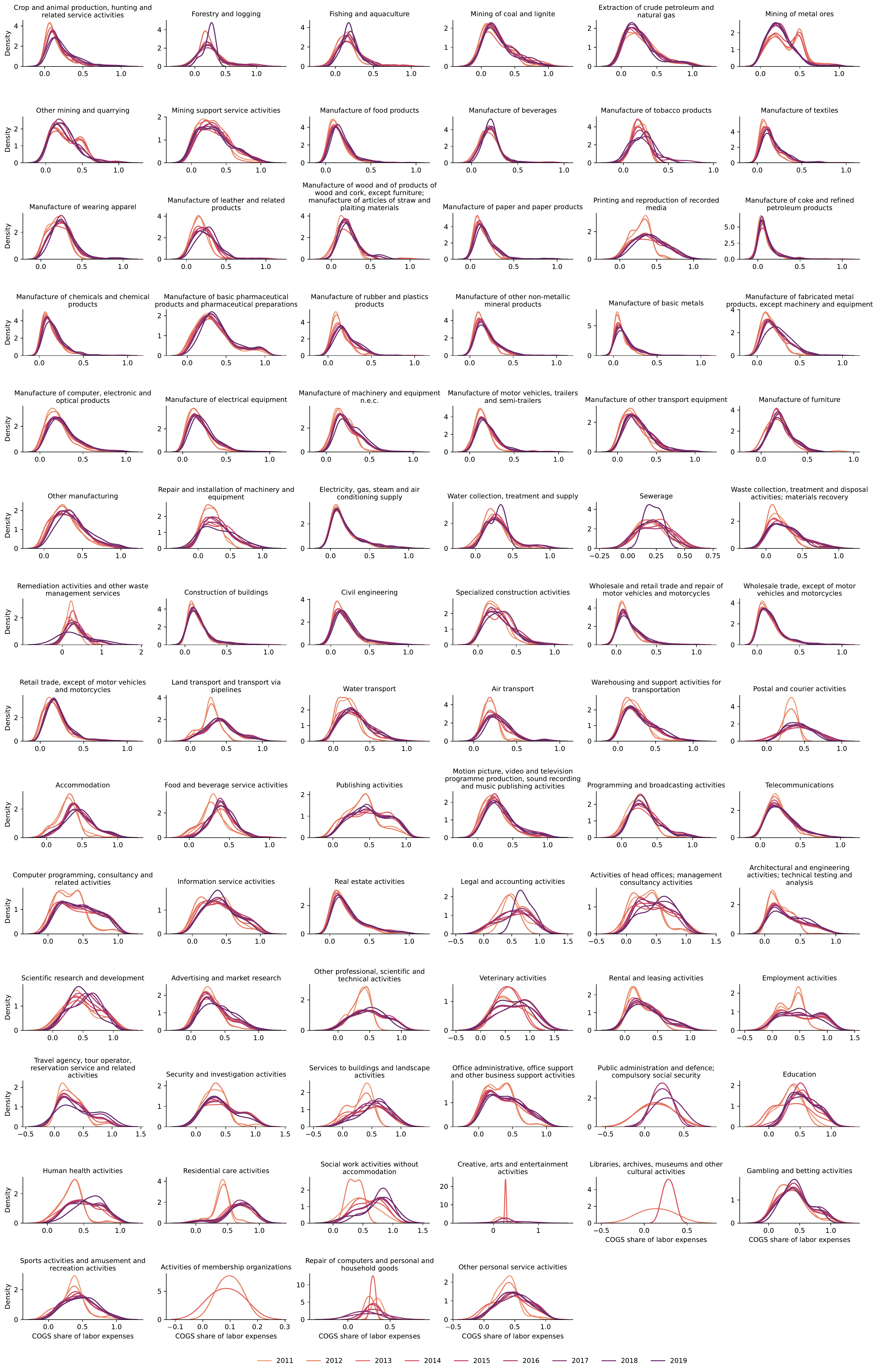}
\caption{Density of the cost of goods sold share of labour expenses for firms in different sectors and over time (2011-2019). The titles show the NACE Rev. 2 sectoral classification at the 2-digit level.}
\label{fig:density_cogs_sahre_laborExp}
\end{figure}

We estimate labour expenses only from 2013 onward and start the rolling window in 2011 because before 2011 the number of firms in each sector was too low. For firms that disclose their labour expenses only for a few years, we extrapolate the missing values of the cost of goods sold share of labour expenses from those that we observe: (1) if we observe labour expenses up to time $t$ and the missing values are only from $t+1$ onward, we use the last value of the cost of goods sold share of labour expenses and propagate it forward; (2) if we observe labour expenses from $t$ onward but do not observe labour expenses form 2013 till $t$, we backpropagate the first available observation; and (3) if we do not observe labour costs in the middle, meaning that we do observe labour costs from 2013 to $t$ and from $t+\tau$ until the end, we estimate the cost of goods sold share of labour expenses with linear interpolation. Table~\ref{tab:Nfirms_labor} reports the percentage of firms that disclose labour costs for each sector average over time (second column) and its standard deviation (last column), while Figure~\ref{fig:density_cogs_sahre_laborExp} shows the density of firms' cost of goods sold share of labour expenses for each sector (NACE Rev. 2 at the 2-digit level) over time.

\begingroup
\small
\def\arraystretch{1.1}
{\centering
\begin{longtable}[c]{p{0.7\textwidth} c c}
\label{tab:Nfirms_labor}\\
\toprule
NACE division description & Av. pct firms & Standard dev. \\
\midrule			
		Accommodation & 0.618 & 0.038 \\
			Activities of head offices; management consultancy activities & 0.528 & 0.022 \\
			Activities of membership organizations & 0.750 & 0.274 \\
			Advertising and market research & 0.575 & 0.026 \\
			Air transport & 0.882 & 0.042 \\
			Architectural and engineering activities; technical testing and analysis & 0.732 & 0.024 \\
			Civil engineering & 0.663 & 0.015 \\
			Computer programming, consultancy and related activities & 0.530 & 0.022 \\
			Construction of buildings & 0.608 & 0.021 \\
			Creative, arts and entertainment activities & 0.944 & 0.167 \\
			Crop and animal production, hunting and related service activities & 0.653 & 0.052 \\
			Education & 0.554 & 0.052 \\
			Electricity, gas, steam and air conditioning supply & 0.639 & 0.018 \\
			Employment activities & 0.541 & 0.029 \\
			Extraction of crude petroleum and natural gas & 0.489 & 0.040 \\
			Fishing and aquaculture & 0.707 & 0.068 \\
			Food and beverage service activities & 0.430 & 0.020 \\
			Forestry and logging & 0.722 & 0.105 \\
			Gambling and betting activities & 0.758 & 0.035 \\
			Human health activities & 0.692 & 0.016 \\
			Information service activities & 0.450 & 0.028 \\
			Land transport and transport via pipelines & 0.562 & 0.060 \\
			Legal and accounting activities & 0.669 & 0.093 \\
			Libraries, archives, museums and other cultural activities & 0.750 & 0.289 \\
			Manufacture of basic metals & 0.677 & 0.023 \\
			Manufacture of basic pharmaceutical products and pharmaceutical preparations & 0.610 & 0.026 \\
			Manufacture of beverages & 0.685 & 0.035 \\
			Manufacture of chemicals and chemical products & 0.603 & 0.031 \\
			Manufacture of coke and refined petroleum products & 0.739 & 0.023 \\
			Manufacture of computer, electronic and optical products & 0.633 & 0.027 \\
			Manufacture of electrical equipment & 0.680 & 0.050 \\
			Manufacture of fabricated metal products, except machinery and equipment & 0.634 & 0.022 \\
			Manufacture of food products & 0.680 & 0.014 \\
			Manufacture of furniture & 0.604 & 0.028 \\
			Manufacture of leather and related products & 0.709 & 0.022 \\
			Manufacture of machinery and equipment n.e.c. & 0.575 & 0.047 \\
			Manufacture of motor vehicles, trailers and semi-trailers & 0.629 & 0.043 \\
			Manufacture of other non-metallic mineral products & 0.657 & 0.021 \\
			Manufacture of other transport equipment & 0.639 & 0.036 \\
			Manufacture of paper and paper products & 0.661 & 0.015 \\
			Manufacture of rubber and plastics products & 0.665 & 0.024 \\
			Manufacture of textiles & 0.776 & 0.027 \\
			Manufacture of tobacco products & 0.698 & 0.062 \\
			Manufacture of wearing apparel & 0.643 & 0.037 \\
			Manufacture of wood and of products of wood and cork, except furniture; manufacture of articles of straw and plaiting materials & 0.656 & 0.047 \\
			Mining of coal and lignite & 0.649 & 0.053 \\
			Mining of metal ores & 0.719 & 0.037 \\
			Mining support service activities & 0.594 & 0.061 \\
			Motion picture, video and television programme production, sound recording and music publishing activities & 0.679 & 0.025 \\
			Office administrative, office support and other business support activities & 0.647 & 0.029 \\
			Other manufacturing & 0.548 & 0.025 \\
			Other mining and quarrying & 0.711 & 0.029 \\
			Other personal service activities & 0.548 & 0.048 \\
			Other professional, scientific and technical activities & 0.612 & 0.038 \\
			Postal and courier activities & 0.813 & 0.089 \\
			Printing and reproduction of recorded media & 0.720 & 0.007 \\
			Programming and broadcasting activities & 0.586 & 0.047 \\
			Public administration and defence; compulsory social security & 0.593 & 0.206 \\
			Publishing activities & 0.549 & 0.026 \\
			Real estate activities & 0.493 & 0.044 \\
			Remediation activities and other waste management services & 0.717 & 0.076 \\
			Rental and leasing activities & 0.515 & 0.033 \\
			Repair and installation of machinery and equipment & 0.681 & 0.022 \\
			Repair of computers and personal and household goods & 1 & 0 \\
			Residential care activities & 0.588 & 0.041 \\
			Retail trade, except of motor vehicles and motorcycles & 0.500 & 0.021 \\
			Scientific research and development & 0.569 & 0.038 \\
			Security and investigation activities & 0.688 & 0.033 \\
			Services to buildings and landscape activities & 0.508 & 0.045 \\
			Sewerage & 0.483 & 0.059 \\
			Social work activities without accommodation & 0.680 & 0.096 \\
			Specialized construction activities & 0.592 & 0.037 \\
			Sports activities and amusement and recreation activities & 0.689 & 0.044 \\
			Telecommunications & 0.690 & 0.019 \\
			Travel agency, tour operator, reservation service and related activities & 0.621 & 0.045 \\
			Veterinary activities & 0.972 & 0.083 \\
			Warehousing and support activities for transportation & 0.690 & 0.031 \\
			Waste collection, treatment and disposal activities; materials recovery & 0.651 & 0.034 \\
			Water collection, treatment and supply & 0.603 & 0.057 \\
			Water transport & 0.577 & 0.030 \\
			Wholesale and retail trade and repair of motor vehicles and motorcycles & 0.658 & 0.019 \\
			Wholesale trade, except of motor vehicles and motorcycles & 0.575 & 0.013 \\
\bottomrule
\caption{Time average of the percentage of firms that disclose labour costs per sector.}
\end{longtable}
}
\endgroup
\pagebreak

Let $w_{i,t}$ be labour expenses of firm $i$ at time $t$ and $g_{i,t}$ its cost of goods sold excluding depreciation and amortisation (for firms employing a \textit{by nature} method in writing their income statements), then the cost of goods sold share of labour expenses of firm $i$ at time $t$ is defined as 
\begin{equation}\label{eq:cogs_share_labour}
    \alpha_{i,t} = \frac{w_{i,t}}{g_{i,t} + w_{i,t}}\;. 
\end{equation}
For some firms, we observed labour expenses for some years but not for other years. Thus, we assumed that their $\alpha_{i,t}$ does not change and kept it constant over time.

\paragraph{Estimation method 1.}
Firstly, we calculate a firm's average cost of goods sold share of labour expenses over a three-year period starting from 2013. For example, $\alpha_{i,2013}$ is computed using a rolling window going from 2011 to 2013, while for $\alpha_{i,2014}$ the rolling window goes from 2012 to 2014.\footnote{
For some firms, picked at random, we look at how their $\alpha$'s change over time; changes are negligible.
} A firm's average cost of goods sold share of labour expenses is given by
\begin{linenomath}
\begin{equation*}
    \bar{\alpha}_{i,t} = \frac{1}{3} \sum_{\tau=t-2}^{t} \alpha_{i,\tau}\;.
\end{equation*} 
\end{linenomath}
Afterwards, for each year $t$, we compute sector $s$'  cost of goods sold share of labour expenses by averaging over the $\alpha_i$'s of the firms in sector $s$ at time $t$:
\begin{linenomath}
\begin{equation*}
    \tilde{\alpha}_{s,t} = \frac{1}{N_{s,t}} \sum_{i \in \mathcal{S}} \bar{\alpha}_{i,t}\;,
\end{equation*}
\end{linenomath}
where $N_{s,t}$ is the number of firms in sector $s$ at time $t$ and $\mathcal{S}$ is the set of firms in sector $s$.

\paragraph{Estimation method 2.}
We calculate a sector-level time-varying cost of goods sold share of labour expenses in the following two ways.
\begin{itemize}
\item[\textbf{2.a}] We compute a sector's cost of good sold share of labour expenses for each time step $t$ by averaging over the $\alpha_i$'s of firms in sector $s$:
\begin{linenomath}
\begin{equation*}
\alpha_{s,t}^{\text{a}}= \frac{1}{N_{s,t}} \sum_{i \in \mathcal{S}} \alpha_{i,t}\;.
\end{equation*}
\end{linenomath}
\item[\textbf{2.b}] We calculate a sector's cost of goods sold share of labour expenses by summing the labour expenses of firms in sector $s$ at time $t$ and dividing these by the sum of firms' labour and intermediate costs:
\begin{linenomath}
    \begin{equation*}
            \alpha_{s,t}^{\text{b}} = \frac{\sum_{i \in \mathcal{S}} w_{i,t}}{\sum_{i \in \mathcal{S}} w_{i,t} + \sum_{i \in \mathcal{S}} g_{i,t}}\;.
    \end{equation*}
    \end{linenomath}
\end{itemize}

\noindent After using either of the two methods above (2.a or 2.b), to compute the sector-level cost of goods sold share of labour expenses at time $t$, we employ a three-year rolling window to calculate the time average:
\begin{linenomath}
\begin{equation*}
    \bar{\alpha}_{s,t}^k = \frac{1}{3} \sum_{\tau=t-2}^{t} \alpha_{s,\tau}^k\;,
\end{equation*}
\end{linenomath}
for $k = \{\text{a}, \text{b}\}$, indicating whether we used method 2.a or 2.b in the first step.

\paragraph{Estimation method 3.}
We start from $ \alpha_{s,t}^a$ calculated using method 2.a and then compute a weighted average using a three-year rolling window:
\begin{linenomath}
\begin{equation*}
    \bar{\alpha}_{s,t}^{*} = \sum_{\tau=t-2}^{t} \theta_{s,\tau} \alpha_{s,\tau}^a\;,
\end{equation*}
\end{linenomath}
the $\theta_{s,t}$'s are weights that sum up to one. $\theta_{s,t}$ is the share of sector $s$ at time $t$ in the total expenditure of the sector on labour and intermediate inputs during the three years. $\theta_{s,t}$ is defined as:
\pagebreak
\begin{linenomath}
\begin{equation*}
    \theta_{s,t} = \frac{\sum_{i\in \mathcal{S}} g_{i,\tau} + w_{i,\tau}}{\sum_{\tau=t-2}^t \sum_{i\in\mathcal{S}} g_{i,\tau} + w_{i,\tau}}\;.
\end{equation*}
\end{linenomath}

\begin{figure}[!htbp]
    \centering
    \includegraphics{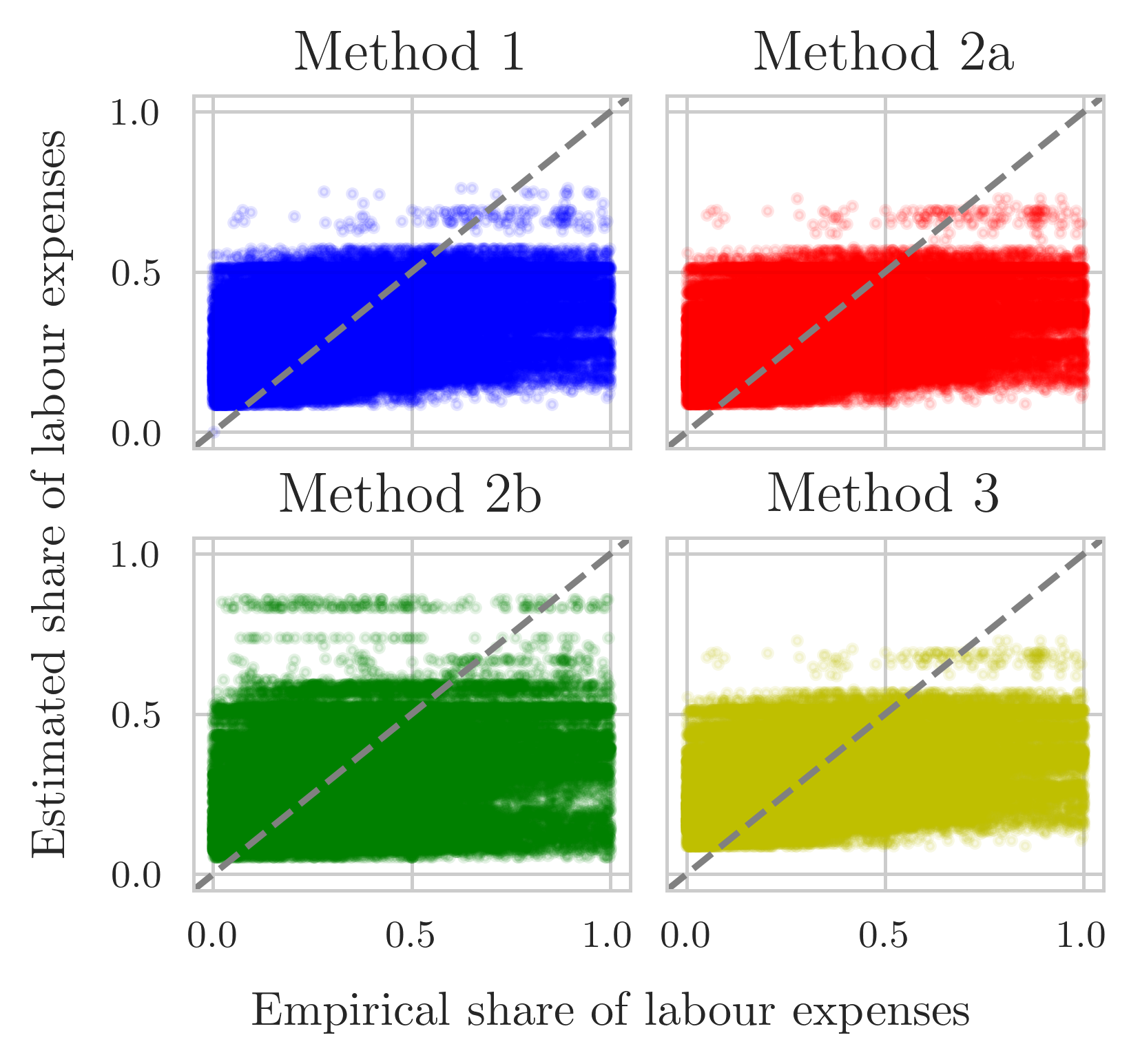}
    \caption{Empirical (\textit{x}-axis) and predicted (\textit{y}-axis) cost of goods sold share of labour expenses for our four estimation methods.}
    \label{fig:asses_alpha_hat}
\end{figure}

\paragraph{Assessing the methods.}
Figure~\ref{fig:asses_alpha_hat} shows the empirical against the predicted cost of goods sold share of labour expenses for our four estimation methods. They all seem to perform very similarly except for method 2b, which has the lowest median absolute error but the highest root mean squared error and median absolute error. Among the others, method 2a and 3 have the same and lowest error metrics (Table~\ref{tab:error_metrics_share_labour_exp}).

\begin{table}[!htbp]
\centering
\begin{tabular}{lrrr}
\toprule
 &  RMSE &   MAE &  MedAE \\
\midrule
Method 1  & 0.184 & 0.138 &  0.106 \\
Method 2a & 0.184 & 0.137 &  0.105 \\
Method 2b & 0.201 & 0.144 &  0.098 \\
Method 3  & 0.184 & 0.137 &  0.105 \\
\bottomrule
\end{tabular}
\caption{Error metrics for the cost of goods sold share of labour expenses for the four estimation methods we devised. RMSE denotes the root mean squared error, MAE the mean absolute error and MedAE the median absolute error.}
\label{tab:error_metrics_share_labour_exp}
\end{table}

\begin{figure}[!htbp]
    \centering
    \includegraphics{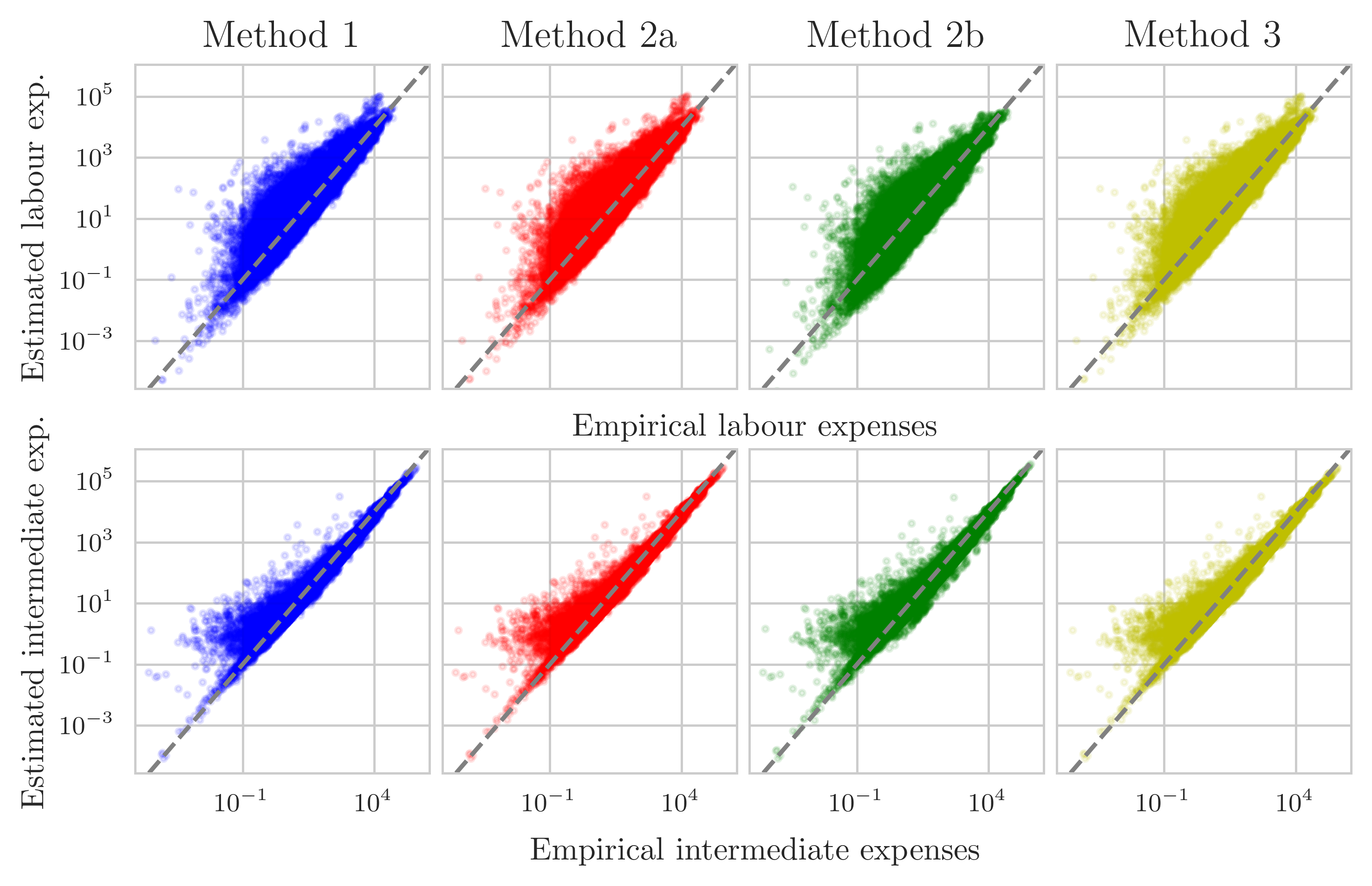}
    \caption{Empirical (\textit{x}-axis) and predicted (\textit{y}-axis) labour expenses (top) and intermediate expenses (bottom) for our four estimation methods.}
    \label{fig:assess_labour_interm_hat}
\end{figure}

Figure~\ref{fig:assess_labour_interm_hat} shows the empirical against the predicted labour expenses (top) and intermediate expenses (bottom) for our four estimation methods. They all seem to perform very similarly. Method 2b has the lowest error metrics (Table~\ref{tab:error_metrics_labour_interm_exp}). The error metrics are the same for the labour and intermediate expenses because they mirror each other. Had we not taken the absolute values, for instance, the error metrics would have opposite sign.

\begin{table}[!htbp]
\centering
\begin{tabular}{l *{7}{r}}
\toprule
 & \multicolumn{3}{c}{Labour exp.} && \multicolumn{3}{c}{Intermediate exp.} \\
 \cmidrule{2-4}  \cmidrule{6-8}
 &  RMSE &   MAE &  MedAE &&  RMSE &   MAE &  MedAE \\
\midrule
Method 1  & 1,153.6 & 126.0 & 8.5  && 1,153.6 & 126.0 &    8.5 \\
Method 2a & 1,163.0 & 126.1 & 8.5  && 1,163.0 & 126.1 &    8.5 \\
Method 2b &  497.0 &  90.2 & 8.4 &&  497.0 &  90.2 &    8.4 \\
Method 3  & 1,162.0 & 126.1 & 8.5 && 1,162.0 & 126.1 &    8.5 \\
\bottomrule
\end{tabular}
\caption{Error metrics for labour and intermediate expenses for the four estimation methods we devised. RMSE denotes the root mean squared error, MAE the mean absolute error and MedAE the median absolute error.}
\label{tab:error_metrics_labour_interm_exp}
\end{table}

\FloatBarrier
\subsubsection{Value-added}\label{app:value_added}
Value-added can be calculated using income statements. It is defined in different ways in the literature. For instance, \citet{jango2019productivity} calculate it as the sum of deflated wages and deflated earnings before interest and taxes. \citet{magerman2016heterogeneous} calculate value-added as the difference between sales and material costs, which is standard in the production network literature (at the firm level). We adopt a third strategy based on \citeauthor{miller2009input}'s (\citeyear{miller2009input}) book on input-output analysis, which also corresponds to the definition in the WIOD (which we use to construct the proxy node).  

Following \cite{miller2009input}, valued-added is the wage bill (labour costs) plus EBITDA (earnings before interest and taxes). Their definition, which follows national and international accounting standards, also includes amortisation and depreciation. We do not observe the wage bill for all companies since labour expenses are disclosed depending on the method followed in writing the income statement. Therefore, for some companies, we estimate labour costs as explained in Section~\ref{app:constr_labor_exp}.

\subsubsection{Cleaning financial statements for the construction of the I-O table}\label{sec:cleaning_financials}

We kept firms with positive sales, intermediate expenses and value-added, and non-negative labour costs (all firms with zero labour costs have disclosed them), EBITDA and amortisation and depreciation.

\paragraph{Sanity check.}
We checked that firms in the dataset respect the accounting identity:
\begin{linenomath}
\begin{equation*}
    \text{sales} = \text{intermediate\_expenditure} + \text{value\_added} + \text{other\_costs}.
\end{equation*}
\end{linenomath}
Since we did not account for other costs, we expected the residual (other\_costs) to be non-negative. However, in 2014, for 30\% of the firms, the residual is negative, meaning that sales were smaller than the sum of the variables on the RHS (excluding other\_costs).

Most firms with a negative residual disclosed their labour expenses: is it possible that we double-counted labour expenses? This would mean that a firm adopts a \textit{by function} method to write the income statement but reports labour costs in a footnote, which would lead FactSet to record the labour expenses. If this were the case, labour costs would already be in the cost of goods sold, leading to counting this expense twice. Thus, we subtract the disclosed labour costs from intermediate expenses for these firms. Is this reasonable? We checked whether this subtraction caused any firm to have negative intermediate expenses. It did, but only for 0.5\% of the observations for which we did the procedure. 

A positive consequence of the exercise just described is that it allows us to identify firms that likely adopt a \textit{by function} method to write the income statement. We propagate this information back in our estimation of the cost of goods sold share of labour expenses in order to adjust the denominator in Equation~\ref{eq:cogs_share_labour}, which for these firms would otherwise be double counting labour costs. After doing this procedure only 1.2\% of the firms in 2014 do not respect the accounting identity. We drop these firms. The cleaning procedure led to all firms in the sample having a strictly positive value-added.
\subsection{FactSet: evaluation of coverage}
First, we show what types of firms are in FactSet. Second, we assess the coverage of global economic activity in FactSet by comparing it to the WIOD. We compare the countries covered and the sectoral composition in FactSet against that of the WIOD. Subsequently, we quantify the percentage of world gross output captured by FactSet and then evaluate the growth rates of gross output.

\subsubsection{Firm types}
\begin{figure}[!htbp]
    \centering
    \includegraphics{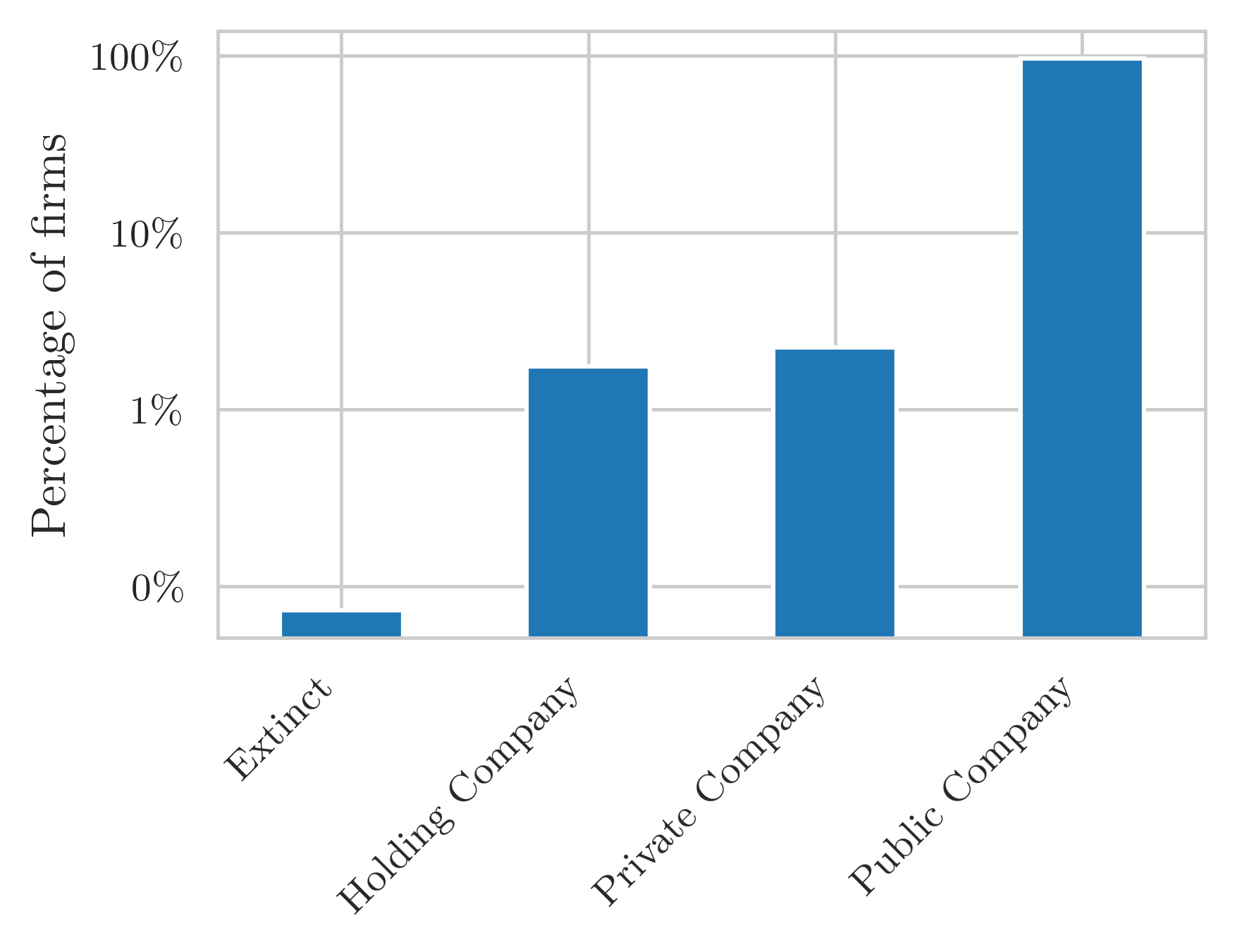}
    \caption{Percentage of firms per type in FactSet in 2014. The firm type is defined by FactSet.}
    \label{fig:PCTfirms_perType}
\end{figure}

Figure~\ref{fig:PCTfirms_perType} shows the percentage of firms for each firm type as defined by FactSet. We plot data for the year 2014 (there are 5,442 firms). In the dataset, there are some big private companies, which also file financial statements, and companies that are extinct by April 2020, when the dataset was downloaded.

\FloatBarrier
\subsubsection{Countries covered in WIOD and FactSet}\label{app:eval_couintries_3DataSets}

Table~\ref{tab:countries_WIOD_factset_copustat} shows the countries that are in the WIOD and the countries where firms in FactSet have their headquarters for the period 2013--2019. The WIOD has a node named ``rest of the world'' (RoW), which captures all the other countries not in the list. The number of overlapping countries is 41. The Czech Republic and Estonia are covered in the WIOD but not in FactSet.

\begin{table}[!htbp]
\Centering
\caption{Countries covered by the WIOD and FactSet.}
\small
\resizebox{\textwidth}{!}{%
\begin{tabular}{p{.4\textwidth}p{.6\textwidth}}
\toprule 
WIOD (43 countries) & FactSet (87 countries)\\
\midrule
 Australia, Austria, Belgium, Brazil, Bulgaria, Canada, China, Croatia, Cyprus, Czechia, Denmark, Estonia, Finland, France, Germany, Greece, Hungary, India, Indonesia, Ireland, Italy, Japan, Korea (the Republic of), Latvia, Lithuania, Luxembourg, Malta, Mexico, Netherlands (the), Norway, Poland, Portugal, Romania, Russian Federation (the), Slovakia, Slovenia, Spain, Sweden, Switzerland, Taiwan (Province of China), Turkey, United Kingdom of Great Britain and Northern Ireland, United States of America (the). 
 
 & Argentina,
 Australia,
 Austria,
 Bahamas (the),
 Bahrain,
 Bangladesh,
 Belgium,
 Bermuda,
 Brazil,
 Bulgaria,
 Canada,
 Cayman Islands (the),
 Chile,
 China,
 Colombia,
 Costa Rica,
 Croatia,
 Cyprus,
 C\^{o}te d'Ivoire,,
 Denmark,
 Egypt,
 Faroe Islands (the),
 Finland,
 France,
 Germany,
 Greece,
 Hong Kong,
 Hungary,
 Iceland,
 India,
 Indonesia,
 Ireland,
 Israel,
 Italy,
 Jamaica,
 Japan,
 Jordan,
 Kenya,
 Korea (the Republic of),
 Kuwait,
 Latvia,
 Lithuania,
 Luxembourg,
 Macao,
 Malaysia,
 Malta,
 Marshall Islands (the),
 Mauritius,
 Mexico,
 Monaco,
 Morocco,
 Netherlands (the),
 New Zealand,
 Nigeria,
 Norway,
 Oman,
 Pakistan,
 Panama,
 Peru,
 Philippines (the),
 Poland,
 Portugal,
 Qatar,
 Republic of North Macedonia,
 Romania,
 Russian Federation (the),
 Saudi Arabia,
 Singapore,
 Slovakia,
 Slovenia,
 South Africa,
 Spain,
 Sri Lanka,
 Sweden,
 Switzerland,
 Taiwan (Province of China),
 Thailand,
 Trinidad and Tobago,
 Tunisia,
 Turkey,
 Ukraine,
 United Arab Emirates (the),
 United Kingdom of Great Britain and Northern Ireland (the),
 United States of America (the),
 Venezuela (Bolivarian Republic of),
 Vietnam,
 Zambia\\
\bottomrule
\end{tabular}
}
\label{tab:countries_WIOD_factset_copustat}
\end{table}

\FloatBarrier
\subsubsection{Sectoral composition}

\begin{figure}[!htbp]
\centering
\includegraphics{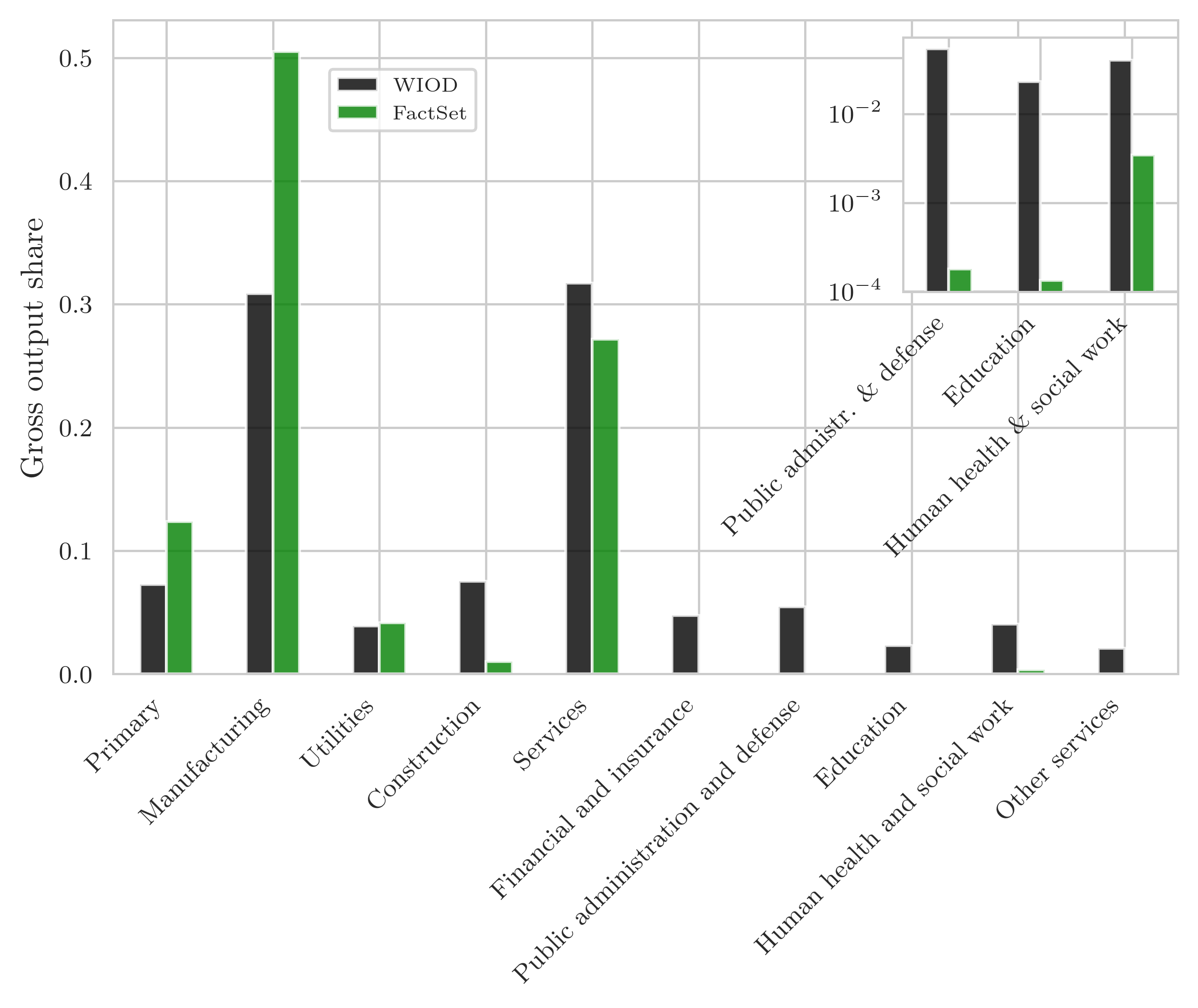}
\caption{Sectoral composition (gross output shares) of the WIOD (black) and FactSet (green) in 2014. For FactSet, we group firms into sectors and use firms' revenues. Sectors are defined as in Table~\ref{tab:isic_macro_class}, where we aggregate ISIC Rev. 4 codes at the 1-digit level into macro classes.}
\label{fig:sectoral_grossO_shares}
\end{figure}

Figure~\ref{fig:sectoral_grossO_shares} shows the sectoral composition in the WIOD (black bars) and in FactSet aggregated at the sector level (green bars) using firms' revenues. We assess the sectoral composition using the sectors' gross output shares (or revenues for firms). We group ISIC Rev.4 codes at the 1-digit level into eight higher-level classes shown in Table~\ref{tab:isic_macro_class}.

\begin{table}[!htbp]
\centering
\resizebox{\textwidth}{!}{%
\begin{tabular}{p{.07\textwidth}p{.62\textwidth}p{.31\textwidth}}
\toprule
ISIC & Description & Macro class \\
\midrule
A & Agriculture, forestry and fishing & Primary \\
B & Mining and quarrying & Primary\\
C & Manufacturing & Manufacturing\\
D & Electricity, gas, steam and air conditioning supply & Utility\\
E & Water supply; sewerage, waste management and remediation activities & Utility\\
F & Construction & Construction\\
G & Wholesale and retail trade; repair of motor vehicles and motorcycles & Services\\
H & Transportation and storage & Services\\
I & Accommodation and food service activities & Services\\
J & Information and communication & Services\\
K & Financial and insurance activities & Financial and insurance\\
L & Real estate activities & Services\\
M & Professional, scientific and technical activities & Services\\
N & Administrative and support service activities & Services\\
O & Public administration and defence; compulsory social security & Public administration and defence\\
P & Education & Education\\
Q & Human health and social work activities & Human health and social work\\
R & Arts, entertainment and recreation & Other services\\
S & Other service activities & Other services\\
T & Activities of households as employers; undifferentiated goods- and services-producing activities of
households for own use & Not included \\
U & Activities of extraterritorial organisations and bodies & Not included\\
\bottomrule
\end{tabular}
}
\caption{Description of ISIC Rev. 4 codes at the 1-digit (left column) and our macro-level classes (right column).}
\label{tab:isic_macro_class}
\end{table}

\FloatBarrier
\subsubsection{Evaluation of gross output}\label{sec:evaluation_grossOutput}
To understand how well FactSet captures global economic activity, we compare gross output levels and growth rates in FactSet to national accounting data (WIOD). We show the evaluation from 2013 to 2018, although we only use 2014.

To compare with world economic activity, we use gross output from the WIOD. The last year for which the WIOD is available is 2014; therefore, we forecast world gross output from 2015 to 2018.\footnote{\label{foot:predict_gross_output}
The WIOD is available from 2003 to 2014 but our firm-level dataset is available from 2014 to 2020. We forecast world gross output from 2015 until 2018 using GDP from the World Bank as follows. We take the ratio of gross output to GDP, which is known to be fairly stable over time, and assume that after 2014 this ratio stays constant. This gives us gross output $q_t$ as a function of GDP $y_t$ and the gross output/GDP ratio $\zeta_t$, thus $q_t = \zeta_{2014} \cdot y_t$ for all $t=2015, \dots, 2020$.
} We assess both the levels and growth rates of gross output.

\begin{figure}[!htbp]
\begin{subfigure}{0.5\linewidth}
\centering
\includegraphics{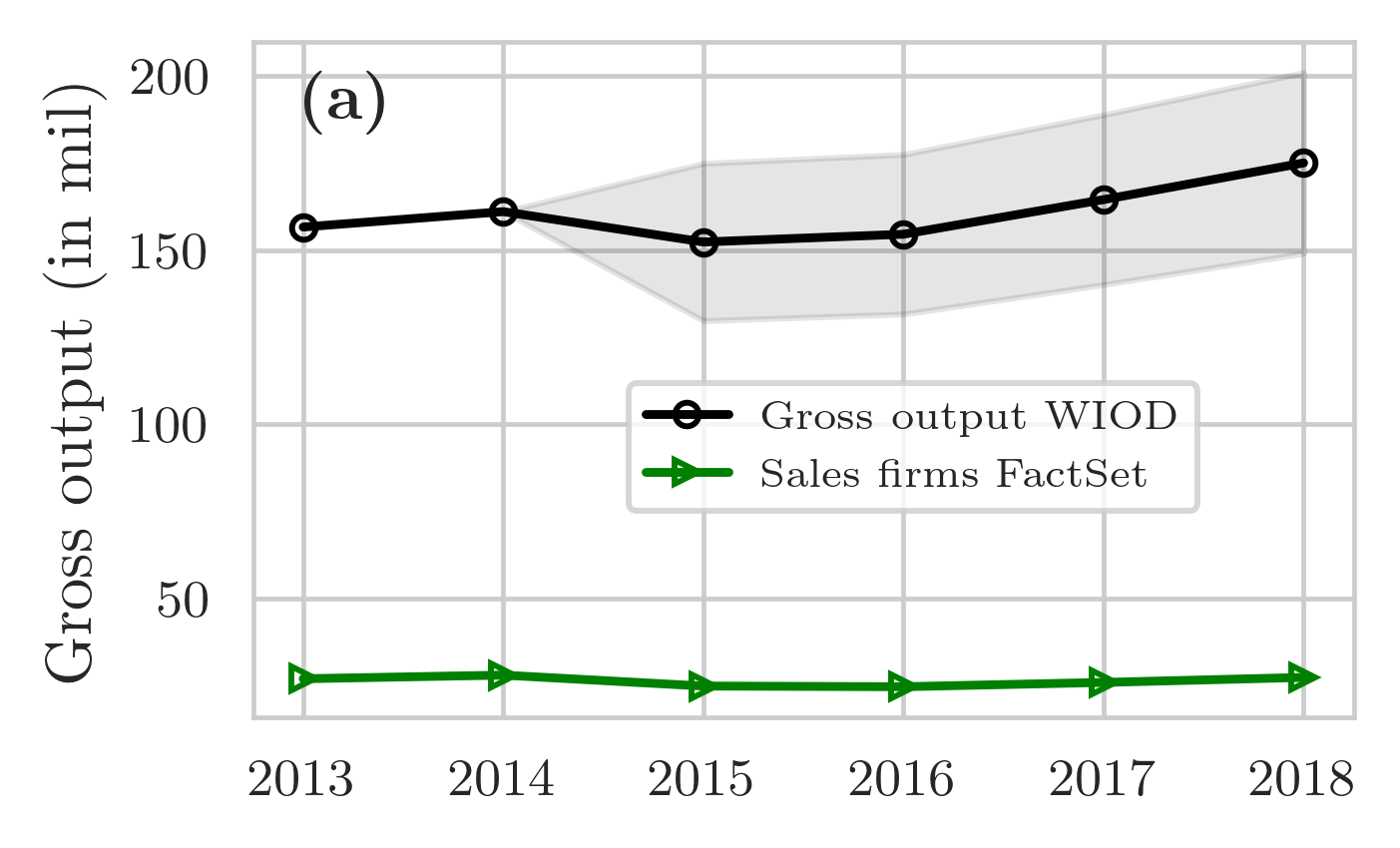}
\end{subfigure}%
\begin{subfigure}{0.5\linewidth}
\centering
\includegraphics{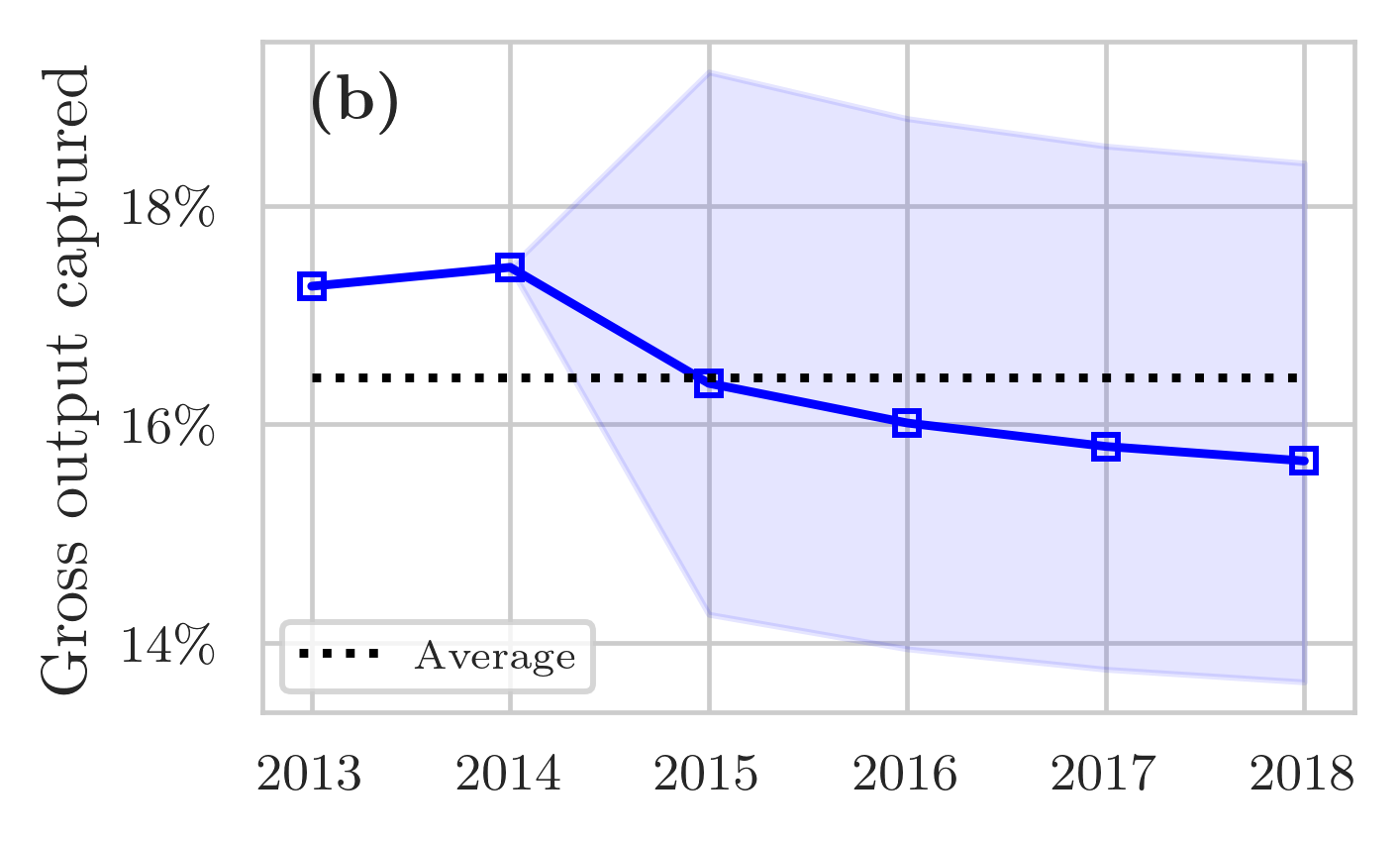}
\end{subfigure}
\caption{\textbf{(a)} World gross output in the WIOD (black dotted line) and cumulative revenues of firms in FactSet (green line with triangles). Values in millions of US dollars. We show error bars for the years for which we forecast gross output (2015--2018); see Footnote~\ref{foot:predict_gross_output} for a description of the forecasting methodology. \textbf{(b)} Percentage captured by FactSet of world gross output (WIOD, Equation~\ref{eq:success_metric_level}). The shaded blue area shows error bounds related to the gross output forecasts. The dashed black line marks the time average.}
\label{fig:grossOutput_WIOD_BEA_factset}
\end{figure}

\paragraph{Levels.}
A comparison of the time series of aggregate firms' revenues and world gross output is shown in Figure~\ref{fig:grossOutput_WIOD_BEA_factset}a. FactSet captures, on average, 16.4\% of world gross output over time. The yearly percentage of world gross output captured by FactSet is given by
\begin{equation}\label{eq:success_metric_level}
    \phi_{t} = \frac{\sum_{i} q_{i,t}}{q_{t}},
\end{equation}
where $q_{t}$ is world gross output at time $t$ and $q_{i,t}$ is firm $i$'s sales at time $t$. 
When forecasting world gross output, besides our central estimate, we also calculate a best and worst case; these yield a lower bound of 15.1\% and an upper bound of 18.3\% on the central estimate.

\begin{figure}[!htbp]
\begin{subfigure}{.5\linewidth}
\centering
\includegraphics{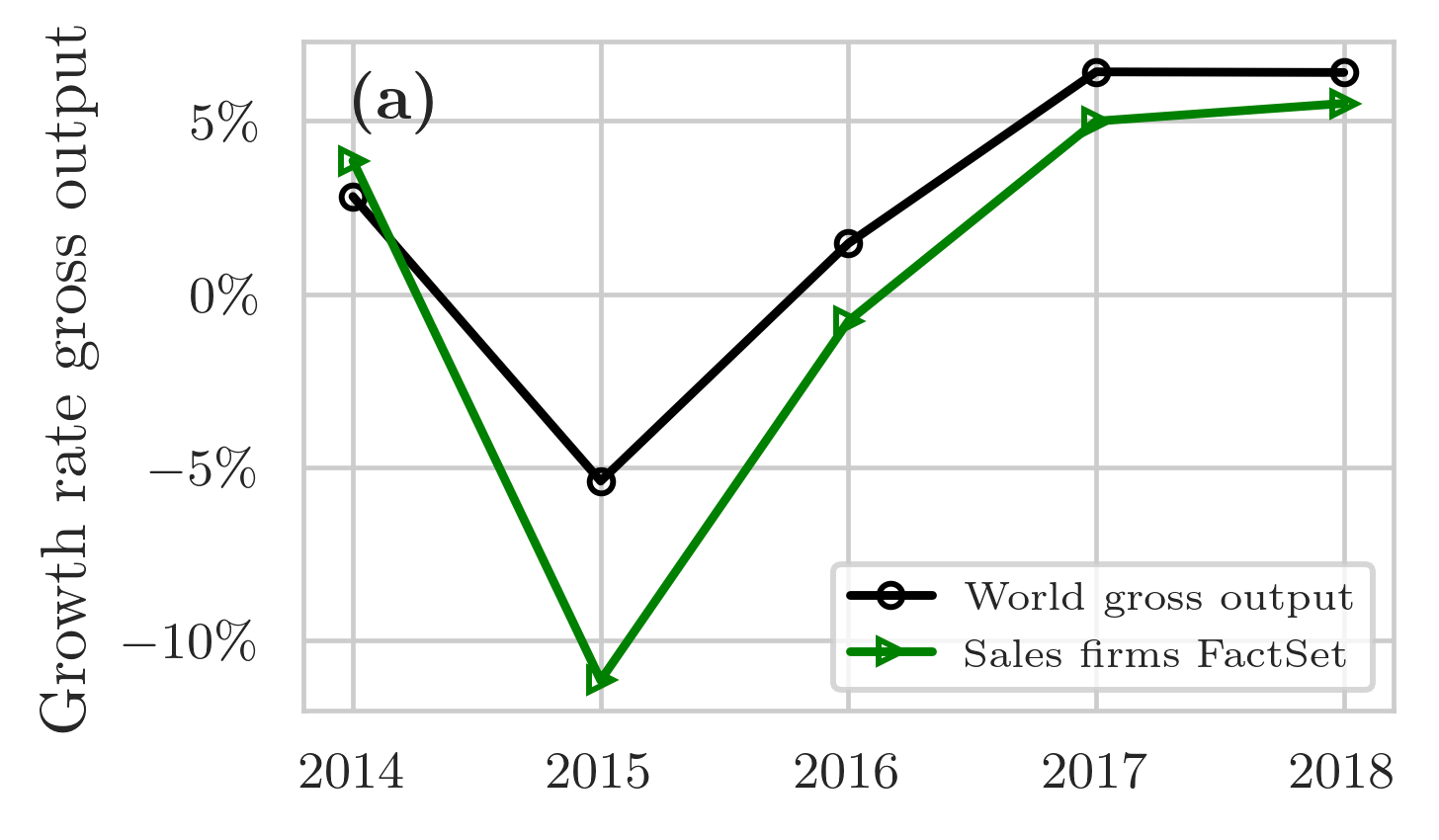}
\end{subfigure}%
\begin{subfigure}{.5\linewidth}
\centering
\includegraphics{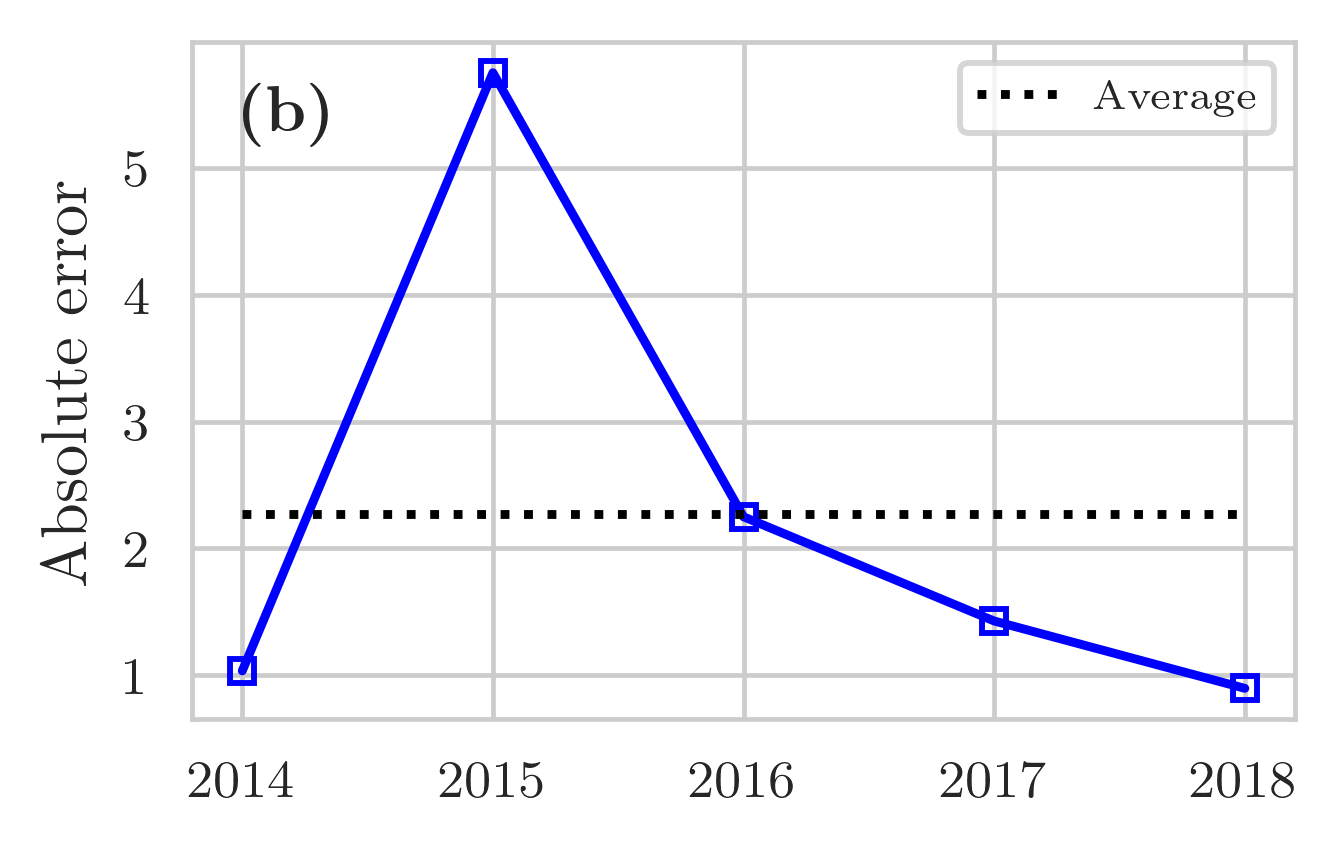}
\end{subfigure}
\caption{\textbf{(a)} Growth rate of world gross output (WIOD, black line with circles) and the growth rates of firms' revenues in FactSet (green line with triangles). As in Figure~\ref{fig:grossOutput_WIOD_BEA_factset}, we forecast world gross output from 2015 to 2018. \textbf{(b)} Absolute error between the growth rate of world gross output (WIOD) and the growth rate of firms' revenues (FactSet). The \textit{y}-axis shows percentage points and the dotted black line shows the time average.}
\label{fig:grossOutput_gr_WIOD_BEA_factset}
\end{figure}

\FloatBarrier
\paragraph{Growth rates.}
For the growth rate of gross output, we carry out the same assessment described above for the levels of gross output. Figure~\ref{fig:grossOutput_gr_WIOD_BEA_factset}a shows the growth rate of gross output in our aggregated FactSet dataset (green line with triangles) and in the WIOD (black line with circles). Over time, FactSet's growth rates differ from world growth rates by 2.3 percentage points on average. To assess how much the growth rates differ, we used the following metric
\begin{equation}\label{eq:success_metric_gr_rate}
\phi_t = | g(\tilde{q}_{t}) - g(q_{t})|,
\end{equation}
where $g(q_{t})$ denotes the growth rate of world gross output at time $t$ and the tilde indicates our aggregate variable constructed from firms' revenues.

\FloatBarrier
\subsection{Merging FactSet with sector-level I-O tables (the WIOD)} \label{sec:proxy_nodes_rebalanc}

Table~\ref{tab:vars_def} describes the variables and associated notation used.

\begin{table}[!htbp]
\centering
\begin{tabular}{p{0.1\textwidth}p{0.8\textwidth}}
\toprule
Notation & Description \\
\midrule
$q_i$ & firm $i$'s total sales \\
$q_s$ & sector $s$' gross output \\

$d_i$ & firm $i$'s intermediate sales \\
$d_s$ & sector $s$' intermediate sales \\

$x_i$ & firm $i$'s intermediate expenditure \\
$x_s$ & sector $s$' intermediate expenditure \\

$f_i$ & firm $i$'s final demand \\
$f_s$ & sector $s$' final demand \\

$k_s$ & sector $s$' gross fixed capital formation \\

$\Delta n_s$ & changes in inventories and valuables of sector $s$\\

$w_i$ & firm $i$'s labour costs\\

$y_i$ & firm $i$'s value-added\\
$y_s$ & sector $s$' value-added\\

$\tau$ & taxes minus subsidies on products of sector $s$\\

$e_s$ & CIF and FOB adjustments on exports of sector $s$\\
$p_s$ & direct purchases abroad by residents and purchases on the domestic territory by non-residents for sector $s$\\
$u_s$ & international transport margins of sector $s$\\

$Z_{r,s}$ & the amount of intermediate inputs sector $s$ buys from sector $r$\\

\bottomrule
\end{tabular}

\caption{Notation and terminology.}
\label{tab:vars_def}
\end{table}
\FloatBarrier

Since FactSet does not cover the entire economy, we supplement it with one proxy node constructed from input-output data at the sector level (we use the WIOD). I-O tables at the sector level capture economic linkages between industries in a country and sometimes also across countries -- clearly, these data are at a more aggregated level than supply chain networks at the firm level. I-O tables at the sector level provide information on the monetary flows related to inputs or consumption expenditure from each industry to each of the other industries, itself and other agents in the economic system such as households and the government sector. 

\begin{figure}[!htbp]
\centering
\includegraphics[scale=.46,keepaspectratio]{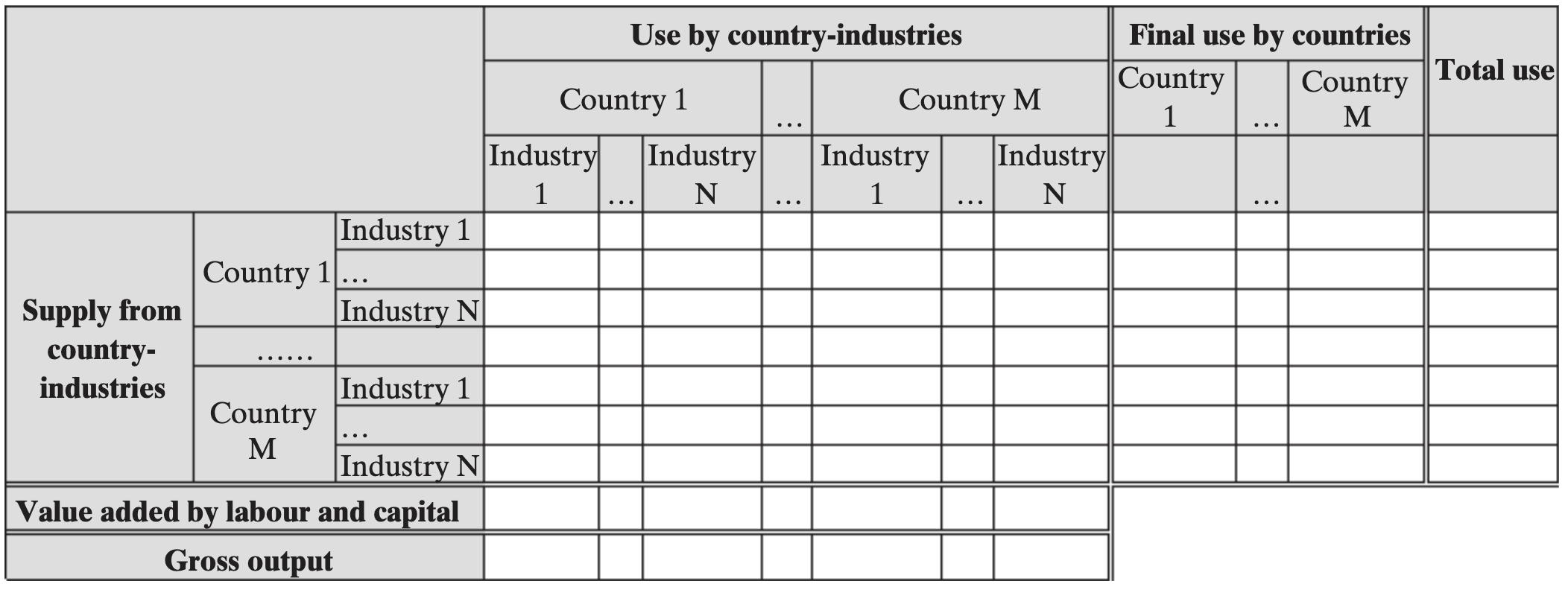}
\caption{Example of a multi-country I-O table; here we show the World Input-Output Database. \textit{Source}: \citet[~p.577]{timmer2015illustrated}.}
\label{fig:WIOD_IO_example}
\end{figure}

Figure~\ref{fig:WIOD_IO_example} shows the structure of the WIOD (the dataset we used), which is standard in aggregate I-O datasets based mostly on national accounts. On the rows, there are the supplying industries grouped by country and on the columns are the customers. Thus, each column portrays an industry production ``recipe'' and the row its customer base. The central block, which involves only trades among industries, is called the inter-industry transaction matrix or intermediate consumption. On its right, there are final uses or final demand, which in the WIOD comprises both consumption and investment demand. At the bottom, there is value-added. The 2016 version of the WIOD covers 43 countries and 56 sectors from 2000 to 2014. 

We construct the I-O table using FactSet and supplement it with a synthetic node constructed using the WIOD. The synthetic, or proxy, node is connected to all the firms (both incoming and outgoing links). The synthetic node accounts for the production, sales and value-added of firms that are not in our network (i.e., the rest of the global economy). While the WIOD is at the country-sector level, we aggregate it into only one node for two reasons. Firstly, the firms in our supply chain data are multinational companies aggregated at the group level, as such their production is scattered around the world, making it difficult to attribute their sales, expenditure and value-added to a specific country. Secondly, while we tried to keep the sector dimension -- which would have allowed us to have one proxy node per sector -- differences between national accounting standards (WIOD) and financial accounting standards (FactSet) prevented us from accurately constructing an I-O table at the firm-sector level. We briefly outline the differences between national accounts and firms' financial accounts in what follows; for a more in-depth discussion, we refer to \cite{bacil2022emprical}.

As shown in Figure~\ref{fig:WIOD_IO_example}, an I-O table is composed of three main parts: the central block (inter-firm or inter-industry transaction matrix), final uses (or final demand) and value-added. The structure of the I-O table implies that there are four main variables to construct and harmonise with our firm-level data when introducing the proxy sector (we discuss this in more detail in the following section, Appendix~\ref{sec:accant_identities_WIOD}). These four variables are total and intermediate sales and costs, value-added and final demand. We build these variables for the proxy sector by subtracting from the WIOD totals the observed firms' cumulative values (for the firms in the cleaned FactSet dataset). While these four variables are relatively easy to identify in sector-level I-O tables, labour expenses and final demand pose challenges in their exact quantification at the firm level.

\subsubsection{Accounting identities in the WIOD}\label{sec:accant_identities_WIOD}

Within a year, the WIOD table is at the country-sector level. Even though we aggregate over countries and sectors, it is helpful to understand the accounting identities underlying it. We show accounting identities by omitting the country and time index, but identities hold for country-sector tables and for one aggregated sector in the same way. Although for one sector, the inter-industry matrix becomes somewhat meaningless.

From the use side, the accounting identity used in the WIOD is
\begin{equation}
    q_s = \sum_{r} Z_{sr} + f_s + k_s + \Delta n_s\;,
\end{equation}
which states that gross output of sector $s$ is the sum of intermediate sales of sector $s$, its final demand $f_s$, gross fixed capital formation $k_s$ (investment) and changes in inventories $\Delta n_s$.

From the expenditure side, the accounting identity is
\begin{equation}
    q_s = \sum_{r} Z_{rs} + \tau_s + e_s + p_s + y_s + u_s\;,
\end{equation}
which states that gross output of sector $s$ (from the expenditure side) equals total expenditure on intermediate inputs, plus taxes minus subsidies $\tau_s$, plus cost, insurance and freight (CIF) and free on board (FOB) adjustments on exports $e_s$, value-added $y_s$ and international transport margins $u_s$.

To construct the proxy sector, the variable definitions in the sector-level I-O table need to match the definition of the variables at the firm level taken from firms' financial statements. We discuss how we harmonise the two datasets in the following two sections.

\subsubsection{Expenditure-side harmonisation}
In this section, we define and describe how we harmonised two main variables that make up the I-O table from the expenditure side: intermediate expenditure and value-added. We do not account for direct purchases abroad by residents and purchases on the domestic territory by non-residents since these are zero for industries. 

\paragraph{Expenditure on intermediate inputs.}
From the expenditure side, we define \textit{sectoral intermediate expenditure} to include also CIF and FOB costs, and international transport margins since these are costs that if firms incur into are in the cost of goods sold:
\begin{equation}
    x_{s} = \sum_{r} Z_{rs} + e_s + u_s\;.
\end{equation}

Firms' intermediate consumption is defined as expenditure on intermediate inputs and services as stated in their invoices and reported as a cumulative figure in their income statements in the cost of goods sold. The cost of goods sold may include labour costs and thus needs to be cleaned for some firms depending on the method they use to write their income statement (see Appendix~\ref{app:methods_income_statem} and~\ref{app:constr_labor_exp}).

\paragraph{Value-added.}
Sectoral \textit{gross value-added} is defined by the System of National Accounts (SNA) as labour and capital compensation, consumption of fixed capital and taxes net of subsidies \citep{SNA2008UN}. The WIOD provides taxes net of subsidies as a separate variable, which thus needs to be added to the definition of value-added.\footnote{
The variable capital compensation that makes up value-added in the WIOD comprises both profit and consumption of fixed capital. It is a residual variable after subtracting labour compensation from value-added. Moreover ``it is the gross compensation for capital, including profits and depreciation allowances. Because of its derivation as a residual, it reflects the remuneration for capital in the broadest sense. This does not include only traditional reproducible assets such as machinery and buildings but also includes non-reproducible assets. Examples are mineral resources and land, intangible assets (such as R\&D knowledge stocks, software, databases, brand names and organisational capital) and financial capital'' \citep[p.~601]{timmer2015illustrated}.
}
Sectoral gross value-added is given by
\begin{equation}
    y_{s}^{\text{gross}} = \tau_s + y_s\;.
\end{equation}

Firms do not disclose the subsidies they received as a separate line item; therefore, we did not include subsidies and simply added taxes (a definition of firms' value-added is provided in Appendix~\ref{app:value_added}).

\subsubsection{Use-side harmonisation}\label{app:sector-firms_demandSide_harm}
In this section, we define and describe how we harmonise the four variables that make up the I-O table from the use side, namely final demand, intermediate sales, gross fixed capital formation and changes in inventories. 

While it is more straightforward to reconcile prices on the expenditure side, it is less so on the use side. On the use side, prices are at the purchaser's price; that is, the amount paid by the purchaser plus trade margins (wholesale and retail), transport margins (if invoiced by the producer) and non-deductible VAT minus deductible VAT. However, at the firm level output (i.e., revenues) is the amount received by the producer for the good or service sold minus VAT (deductible and non) and subsidies. While transport and trade margins may be in the firm's revenues, VAT and subsidies are not. We cannot take care of these mismatches because we lack the data to do so.

\paragraph{Final demand.}

Sectoral \textit{final demand} is defined as in the WIOD and it comprises the consumption of households, the government and non-profit organisations.

Firms do not disclose sales to final demand nor sales to other firms; hence revenues include sales to other firms in the network and sales to final demand. To infer how much a firm sells to final demand, we use the share of final demand satisfied by the sector the firm is in. More formally, let $q_s$ be gross output of sector $s$, $f_s$ be final demand of sector $s$ and $q_i$ be firm $i$'s revenues (or total sales), then final demand of firm $i$ is given by
\begin{equation}\label{eq:firms_final_demand}
    f_i = q_i \frac{f_s}{q_s}\;.
\end{equation}

\paragraph{Gross fixed capital formation.}
Gross fixed capital formation (GFCF) is defined in the SNA as ``the value of [...] acquisitions less disposals of fixed assets'' \citep{SNA2008UN}; it is a measure of investment. In the WIOD, GFCF corresponds to the part of a sector's output that ends up as an investment. It also includes some intangible assets when they are part of the SNA, but not all intangibles are covered.

In firm-level data, firms making capital goods disclose their customers and sales of capital goods are accounted for in the disclosed revenues. However, we cannot distinguish these types of transactions. Therefore, we apply a correction factor to a firm's revenues similar to \cite{magerman2016heterogeneous}. As for final demand, we estimated a firm's GFCG as follows
\begin{equation}
    k_i = q_i \frac{k_s}{q_s}\;,
\end{equation}
where $k$ is gross fixed capital formation either of firm $i$ or sector $s$. A firm's expenditure on capital goods is not included in the cost of goods sold, thus no adjustment is needed on the expenditure side. 

In the WIOD, the ratio $k_s/q_s$ is negative for ISIC code E37-E39 `` sewerage; waste collection, treatment and disposal activities; materials recovery; remediation activities and other waste management services''. We assume that firms in this sector do not provide investment goods and hence have zero GFCF.

\paragraph{Intermediate sales.}
Sectoral \textit{intermediate sales} are sales of intermediate inputs and services to other sectors in the economy. We use the same definition for firms, with the caveat that to get intermediate sales, we have to subtract from revenues the sales to final demand and GFCF.

\paragraph{Changes in inventories.}
We exclude changes in inventory since, in the WIOD, it is a column used for adjustments. Suppose we were to include changes in inventories. In that case, it should go with intermediate expenditure since the variable to which inventory changes correspond to at the firm level is the cost of goods sold, which is defined as beginning inventory minus ending inventory plus purchases during the period. 

\subsection{Sectoral codes}\label{app:naics_descr}

Table~\ref{tab:nace_descr} and~\ref{tab:isic_descr} describe, respectively, NACE Rev.2 codes at the section level (1-digit) and ISIC Rev. 4 codes at the 2-digit level.

\begin{table}[!htbp]
\centering
\resizebox{\textwidth}{!}{%
\begin{tabular}{p{0.1\textwidth}p{0.70\textwidth}p{0.1\textwidth}}
\toprule
Section (1-digit) & Description & Divisions (2-digit) \\
\midrule
A & Agriculture, forestry and fishing & 01 – 03 \\
B & Mining and quarrying & 05 – 09\\
C & Manufacturing & 10 – 33\\
D & Electricity, gas, steam and air conditioning supply & 35\\
E & Water supply; sewerage, waste management and remediation activities & 36 – 39\\
F & Construction & 41 – 43\\
G & Wholesale and retail trade; repair of motor vehicles and motorcycles & 45 – 47\\
H & Transportation and storage & 49 – 53\\
I & Accommodation and food service activities & 55 – 56\\
J & Information and communication & 58 – 63\\
K & Financial and insurance activities & 64 – 66\\
L & Real estate activities & 68\\
M & Professional, scientific and technical activities & 69 – 75\\
N & Administrative and support service activities & 77 – 82\\
O & Public administration and defence; compulsory social security & 84\\
P & Education & 85\\
Q & Human health and social work activities & 86 – 88\\
R & Arts, entertainment and recreation & 90 – 93\\
S & Other service activities & 94 – 96\\
T & Activities of households as employers; undifferentiated goods- and services-producing activities of
households for own use & 97 – 98\\
U & Activities of extraterritorial organisations and bodies & 99\\
\bottomrule
\end{tabular}
}
\caption{Description of NACE Rev.2 at the 1-digit (first column) and 2-digit level (last column).}
\label{tab:nace_descr}
\end{table}

\begingroup
\footnotesize
{\setlength\tabcolsep{.1pt}
\centering
\begin{longtable}[c]{p{0.12\textwidth}p{0.88\textwidth}}
\label{tab:isic_descr}\\
\toprule
ISIC code & Description \\
\midrule
A01	& Crop and animal production, hunting and related service activities\\
A02	& Forestry and logging\\
A03	& Fishing and aquaculture\\
B	& Mining and quarrying\\
C10-C12	& Manufacture of food products, beverages and tobacco products\\
C13-C15	& Manufacture of textiles, wearing apparel and leather products\\
C16	& Manufacture of wood and of products of wood and cork, except furniture; manufacture of articles of straw and plaiting materials\\
C17	& Manufacture of paper and paper products\\
C18	& Printing and reproduction of recorded media\\
C19	& Manufacture of coke and refined petroleum products \\
C20	& Manufacture of chemicals and chemical products \\
C21	& Manufacture of basic pharmaceutical products and pharmaceutical preparations\\
C22	& Manufacture of rubber and plastic products\\
C23	& Manufacture of other non-metallic mineral products\\
C24	& Manufacture of basic metals\\
C25	& Manufacture of fabricated metal products, except machinery and equipment\\
C26	& Manufacture of computer, electronic and optical products\\
C27	& Manufacture of electrical equipment\\
C28	& Manufacture of machinery and equipment n.e.c.\\
C29	& Manufacture of motor vehicles, trailers and semi-trailers\\
C30	& Manufacture of other transport equipment\\
C31-C32	& Manufacture of furniture; other manufacturing\\
C33	& Repair and installation of machinery and equipment\\
D35	& Electricity, gas, steam and air conditioning supply\\
E36	& Water collection, treatment and supply\\
E37-E39	& Sewerage; waste collection, treatment and disposal activities; materials recovery; remediation activities and other waste management services \\
F	& Construction\\
G45	& Wholesale and retail trade and repair of motor vehicles and motorcycles\\
G46	& Wholesale trade, except of motor vehicles and motorcycles\\
G47	& Retail trade, except of motor vehicles and motorcycles\\
H49	& Land transport and transport via pipelines\\
H50	& Water transport\\
H51	& Air transport\\
H52	& Warehousing and support activities for transportation\\
H53	& Postal and courier activities\\
I	& Accommodation and food service activities\\
J58	& Publishing activities\\
J59-J60	& Motion picture, video and television programme production, sound recording and music publishing activities; programming and broadcasting activities\\
J61	& Telecommunications\\
J62-J63	& Computer programming, consultancy and related activities; information service activities\\
K64	& Financial service activities, except insurance and pension funding\\
K65	& Insurance, reinsurance and pension funding, except compulsory social security\\
K66	& Activities auxiliary to financial services and insurance activities\\
L68	& Real estate activities\\
M69-M70	& Legal and accounting activities; activities of head offices; management consultancy activities\\
M71	& Architectural and engineering activities; technical testing and analysis\\
M72	& Scientific research and development\\
M73	& Advertising and market research\\
M74-M75	& Other professional, scientific and technical activities; veterinary activities\\
N	& Administrative and support service activities\\
O84	& Public administration and defence; compulsory social security\\
P85	& Education\\
Q	& Human health and social work activities\\
R-S	& Other service activities\\
T	& Activities of households as employers; undifferentiated goods- and services-producing activities of households for own use\\
U	& Activities of extraterritorial organizations and bodies\\
\bottomrule
\caption{Description of ISIC Rev. 4 at the 2-digit level.}
\end{longtable}
}
\endgroup

\section{Derivations}
\subsection{The conditional maximum entropy reconstruction method of Parisi et al. (2020)}\label{app:derivation_parisi}

The maximum entropy method developed by \cite{parisi2020faster} aims to reconstruct an ensemble of likely weighted networks that are consistent with prior information about the binary topology and that satisfy (in expectation) a set of constraints on observed network quantities -- in our case, the in- and out-strengths. To reconstruct the ensemble of weighted networks, the method infers the probability distribution over the likely weighted network configurations by maximising the conditional (on the binary topology) Shannon entropy subject to imposed constraints. 

The reconstruction procedure consists of two steps. The first step involves a deterministic maximum entropy procedure, known as \textit{MaxEnt} in the literature, from which we derived a value for each link weight. Since MaxEnt has been found to be the best performing in reconstructing weights \citep{anand2015filling,lebacher2019search}, the weights found in the first step are then used in the second step to constraint the expected value of each link weight.\footnote{
While we use the MaxEnt prescription, one could choose any value, including an observed value, when available. This value then needs to be subtracted from the in- and out-strengths.
}
This procedure yields a computational benefit because, as we will see below, we have to solve $m$ (the number of links) decoupled equations. Instead, had we imposed constraints on the in- and out-strengths, we would have had to solve 2$N$ coupled equations, where $N$ is the number of nodes (see \cite{parisi2020faster} for the derivation).

\paragraph{First step.}
We derive a value for the expected weights using the MaxEnt method. MaxEnt is derived by solving a deterministic maximum entropy problem, which maximises an entropy-like functional subject to constraints on intermediate sales and costs of each firm:

\begin{equation}\label{eq:MaxEnt_primal}
\begin{aligned}
& \underset{\{W_{ij}\}}{\text{maximise}} & & S(\bm{W}) = - \sum_{ij} W_{ij} \log W_{ij}\\
& \text{subject to} & & \sum_{j} W_{ij} = s_i^{\text{out}^*}\; i = 1,\dots, N\\
&&& \sum_{j} W_{ji} = s_i^{\text{in}^*}\; i = 1,\dots, N\;,
\end{aligned}
\end{equation}
which yields
\begin{equation}\label{eq:MaxEnt_sol}
    W_{ij}^{\text{ME}} = \frac{s_i^{\text{out}^*} s_j^{\text{in}^*}}{W^{\text{tot}^*}}\;,
\end{equation}
where $W_{ij}$ is the money flow from firm $j$ to firm $i$, $s_i^{\text{out}^*}$ are the observed total intermediate sales (out-strength) of firm $i$, $s_i^{\text{in}^*}$ are the observed total intermediate costs (in-strength) of firm $i$ and $W^{\text{tot}^*} = \sum_{i}s_i^{\text{out}^*} = \sum_{i}s_i^{\text{in}^*}$ is the total weight of the empirical network. Note that MaxEnt generates a fully connected network.

\paragraph{Second step.}
We solve the conditional maximum entropy problem to derive a value for the expected weights given that there is a link between $i$ and $j$. Let $\bm{W} \in \mathbb{W}$ be a weighted adjacency matrix, which is a realisation of the random variable $\mathcal{W}$. Similarly, let $\bm{A} \in \mathbb{A}$ be a binary adjacency matrix, which is a realisation of the random variable $\mathcal{A}$.  The method maximises the conditional entropy $S(\mathcal{W}\mid \mathcal{A})$ defined over the probability density function of the weighted network configurations compatible with the binary topology and satisfying the constraints on the expected weights given by the MaxEnt prescription (Equation~\ref{eq:MaxEnt_sol}):
\begin{equation}\label{eq:CReMb_primal_app}
\begin{aligned}
& \underset{\{Q(\bm{W}\mid \bm{A})\}}{\text{maximise}} & & S(\mathcal{W}\mid \mathcal{A}) = - \sum_{\bm{A}\in\mathbb{A}} P(\bm{A})\int_{\mathbb{W}_{\bm{A}}} Q(\bm{W}\mid\bm{A}) \log Q(\bm{W}\mid\bm{A}) \text{d}\bm{W}\\
& \text{subject to} & & \expval{W_{ij}} = \sum_{\bm{A}\in\mathbb{A}} P(\bm{A})\int_{\mathbb{W}_{\bm{A}}}  W_{ij}(\bm{W}) Q(\bm{W}\mid\bm{A})\text{d}\bm{W} =  W_{ij}^{\text{ME}}\;\; \forall (i, j) \in \mathcal{E}\\
&&& \int_{\mathbb{W}_{\bm{A}}} Q(\bm{W}\mid\bm{A}) \text{d}\bm{W} = 1,\; \forall\bm{A}\in \mathbb{A}\;,
\end{aligned}
\end{equation}
where $\mathbb{W}_{\bm{A}}$ is the set of weighted configurations compatible with the binary topology, $Q(\bm{W}\mid\bm{A})$ is the conditional probability of generating a weighted network $\bm{W}$ consistent with adjacency matrix $\bm{A}$ and $\mathcal{E}$ is the edge set. The information on the binary network enters in probabilistic terms through $P(\bm{A})$, the probability of observing the binary topology $\bm{A}$. $P(\bm{A})$ could be estimated, in a prior step using any suitable method. However, we supply the observed topological structure $\mathcal{A} = \bm{A}^*$, thus $P(\bm{A}^*) = 1$ and $P(\bm{A}) = 0\;, \forall \bm{A}\in\mathbb{A}, \bm{A}\neq \bm{A}^*$.

Solving the above maximisation problem yields the Hamiltonian $H(\bm{W})=\sum_{i\neq j} \lambda_{ij} W_{ij}$ and the graph probability $Q(\bm{W}|\bm{A}) = \prod_{i\neq j} Q_{ij}(W_{ij}\mid A_{ij})$ with
\begin{equation}\label{eq:CReMb_prob_functional_form_app}
    Q_{ij}(W_{ij}\mid A_{ij}) = \begin{dcases*}
     \lambda_{ij} e^{- \lambda_{ij} W_{ij}} &  $W_{ij} > 0$\;, \text{if}\; $A_{ij} = 1$\;,\\
    0 & \text{otherwise}.
\end{dcases*}
\end{equation}
Therefore, $Q_{ij}(W_{ij}|A_{ij}=1)$ is of exponential form with parameter $\lambda_{ij}$, the Lagrange multiplier associated to the constraint on $\expval{W_{ij}}$. It follows that the generalised likelihood for the given set of constraints is
\begin{equation}
    g(\bm{\lambda}) = - \sum_{i\neq j} \lambda_{ij}W_{ij}^{\text{ME}} + \sum_{i\neq j} \sum_{\bm{A}\in\mathbb{A}}P(\bm{A})A_{ij} \log\lambda_{ij}\;,
\end{equation}

\noindent with first order conditions given by
\begin{equation}
    \expval{W_{ij}} = \frac{p_{ij}}{\lambda_{ij}} = W_{ij}^{\text{ME}},\;  \forall i\neq j,
\end{equation}
where $p_{ij} = \sum_{\bm{A}\in\mathbb{A}}P(\bm{A})A_{ij}$ is the probability that a link going from $i$ to $j$ exists. Substituting the MaxEnt prescription (Equation~\ref{eq:MaxEnt_sol}), the Lagrange multipliers are given by
\begin{equation}
    \lambda_{ij}^* = p_{ij}\frac{W^{\text{tot}^*}}{s_i^{\text{out}^*} s_j^{\text{in}^*}}\;.
\end{equation}
Since we consider a deterministic binary topology, the Lagrange multipliers simplify to
\begin{equation}
    \lambda_{ij}^* = \frac{W^{\text{tot}^*}}{s_i^{\text{out}^*} s_j^{\text{in}^*}} \quad \text{for}\; (i, j) \in \mathcal{E}\;.
\end{equation}

The MexEnt procedure assumes that the network is fully connected; however, the generalised maximum entropy method we employ assumes no self-loops. Additionally, we know where the links are. Therefore, we need to redistribute all the weights corresponding to $A_{ij} = 0, \forall (i, j) \notin\mathcal{E}$. To do so, we employ the IPF algorithm that redistributes the weights in an iterative procedure such that at the $n$-th iteration the weight is given by
\begin{equation}
    W_{ij}^{n} = s_i^{\text{out}^*}\frac{W_{ij}^{(n-1)}}{\sum_{k\neq i}W_{ki}^{(n-1)}}\;\; \text{and}\;\;   W_{ij}^{(n+1)} = s_j^{\text{in}^*}\frac{W_{ij}^n}{\sum_{k\neq j}W_{kj}^n}\;.
\end{equation}

\paragraph{Confidence interval on the expected link weight.}
One of the benefits of \citeauthor{parisi2020faster}'s method is that one can calculate a confidence interval $[w^-, w^+]$ for each of the expected weights. The lower bound is given by
\begin{equation}\label{eq:lower_bound_weights_app}
    w^- = - \frac{\ln[e^{-1} + q^-]}{\lambda_{ij}^*}\;,
\end{equation}
where $q^-$ is a desired confidence level. The upper bound is given by
\begin{equation}\label{eq:upper_bound_weights_app}
    w^+ = - \frac{\ln[e^{-1} - q^+]}{\lambda_{ij}^*}\;.
\end{equation}
We refer to Appendix E in \cite{parisi2020faster} for the whole derivation.

\section{Additional results}
\subsection{Weights}

\begin{figure}[H]
    \centering
    \includegraphics{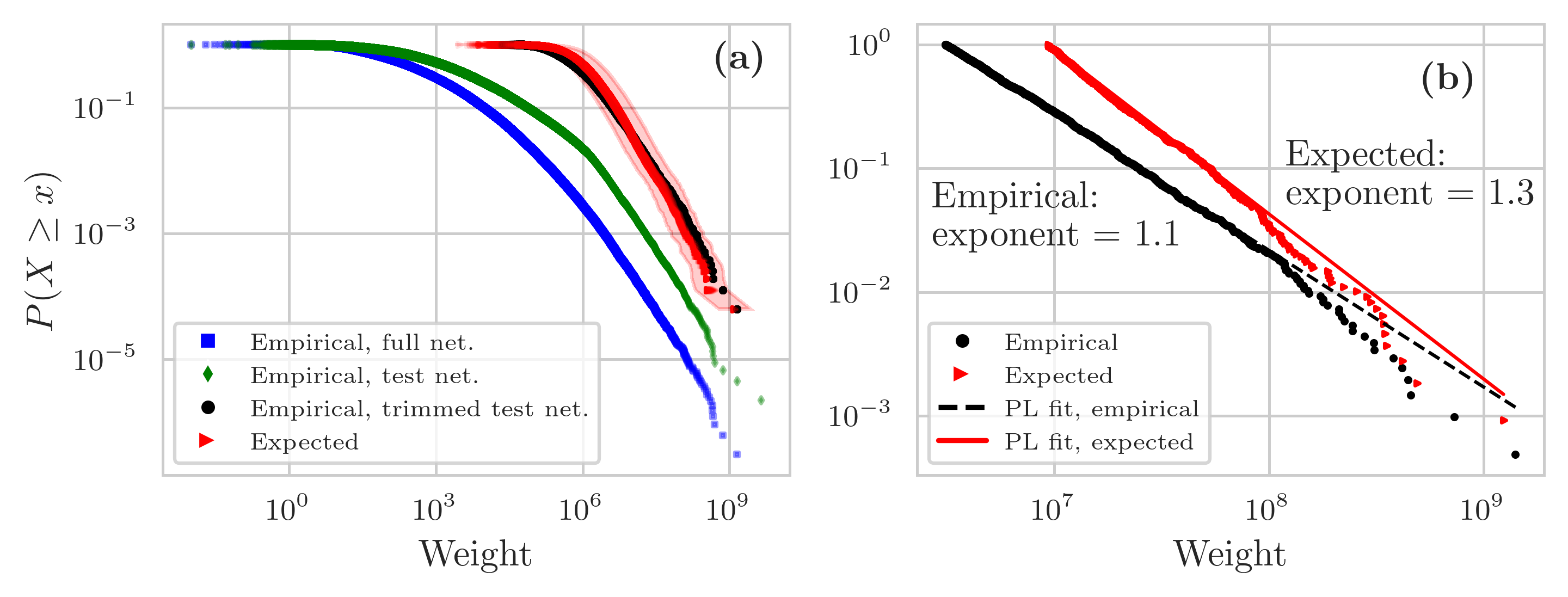}
    \caption{\textbf{(a)} CCDF of the weights of the empirical full network (all firms in the Ecuadorian economy; blue squares), the empirical test network (green diamonds), the empirical trimmed test networks (black dots) and the reconstructed networks (red triangles); the shaded area indicates the 50\% confidence interval. We show all 50 randomised test networks. \textbf{(b)} Power-law fit to the empirical (black dots) and expected (red triangles) weight distribution. The black dashed line shows the power-law fit to the empirical distribution, while the red solid line the power-law fit of the reconstructed distribution; we show results for one of the reconstructions.}
    \label{fig:weight_ccdf_ec}
\end{figure}

\paragraph{Ecuador.}
Figure~\ref{fig:weight_ccdf_ec}a shows the CCDF of the empirical weights of the full network (blue squares), the test network (green diamonds), the trimmed test networks (black dots) and that of the reconstructed networks (red triangles). All three weight distributions display heavy tails. Figure~\ref{fig:weight_ccdf_ec}b shows the power-law fit for one of the randomised networks. The distribution of the empirical weights has a power-law exponent of 1.1, while that of the reconstruction is higher and equal to 1.3. The cut-off point is also higher for the reconstructed weight distribution.

\begin{figure}[H]
    \centering
    \includegraphics{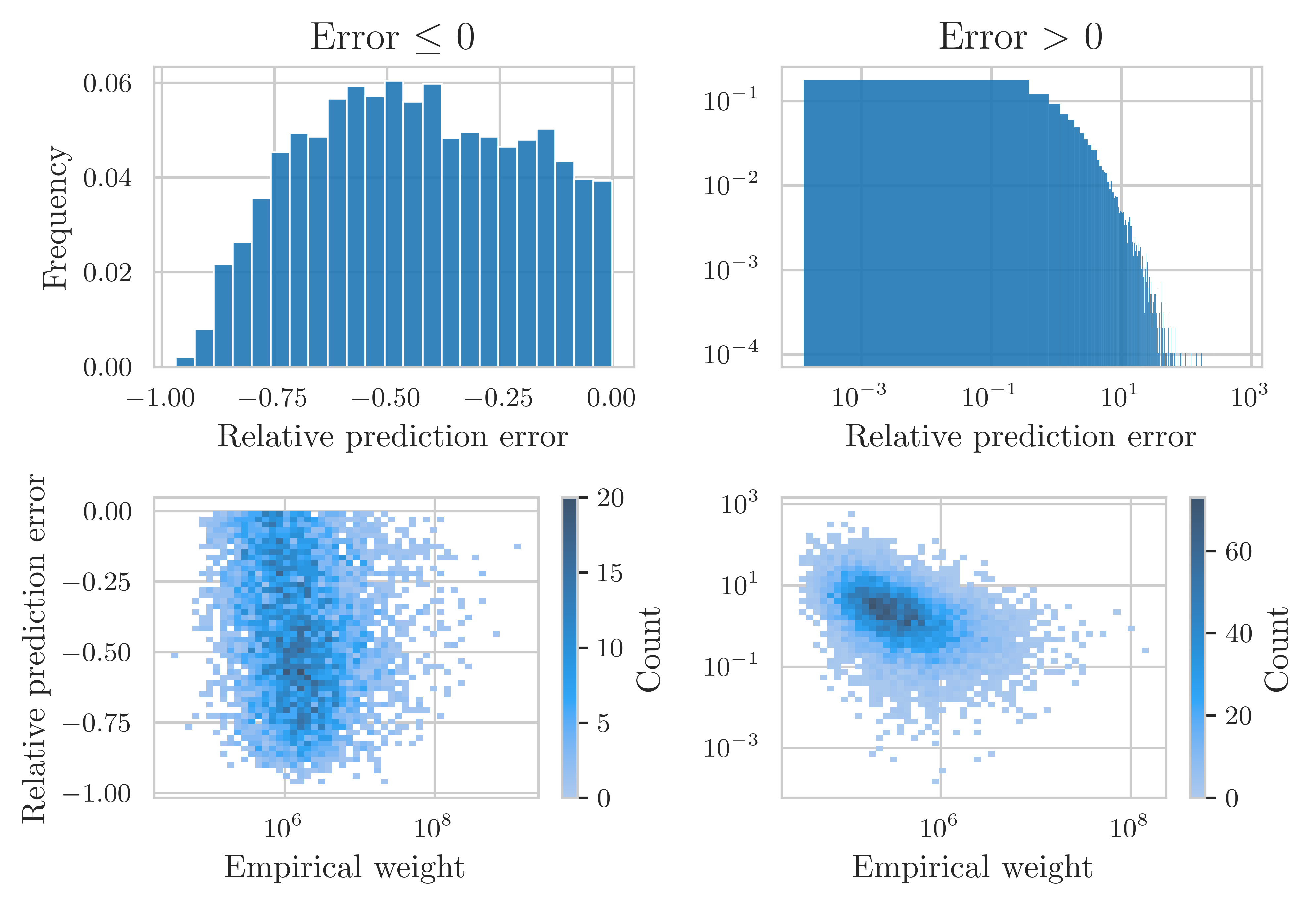}
    \caption{\textbf{Top:} Histogram of the relative prediction errors of the weights for Ecuador. We plot separately the errors that are non-posite (left) and the errors that are positive (right). \textbf{Bottom:} Empirical weights (\textit{x}-axis) against the relative prediction error (\textit{y}-axis) for Ecuador. We plot the error terms that are non-positive on the left and those that are positive on the right. We divide both axes into 50 log-spaced bins and then count the number of data points falling in each square.}
    \label{fig:pred_error_weights_ec}
\end{figure}

The panels in the top row of Figure~\ref{fig:pred_error_weights_ec} show the histograms of the relative prediction errors of Ecuador's weights for one of the randomised reconstructions. The top left panel shows the histogram of the non-positive error terms and the panel on the right the positive errors. The relative prediction error is defined as $\epsilon_i = (\bm{W}_{ij} - \bm{W}_{ij}^*)/ \bm{W}_{ij}^*$, where $\bm{W}_{ij}^*$ is the empirical value and $\bm{W}_{ij}$ is the reconstructed weight. The mean relative prediction error is 285.7\%. Such a high mean is driven by a big outlier, which is 674,065\%; the median is 40\%.

The panels in the bottom row of Figure~\ref{fig:pred_error_weights_ec} show the empirical weights on the \textit{x}-axis and the relative prediction errors on the \textit{y}-axis; we plot non-positive errors on the left and positive errors on the right. The reconstruction method consistently under-predicts weights with values above $\sim 10^8$. 

\begin{figure}[H]
    \centering
    \includegraphics{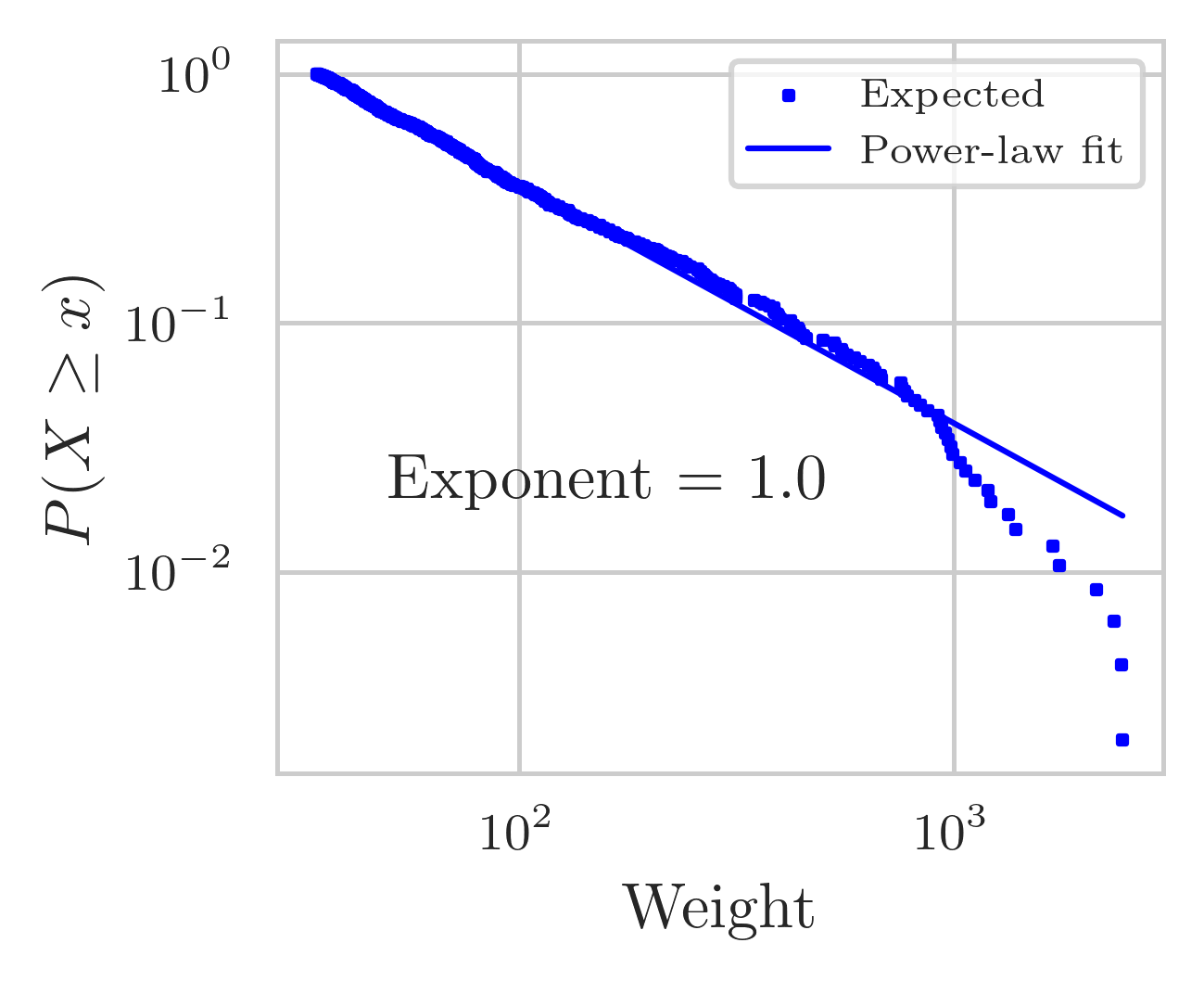}
    \caption{CCDF of the expected weights (blue squares) and its power-law fit (solid blue line) for FactSet.}
    \label{fig:weight_ccdf_factset}
\end{figure}

\FloatBarrier
\paragraph{FactSet.}
Figure~\ref{fig:weight_ccdf_factset} shows the CCDf of the expected weight distribution (blue squares) and its power-law fit (solid blue line) for FactSet. The estimated exponent equals 1.

\FloatBarrier
\subsection{Technical and allocation coefficients}

\begin{figure}[H]
    \centering
    \includegraphics{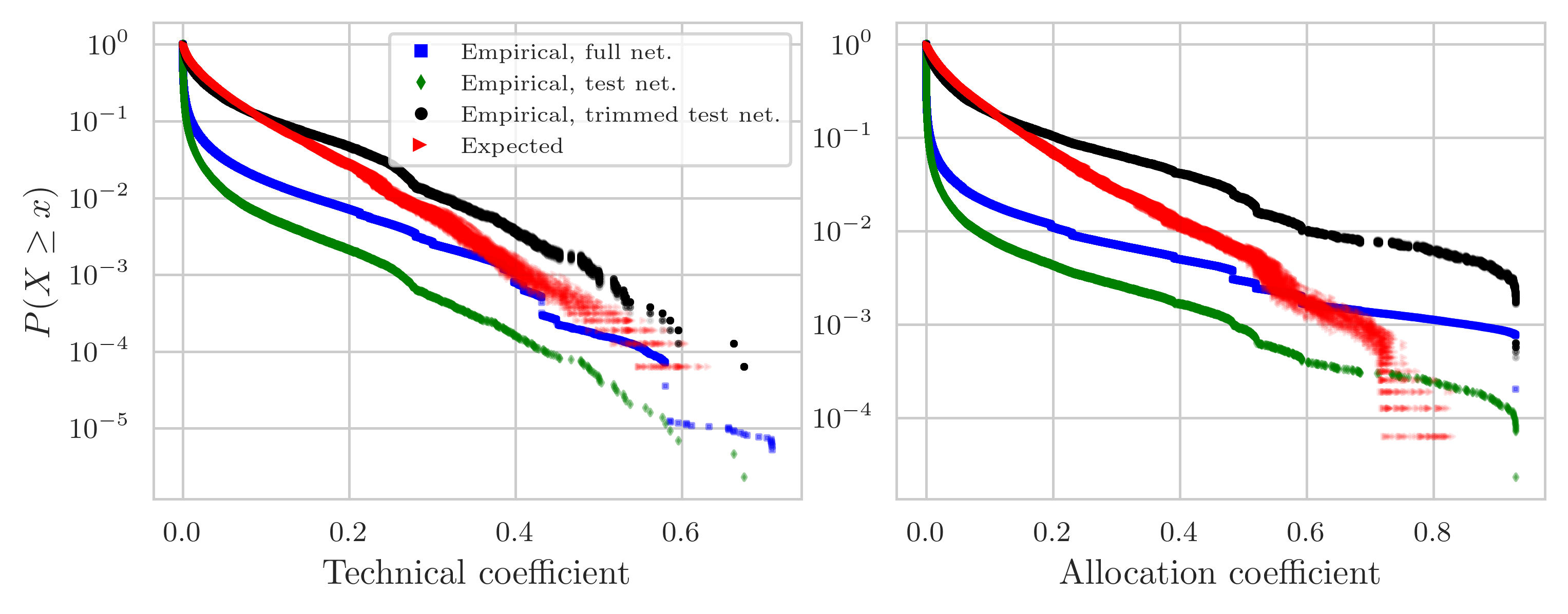}
    \caption{CCDF of the technical (left) and allocation coefficients (right) of the empirical full network (blue squares), the empirical test network (green diamonds), the empirical trimmed test networks (black dots) and of the reconstructed networks (red triangles). We show all 50 randomised networks.}
    \label{fig:coeff_ccdf_ec}
\end{figure}

\paragraph{Ecuador.}
Figure~\ref{fig:coeff_ccdf_ec} shows the distribution of the empirical technical (left) and allocation coefficients (right) of the full network (blue squares), the test network (green diamonds), the trimmed test networks (black dots) and reconstructed networks (red triangles). The three distributions are right-skewed but do not display heavy tails.

\FloatBarrier
\subsection{Multipliers}
\subsubsection{Influence vector}\label{app:infl_vec}

\begin{figure}[H]
    \centering
    \includegraphics{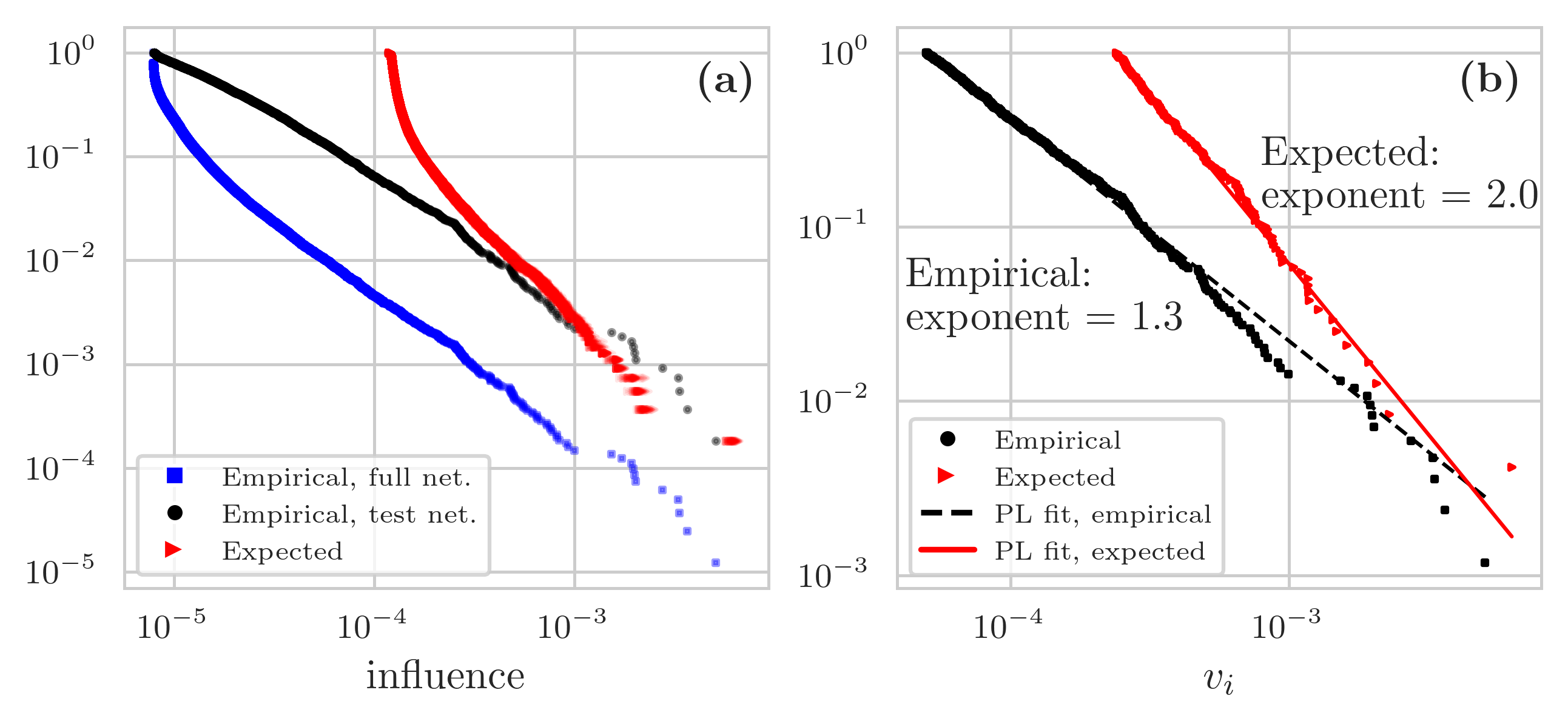}
    \caption{\textbf{(a)} CCDF of the influence vector of the empirical full network (all firms in the Ecuadorian economy; blue squares), the empirical (trimmed) test network (black dots) and the reconstructed networks (red triangles). We show all 50 randomised test networks. \textbf{(b)} Power-law fit to the empirical (black dots) and expected (red triangles) influence vector distribution. The black dashed line shows the power-law fit to the empirical distribution, while the red solid line the power-law fit of the reconstructed distribution; we show results for one of the reconstructions.}
    \label{fig:inflvec_ccdf_ec}
\end{figure}

\paragraph{Ecuador.}
Figure~\ref{fig:inflvec_ccdf_ec}a shows the distribution of the empirical influence vector of the full network (blue squares), our test networks comprising 5,440 firms (black dots) and the reconstructed networks (red triangles). We compute the empirical influence vector for the firms in the (trimmed) test network using the network comprising all firms and then plot the distribution of only the 5,440 firms included in the (trimmed) test network. Therefore, the firms in the trimmed test network and the test network have the same influence. Figure~\ref{fig:inflvec_ccdf_ec}b shows the power-law fit for one of the reconstructions. The empirical distribution has a power-law exponent equal to 1.3, while the expected has a higher exponent, equal to 2. The CCDF of the expected influence vector also has a higher cut-off.

\begin{figure}[H]
    \centering
    \includegraphics{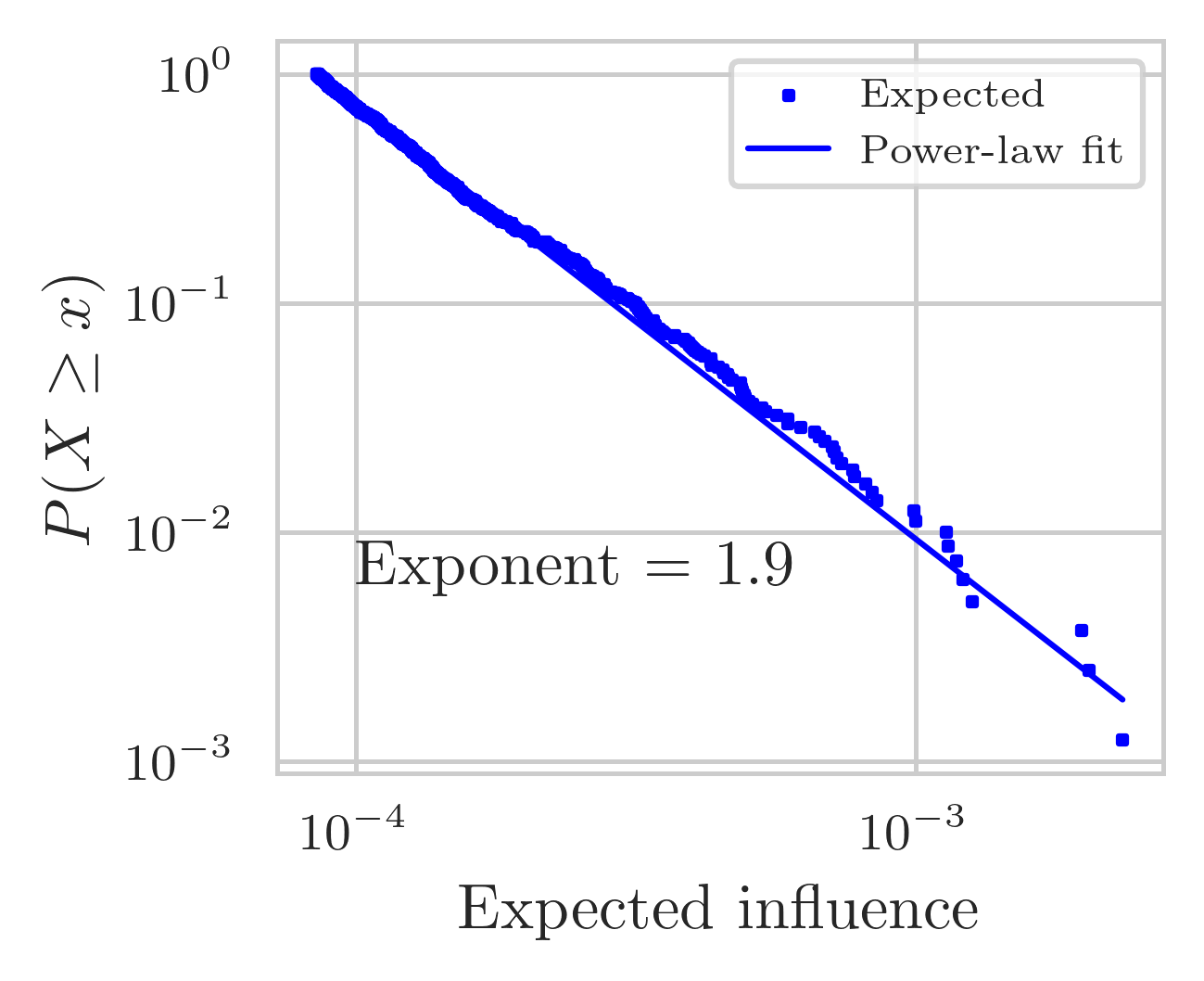}
    \caption{CCDF of the expected influence vector (blue squares) and its power-law fit (solid blue line) for FactSet.}
    \label{fig:inflvec_PLfit_facstet}
\end{figure}

\paragraph{FactSet.}
Figure~\ref{fig:inflvec_PLfit_facstet} shows the distribution of the expected influence vector (blue squares) and its power-law fit (solid blue line) for FactSet. The power-law exponent equals 1.9.

\subsubsection{Prediction errors}\label{app:multiplier_error}

\begin{figure}[H]
    \centering
    \includegraphics{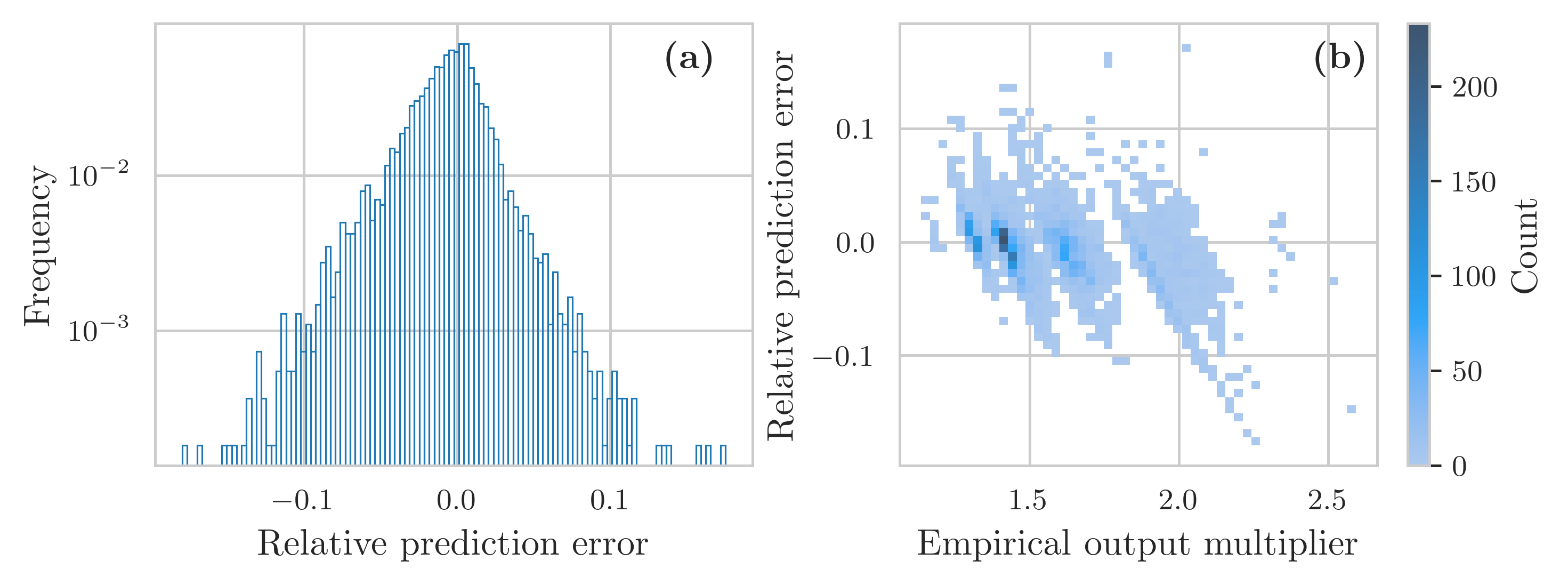}
    \caption{\textbf{(a)} Histogram of the relative prediction errors of the output multipliers for Ecuador. \textbf{(b)} 2D histogram of the empirical output multipliers (\textit{x}-axis) and the prediction errors (\textit{y}-axis). We use 50 log-spaced bins for both axes and count the member of points falling into each square.}
    \label{fig:pred_error_outputM_ec}
\end{figure}

\begin{figure}[H]
    \centering
    \includegraphics{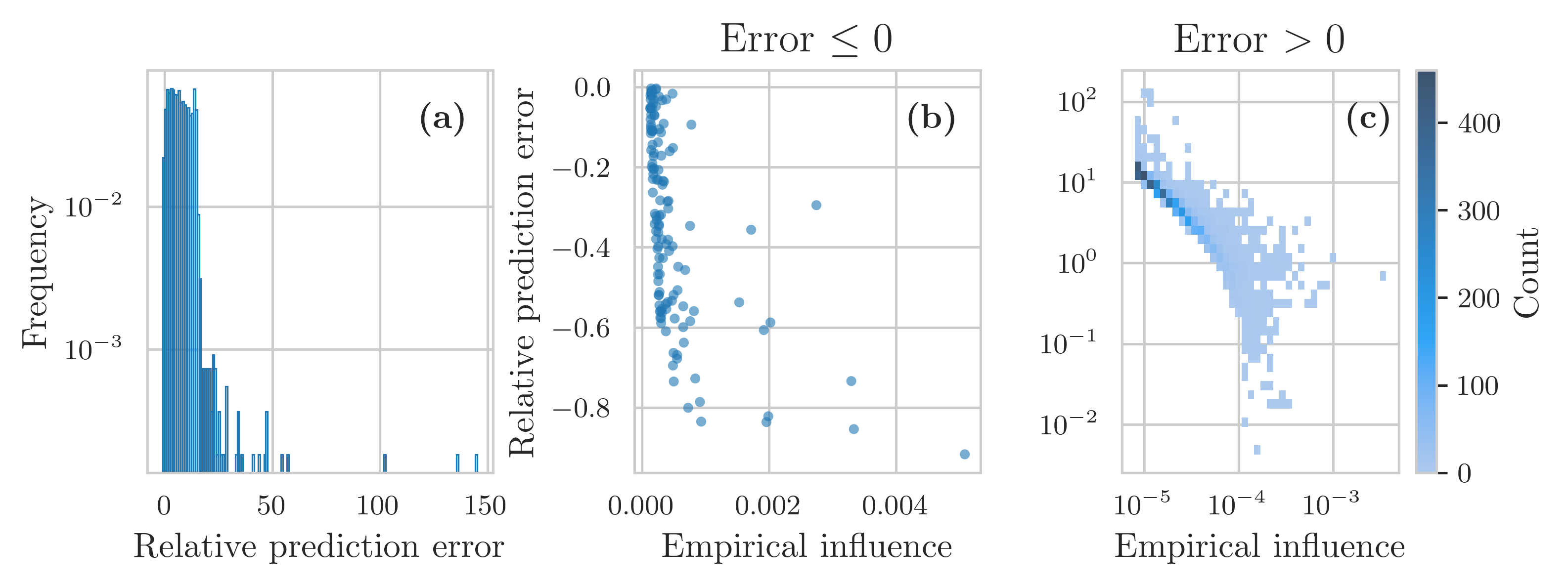}
    \caption{\textbf{(a)} Histogram of the relative prediction errors of the influence vector for Ecuador. \textbf{(b)}, \textbf{(c)} 2D histogram of the empirical influence vector (\textit{x}-axis) and the relative prediction errors (\textit{y}-axis) for errors that are non-positive and positive, respectively. We use 50 log-spaced bins for both axes and count the number of points falling into each square.}
    \label{fig:pred_error_inflVec_ec}
\end{figure}

Figure~\ref{fig:pred_error_outputM_ec}a shows the histogram of the relative prediction errors for the output multipliers, while Figure~\ref{fig:pred_error_inflVec_ec}a for the influence vector, both are for Ecuador and for one of the 50 randomised networks. For the output multipliers, the mean relative prediction error is -0.7\% and the median is -0.4\%. For the influence vector, the mean relative prediction error is 735.1\% and the median is 679.4\%.

Figure~\ref{fig:pred_error_outputM_ec}b shows the empirical output multipliers against the relative prediction errors for Ecuador, while Figure~\ref{fig:pred_error_inflVec_ec}b and~\ref{fig:pred_error_inflVec_ec}c for the influence vector. Figure~\ref{fig:pred_error_inflVec_ec}b shows observations for which the error term is non-positive, while Figure~\ref{fig:pred_error_inflVec_ec}c for observations for which the error term is positive.

\subsubsection{Empirical and expected multipliers by sector}\label{sec:multipliers_sec}

\begin{figure}[H]
    \centering
    \includegraphics{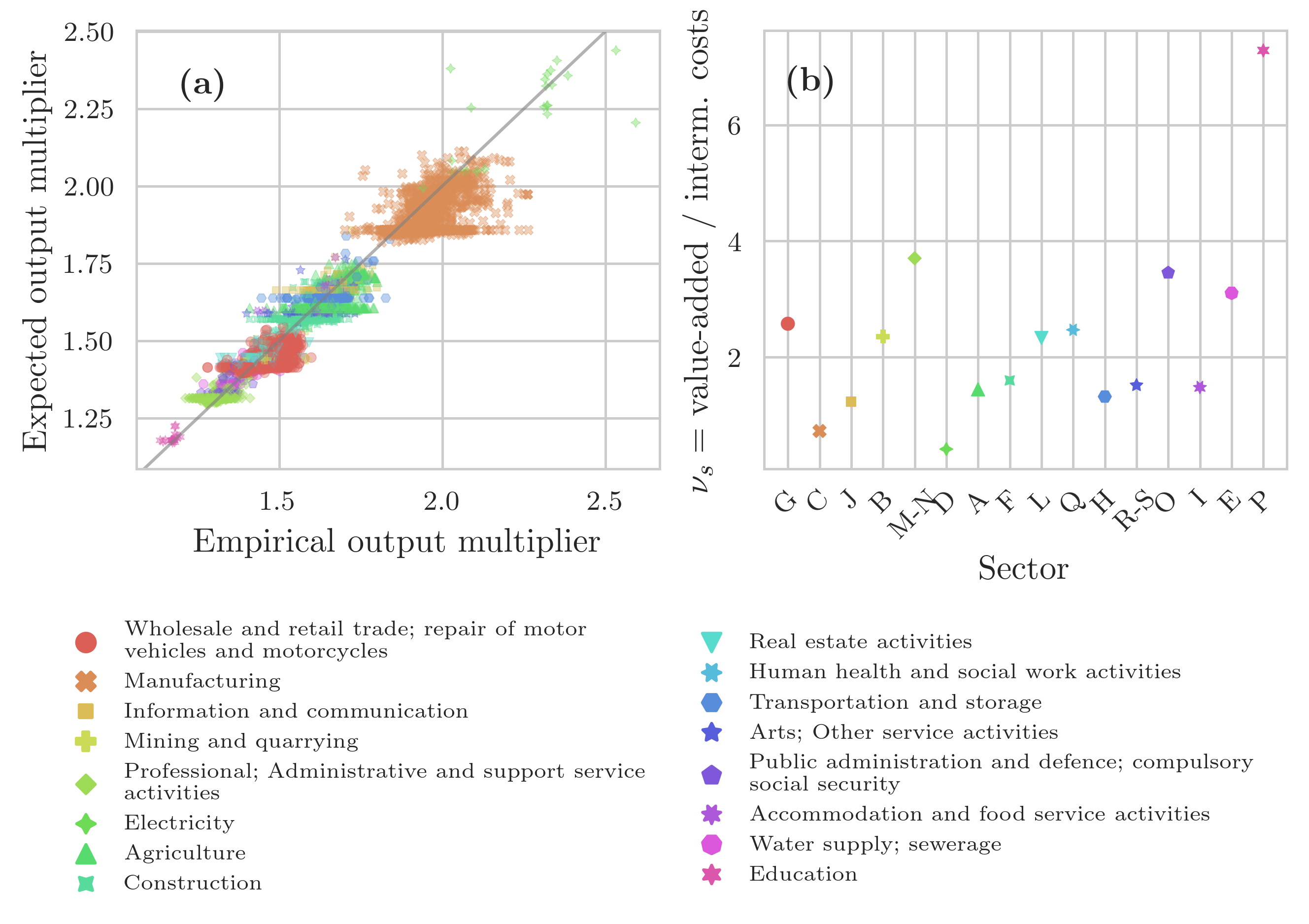}
    \caption{\textbf{(a)} Empirical (\textit{x}-axis) and expected (\textit{y}-axis) output multipliers for Ecuador. Perfect prediction is achieved when all points lie on the 45-degree line (solid grey line). \textbf{(b)} $\nu_s = y_s / s_s^{\text{in}}$, i.e., the ratio of value-added and total intermediate expenses by sector in Ecuador's I-O table. Different colours and shapes correspond to the sectors in the I-O table in which firms are.}
    \label{fig:output_multi_ECSec}
\end{figure}

\paragraph{Output multipliers.} Figure~\ref{fig:output_multi_ECSec} shows the empirical output multipliers on the \textit{x}-axis against the expected output multipliers on the \textit{y}-axis for Ecuador. Points are coloured based on the sector firms are in. Data points tend to organise in clusters, which form because the calculation of the technical coefficients requires knowing the value-added of each firm. Since we did not have access to firms' value-added, we used sector-level I-O tables to infer firms' value-added. The value-added of firm $i$ is $y_i = \nu_s \cdot s_i^{\text{in}}$, where $\nu_s = y_s / s_s^{\text{in}}$ is the ratio of value-added and total intermediate expenses of sector $s$. We show $\nu_s$ in Figure~\ref{fig:output_multi_ECSec}b, also colour-coded by sector.

\begin{figure}[H]
    \centering
    \includegraphics{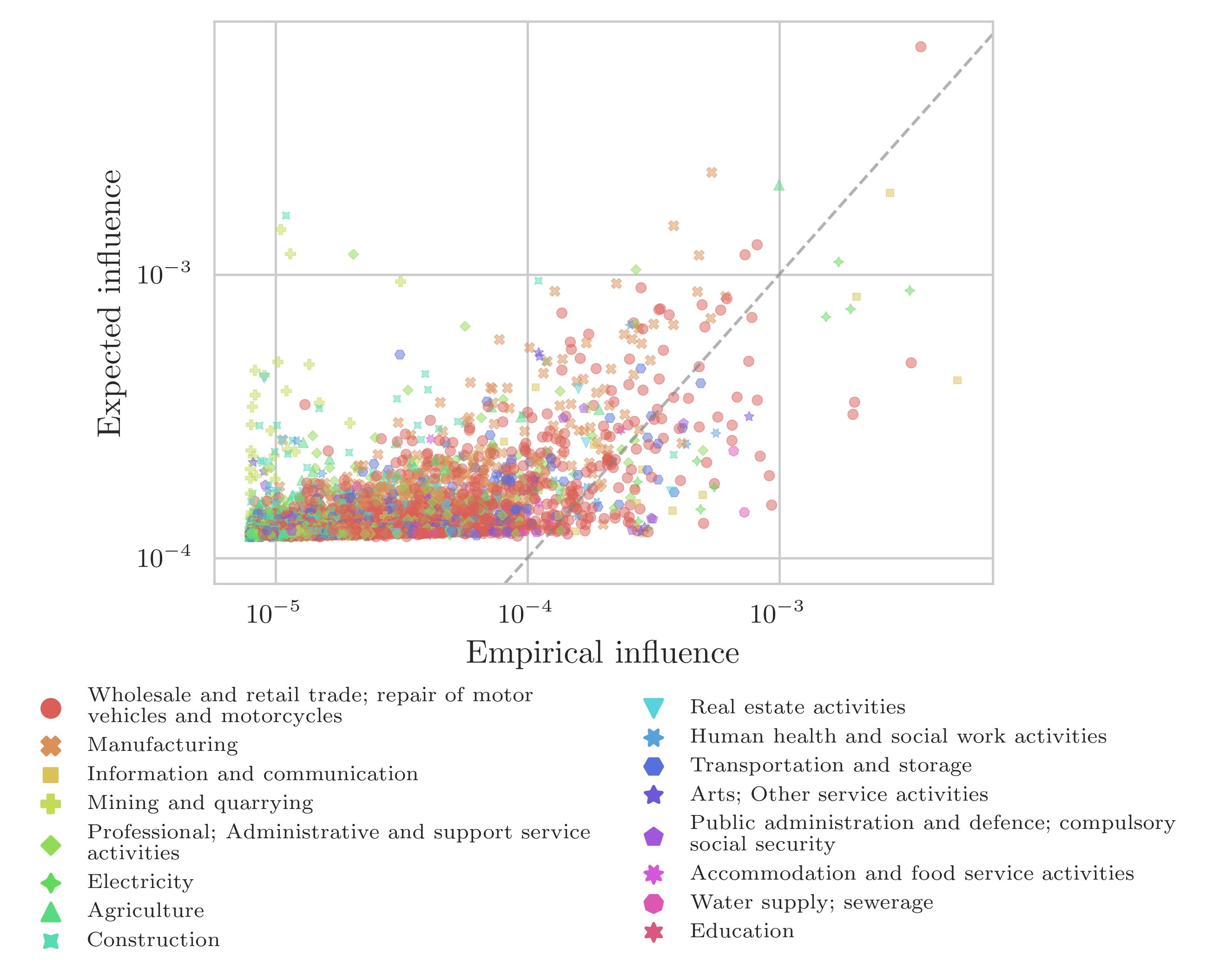}
    \caption{Empirical (\textit{x}-axis) and expected (\textit{y}-axis) influence vector for Ecuador. Perfect prediction is achieved when all points lie on the 45-degree line (dashed grey line). The colours and shapes correspond to different sectors.}
    \label{fig:inflvec_multi_ECSec}
\end{figure}

\paragraph{Influence vector.} Figure~\ref{fig:inflvec_multi_ECSec} shows the empirical influence vector on the \textit{x}-axis against the expected influence vector on the \textit{y}-axis for Ecuador. Points are coloured based on the sectors in the I-O table in which firms are. 

\FloatBarrier
\subsubsection{The effect of the proxy node on the centrality measures}\label{app:centrility_proxy_node}
\begin{figure}[!htbp]
\centering    \includegraphics{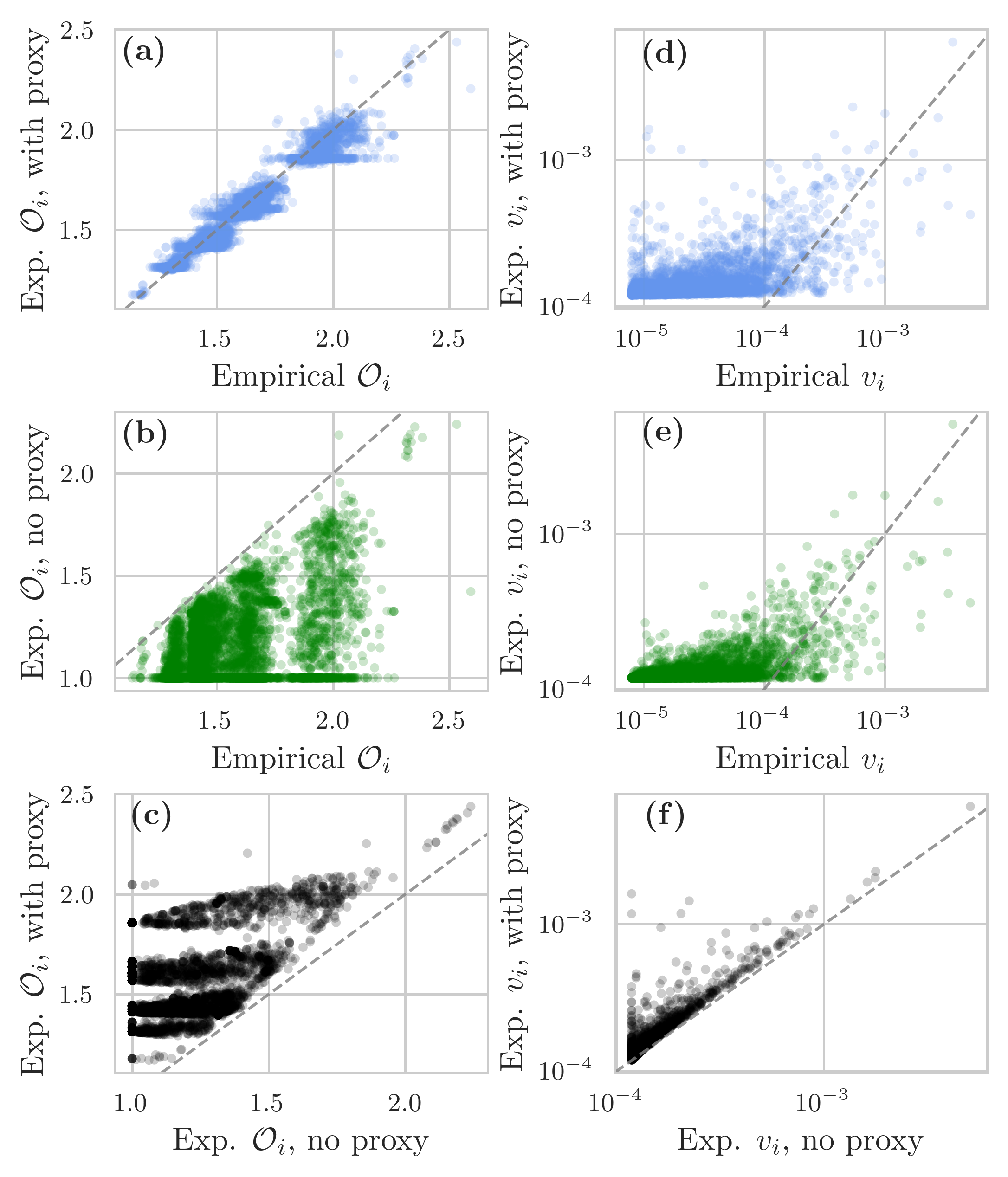}
\caption{\textbf{(a)}, \textbf{(d)} Empirical centrality measure (\textit{x}-axis) and the expected calculated including the proxy node (\textit{y}-axis) for the output multipliers ($\mathcal{O}_i$) and the influence vector ($v_i$), respectively. \textbf{(b)}, \textbf{(e)} Empirical centrality measure (\textit{x}-axis) and the expected calculated excluding the proxy node (\textit{y}-axis) for the output multipliers ($\mathcal{O}_i$) and the influence vector ($v_i$), respectively.  \textbf{(c)}, \textbf{(f)} Expected centrality measure calculated excluding the proxy node (\textit{x}-axis) and the expected centrality measure calculated including the proxy node (\textit{y}-axis) for the output multipliers ($\mathcal{O}_i$) and the influence vector ($v_i$), respectively. The dashed grey line marks the identity line.}
\label{fig:centrility_proxy_node}
\end{figure}

Figure~\ref{fig:centrility_proxy_node}a and Figure~\ref{fig:centrility_proxy_node}d show the empirical multipliers against the reconstructed multipliers including the proxy node, while Figure~\ref{fig:centrility_proxy_node}b and Figure~\ref{fig:centrility_proxy_node}e excluding the proxy node from the calculations. Figure~\ref{fig:centrility_proxy_node}c and Figure~\ref{fig:centrility_proxy_node}f compare the expected multipliers that include to proxy node to those that exclude the proxy node from the calculations. 

The influence vector is not that much affected by the inclusion, or not, of the proxy node (Figure~\ref{fig:centrility_proxy_node}d and Figure~\ref{fig:centrility_proxy_node}e). The inclusion of the proxy node increases the influence of all the nodes compared to when it is not included (Figure~\ref{fig:centrility_proxy_node}f). While the error metrics are only slightly lower when the proxy node is excluded, the cosine similarity is slightly higher (0.56 vs 0.55) and the power-law exponent slightly smaller (2.0 vs 2.1). The mean and median are off for both and we can only recover the variance if we include the proxy node, otherwise it is smaller. The proxy node is always the one with the higher influence.

On the contrary, the output multipliers are highly affected by the inclusion, or not, of the proxy node (Figure~\ref{fig:centrility_proxy_node}a and Figure~\ref{fig:centrility_proxy_node}b). When the proxy node is not included the output multipliers are underestimated (Figure~\ref{fig:centrility_proxy_node}b). The proxy node is never the one with the highest centrality. When the proxy node is included, we can recover the mean, median and standard deviation, while when it is excluded, these are all underestimated (especially the mean and median). 

\FloatBarrier

Before delving into the details of why this is happening, it is useful to define again these two centrality measures.The output multiplier is defined as 
\begin{linenomath}
\begin{equation*}
\mathcal{O} \equiv (\bm{I} - \bm{T}^\top)^{-1} \bm{1}\;,
\end{equation*}
or in scalar form 
\begin{equation*}
\mathcal{O}_j = \sum_i \frac{W_{ij}}{q_j} \mathcal{O}_i + 1\;.
\end{equation*}
\end{linenomath}

\noindent The output multipliers measure the centrality of a node by assigning a centrality score for each incoming edge -- of course, these edges are weighted and normalised by the total cost of the node for which we are measuring the multiplier.  The influence vector is defined as 
\begin{linenomath}
\begin{equation*}
\bm{v} \equiv \frac{\alpha}{N}[\bm{I} - (1 - \alpha)\bm{\Omega}]^{-1}\bm{1}\;, 
\end{equation*}
or in scalar form 
\begin{equation*}
v_i = \sum_j (1 - \alpha) \frac{W_{ij}}{s_j^{in}} v_j + \frac{\alpha}{N}\;. 
\end{equation*}
\end{linenomath}
The influence vector, instead, assigns a centrality score for each node the focal node points to. So nodes that point to more nodes are more central -- again, edges are weighted and normalised by the in-strength of the nodes to which the focal node points to. In both centrality measures, a node can be very influential because it has many incoming or outgoing links with nodes of mild influence, for the output multiplier and the influence vector, respectively, or because the node has few incoming or outgoing links with highly influential nodes, for the output multiplier and the influence vector.

Since the only thing that changes in the calculated multipliers is the inclusion or not of the proxy node, let us focus on the proxy node's role in the multipliers:
\begin{linenomath}
\[
\mathcal{O}_j \sim 1 + \frac{W_{pj}}{q_j} \mathcal{O}_p + \dots\;, \text{and}
\]
\[ 
v_i \sim \frac{\alpha}{N} + (1 - \alpha)\frac{W_{ip}}{s_{p}^{in}} v_p + \dots\;,
\]
\end{linenomath}
where $p$ indexes the proxy node. Each link of the proxy node matters more in the output multipliers since, empirically, $W_{pj}/q_j > (1 - \alpha) W_{ip}/s_{p}^{in}$, for all but the link the proxy node has with itself and with one another node.

\FloatBarrier
\subsubsection{Reconstructed and benchmark influence vector}\label{app:unif_reconstr_influence}
\begin{figure}[!htbp]
    \centering
    \includegraphics{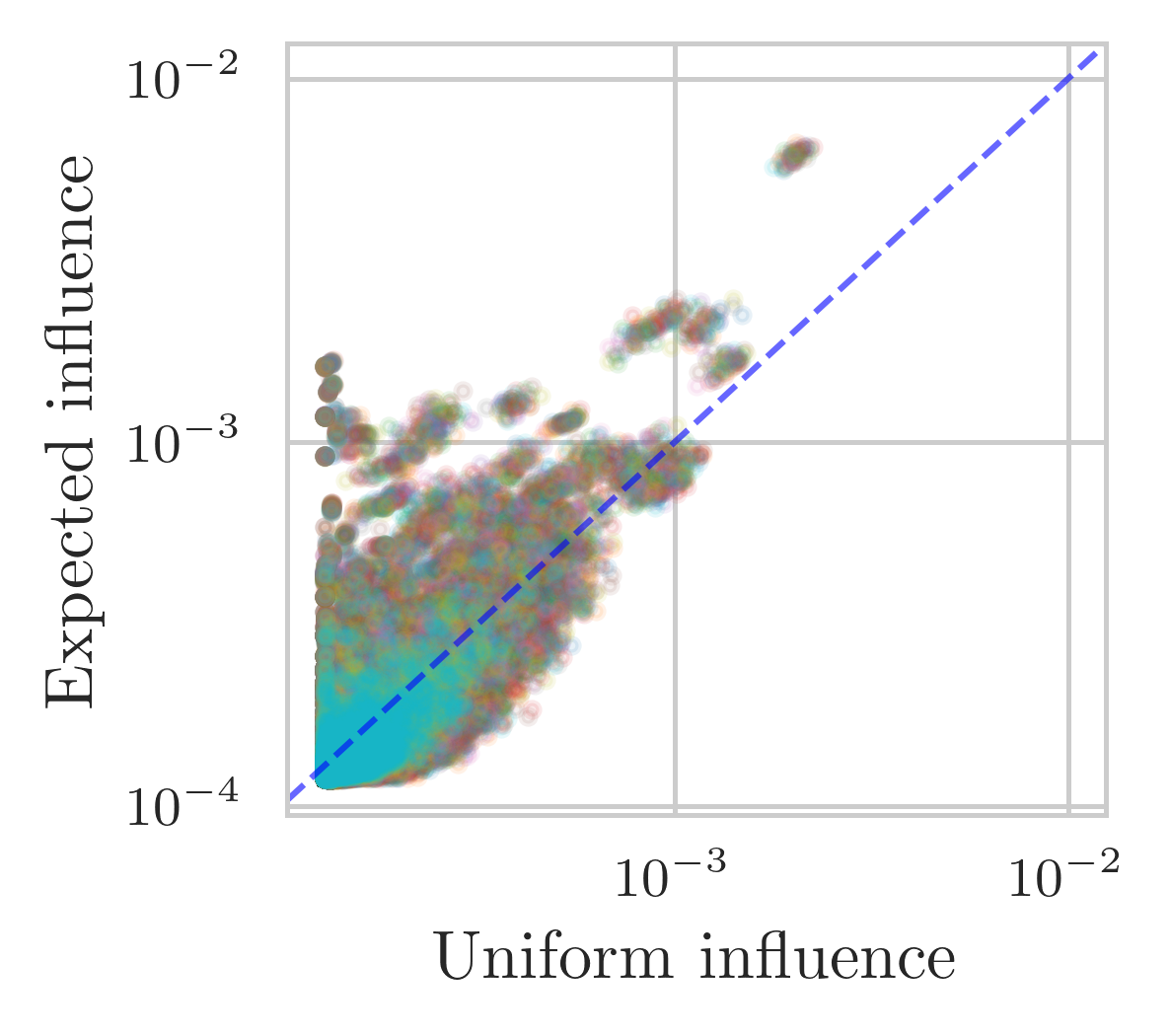}
    \caption{``Uniform'' influence on the \textit{x}-axis and expected influence on the \textit{y}-axis for the 50 randomised networks. The ``uniform'' influence is calculated by assuming that each firm buys inputs from its suppliers in the same proportion; this is the influence vector we use in our benchmark for aggregate fluctuations. The dashed blue line marks the identity line.}
    \label{fig:unif_reconstr_influence}
\end{figure}

Figure~\ref{fig:unif_reconstr_influence} shows the ``uniform'' influence, which we use to benchmark aggregate volatility, against the reconstructed influence vector for the 50 randomised networks. It can be seen that the reconstructed influences tend to be bigger than those of the benchmark. However, the error metrics of the two influence vectors have very similar values, except for the cosine similarity, which is lower for the uniform influence vector (Table~\ref{tab:errors_inflVec_inflVecUniform}).

\begin{table}[!htbp]
\small
\centering
\def\arraystretch{1.1}
\begin{tabular}{l *{3}{c} r}
\toprule
Type & RMSE & MAE & MedAE & Cosine similarity\\ 
\midrule
Influence vector & 2$\times 10^{-4}$ & 1$\times 10^{-4}$ & 1$\times 10^{-4}$ & 0.560 \\
 & (9$\times 10^{-7}$) & (1$\times 10^{-7}$) & (5$\times 10^{-8}$) & (3$\times 10^{-3}$) \\
Uniform influence vector & 2$\times 10^{-4}$ & 1$\times 10^{-4}$ & 1$\times 10^{-4}$ & 0.478 \\
 & (4$\times 10^{-7}$) & (1$\times 10^{-7}$) & (4$\times 10^{-8}$) & (4$\times 10^{-3}$) \\

\bottomrule
\end{tabular}
\caption{Comparison metrics for the reconstructed and uniform influence vector for Ecuador. RMSE denotes the root mean squared error, MAE the mean absolute error and MedAE the median absolute error. For each metric, we show its mean value across the 50 randomised reconstructions. Below the mean value, the standard deviation in parenthesis. We excluded the proxy node from the calculations.}
\label{tab:errors_inflVec_inflVecUniform}
\end{table}

\FloatBarrier
\subsection{Different numbers of unknown links}
\subsubsection{In- and out-degrees}\label{app:in-out-degs}
\begin{figure}[!htbp]
    \centering
    \includegraphics{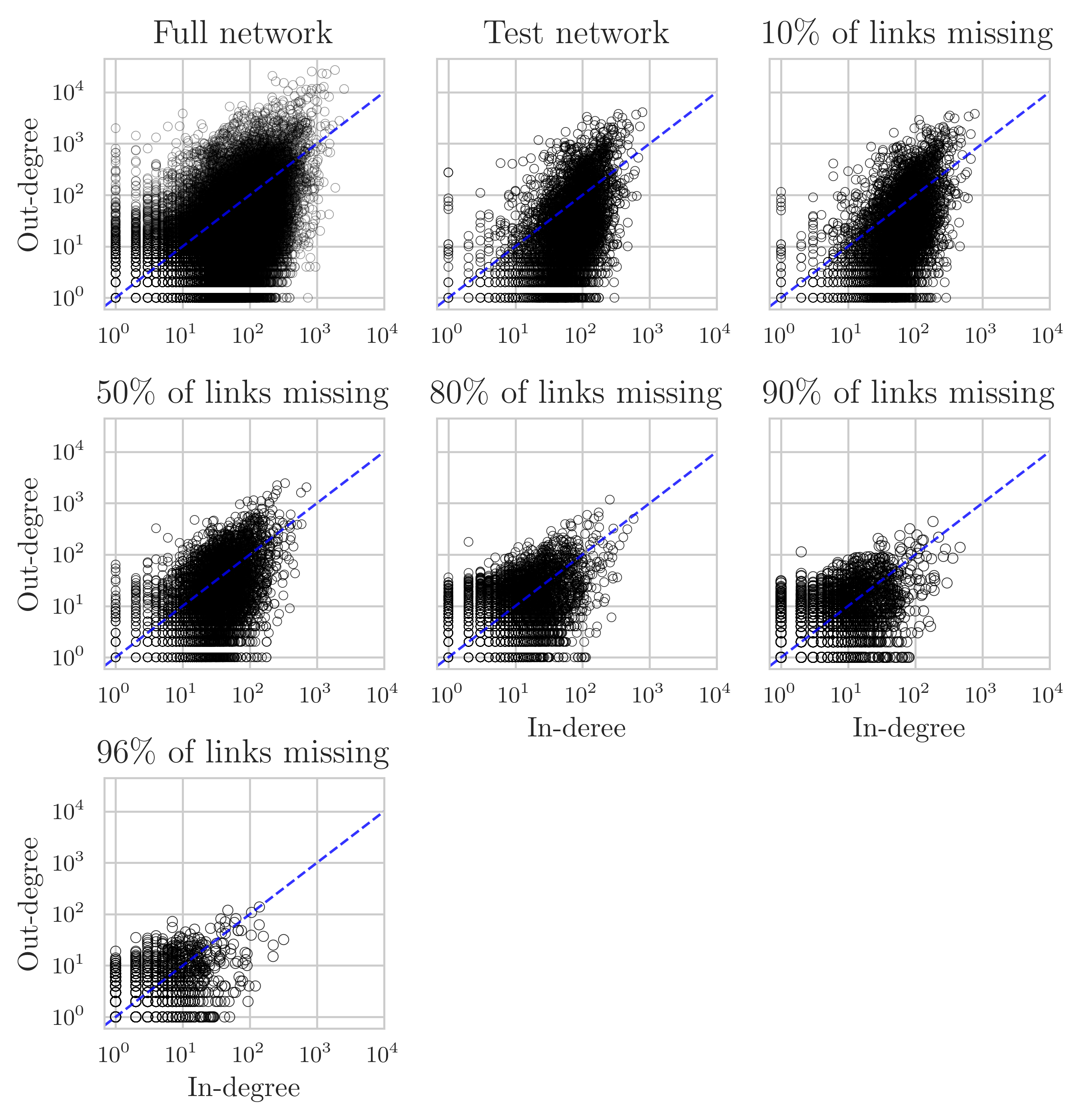}
    \caption{Number of suppliers (in-degree) on \textit{x}-axis and number of customers (out-degree) on \textit{y}-axis for the full network, the test network and for test networks with different numbers of unknown links. The blue dashed line marks the identity line.}
    \label{fig:in-out-degs}
\end{figure}

Figure~\ref{fig:in-out-degs} shows the number of suppliers (in-degree) against the number of customers (out-degree) for different networks. The first panel on the top left-most corner shows the in- and out-degrees for the full network, the second panel for the test network and the consecutive panels for test networks with different numbers of unknown links. The last panel, at the bottom right-most corner, shows results for the trimmed test network that we discuss throughout the paper, which has the same average degree as FactSet. It can be noticed that in the full network firms tend to have more customers than suppliers, something also observed by \cite{bacil2022emprical}. In creating our test network, firms lose more customers than suppliers. The maximum out-degree is 27,030 in the full network and decreases almost 7 folds in the test network (96\% of missing links), reaching a value of 4,081, while the maximum in-degree is 2,543 in the full network and lowers to 781 in the test network (96\% of missing links), which is a 3-fold decrease. 

\FloatBarrier
\subsubsection{Comparison of empirical and expected quantities}\label{app:comparison_Nlinks_unknown}

\begin{figure}[!htbp]
    \centering
    \includegraphics{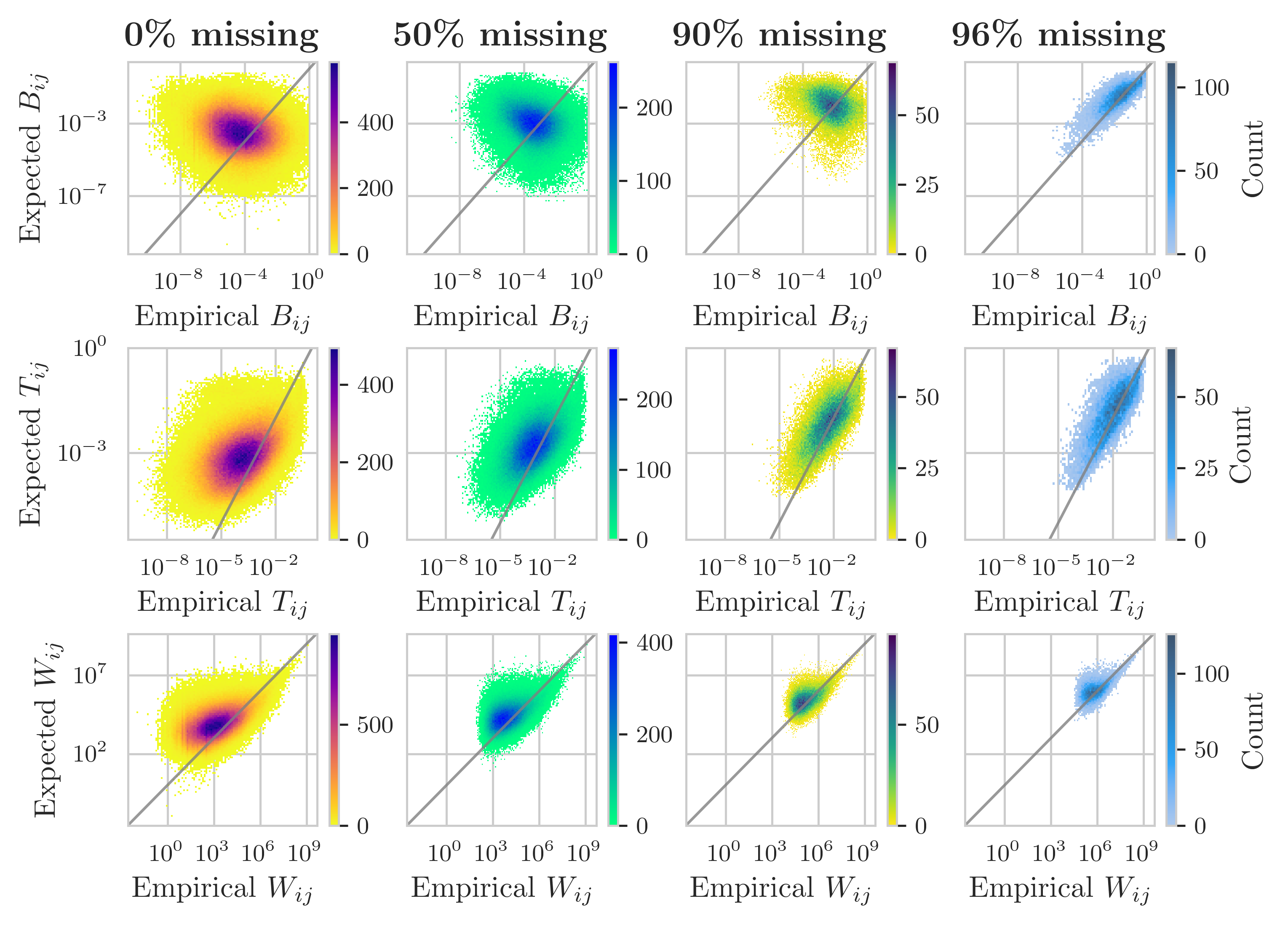}
    \caption{2D histograms for the empirical values on the \textit{x}-axis and the expected values on the \textit{y}-axis for different numbers of unknown links (0\%, 50\%, 90\% and 96\%) for the allocation coefficients ($B_{ij}$), the technical coefficients ($T_{ij}$) and the weights ($W_{ij}$). We bin each axis into 100 log-spaced bins and count the number of data points that fall in each square.}
    \label{fig:scatters_differentUnkns_micro}
\end{figure}

\paragraph{Microscale quantities.}
Figure~\ref{fig:scatters_differentUnkns_micro} shows the empirical values against the expected values for the allocation coefficients (top panels), the technical coefficients (2$^{nd}$ row), and the weights (bottom panels) for different numbers of unknown links. The allocation coefficients display a transition when the number of unknown links rises from 90\% to 96\%, something that the other two quantities do not experience. It also highlights the sampling process used for the deletion of links: as the number of unknown links increases, higher and higher weights are eliminated. However, the shape of the joint density does not vary, it is only re-scaled.

\begin{figure}[!htbp]
    \centering
    \includegraphics{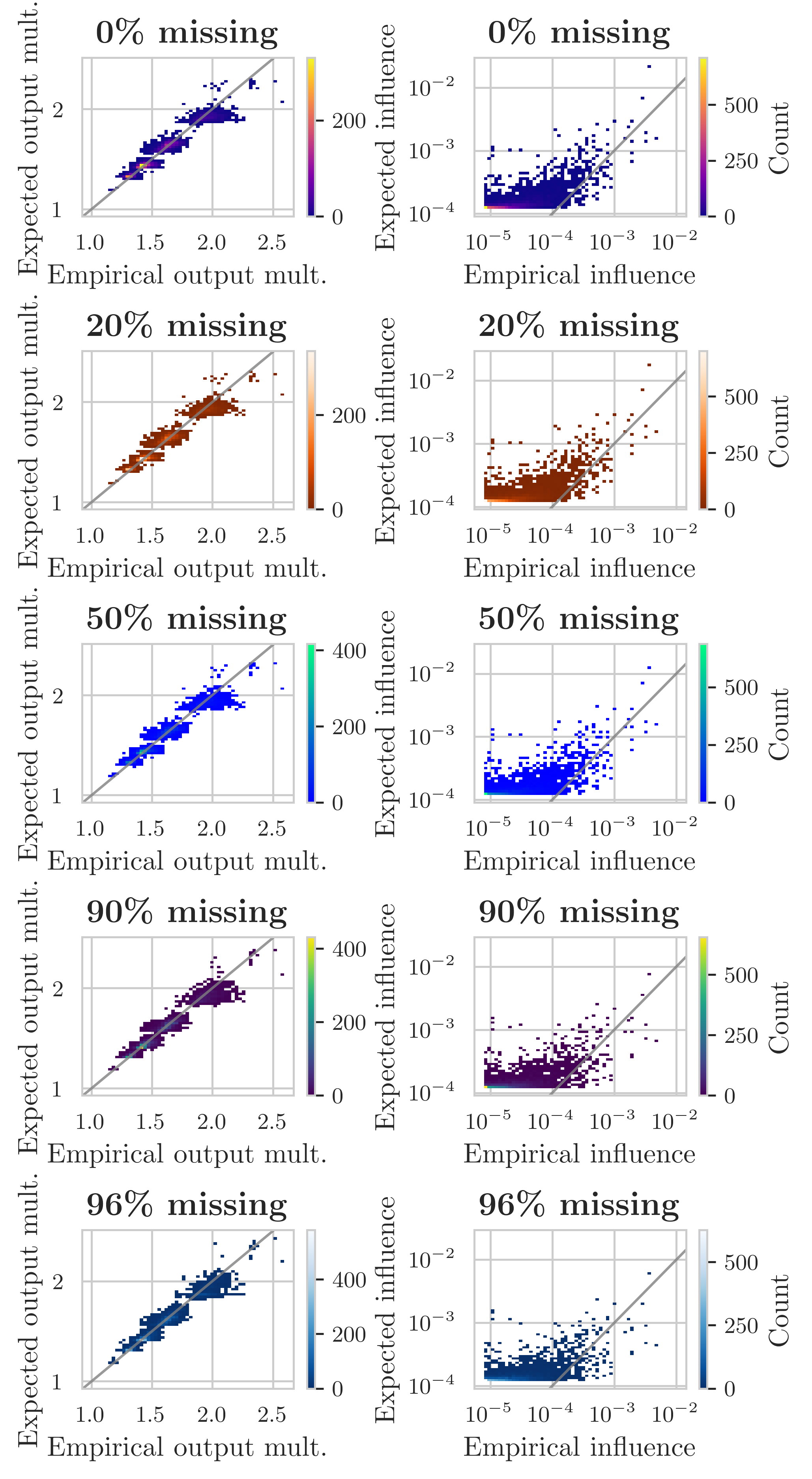}
    \caption{2D histograms for the empirical values on the \textit{x}-axis and the expected values on the \textit{y}-axis for different numbers of unknown links (0\%, 20\%, 50\%, 90\% and 96\%) for the output multipliers (left column) and the influence vector (right column). We bin each axis into 50 log-spaced bins and count the number of data points that fall in each square.}
    \label{fig:scatters_differentUnkns_higherOr}
\end{figure}

\begin{figure}[!htbp]
    \centering
    \includegraphics{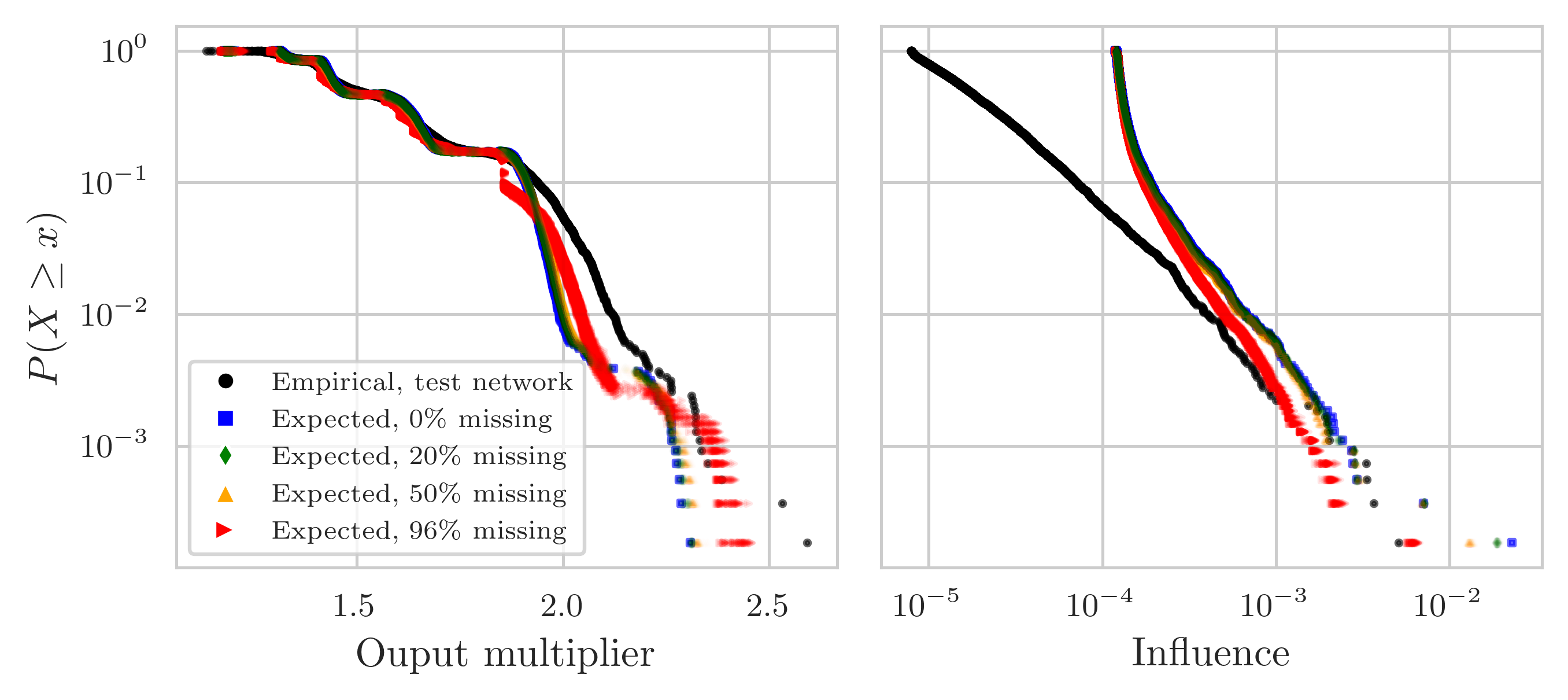}
    \caption{CCDF of the output multipliers (left) and influence vector (right) for different numbers of unknown links (0\% blue squares, 20\% green diamonds, 50\% yellow triangles and 96\% red triangles) and the empirical (black dots).}
    \label{fig:CCDF_differentUnkns_higherOr}
\end{figure}

\FloatBarrier
\paragraph{Higher-order quantities.}
Figure~\ref{fig:scatters_differentUnkns_higherOr} shows the empirical values against the expected values for the output multipliers (left column) and the influence vector (right column) for different numbers of unknown links (0\%, 20\%, 50\%, 90\% and 96\%). Figure~\ref{fig:CCDF_differentUnkns_higherOr} shows the empirical CCDF (black dots) for the output multipliers (left) and the influence vector (right). It also shows the CCDF of the reconstructed multipliers for different numbers of unknown links and for the 50 randomised networks.

\end{appendices}

\end{document}